\documentclass[10pt,a4paper]{article} 
\pdfoutput=1
\usepackage{jcappub}

\allowdisplaybreaks

\usepackage{hyperref}
\usepackage{ifthen}
\usepackage{amsmath}
\usepackage{amssymb}
\usepackage{color}
\usepackage{graphicx}

\newcommand*{\VEC}[1]  {\textbf{#1}}
\newcommand*{\pp}  {\parallel}

\newcommand*{\df}  {\delta}
\newcommand*{\tf}  {\theta}

\begin{document}

\title{Distribution function approach to redshift space distortions. Part IV: perturbation theory applied to dark matter}
\author[a]{Zvonimir Vlah,}
\emailAdd{zvlah@physik.uzh.ch}

\author[a,b,c,d]{Uro\v{s} Seljak,} 
\emailAdd{seljak@physik.uzh.ch}
\author[c,e]{Patrick McDonald,}
\author[d]{Teppei Okumura,} 
\author[a]{and Tobias Baldauf} 

\affiliation[a]{Institute for Theoretical Physics, University of Z\"{u}rich, Winterthurerstrasse 190, Z\"{u}rich, Switzerland}
\affiliation[b]{Physics, Astronomy Department, University of California, Berkeley, California, USA.} 
\affiliation[c]{Lawrence Berkeley National Laboratory, Berkeley, CA, USA}
\affiliation[d]{Institute for Early Universe, Ewha University, Seoul, S. Korea}
\affiliation[e]{Physics Dept., Brookhaven National Laboratory, Upton, NY, USA}



\abstract{
  We develop a perturbative approach to redshift space distortions (RSD)
 using the phase space
  distribution function approach 
  and apply it to the dark matter redshift space power spectrum and its moments. 
  RSD can be written as a sum over density weighted velocity moments 
correlators, 
  with the lowest order being density, momentum density and
  stress energy density. 
  We use standard and extended perturbation theory (PT) to determine their auto and cross correlators, 
  comparing them to N-body simulations. 
  We show which of the terms can be modeled well with the standard PT and which
need additional terms that include higher order corrections which cannot be 
modeled in PT. Most of these additional terms are related to the small scale 
velocity dispersion effects, the so called finger of god (FoG) effects,
which affect some, but not all, of the terms in this expansion, and which can 
be approximately 
modeled using a simple physically motivated ansatz such as the halo model. 
We point out that there are several velocity dispersions that enter into the
detailed RSD analysis with very different amplitudes, which 
can be approximately predicted by the halo model. 
In contrast to previous models 
our approach systematically includes all of the terms
at a given order in PT and provides a physical interpretation for the small scale dispersion values. 
  We investigate RSD power spectrum as a function of 
$\mu$, the cosine of the angle between the Fourier mode and line of sight, 
  focusing on the lowest order powers of $\mu$ and multipole moments
which dominate the observable RSD power spectrum. 
Overall we find considerable success in modeling many, but not all, of the 
terms in this expansion. 
This is similar 
to the situation in real space, but predicting power spectrum in redshift space
is more difficult because of the explicit influence of small scale dispersion 
type effects in RSD, which extend to very large scales. 

}


\keywords{cosmological perturbation theory, power spectrum, redshift surveys}
\arxivnumber{1207.0839}


\maketitle

\section{Introduction}
\label{sec:intro}

Galaxy clustering surveys are one of the most important venues to extract cosmological 
information today. This is because by measuring the 3 dimensional distribution
of galaxies we can in principle relate it to the 3 dimensional distribution of 
the underlying dark matter. The dark matter distribution is sensitive to many 
of 
the cosmological parameters. The growth of dark matter structures in time also 
provides important 
constraints on the models, such as the nature and amount of dark energy. 

Since galaxies are not perfect tracers of dark matter, their clustering 
is biased relative to the dark matter. This means that galaxy surveys cannot 
determine 
the rate of growth of structure unless this biasing is determined. Fortunately, galaxy redshift 
surveys provide additional information, because the observed redshift is a sum 
of the 
radial distance to the galaxy and its peculiar velocity (Doppler shift). 
Galaxies are expected to follow the 
same gravitational potential as the dark matter and thus they are
expected to have the same velocity (in a large-scale average at least). 
This leads to a clustering strength that depends on the angle between the 
galaxy pairs and the 
line of sight, which is 
referred to as redshift space distortions (RSD). In linear theory it can be 
easily related to the dark matter clustering \cite{Kaiser:1987qv,Hamilton:1997zq}. 
These distortions thus make the galaxy clustering in redshift space more 
complex, but at the same
time provide an opportunity to extract important information on the
dark matter clustering directly from the redshift surveys. To what extent this 
is possible is a matter of considerable debate: there are significant nonlinear 
effects that spoil this simple picture, once one goes beyond very large scales, 
as will also be seen in this paper. 

It is worth pursuing how far we can understand RSD for the simple reason that 
RSD offer a unique way to measure growth rate of structure formation \cite{Cole:1993kh}, 
and also can provide tests of dark energy models and general relativity 
\cite{White:2008jy, McDonald:2008sh, Bernstein:2011ju}.  
Generically, if one had a good understanding of the nonlinear effects,
RSD would be the most powerful technique for these studies because redshift surveys provide 
3-dimensional information, while other methods, such as weak lensing, provide 2-dimensional 
information (or slightly more if the so-called tomographic information is used \cite{Amara:2006kp,Casarini:2012qj}). The most
problematic part of RSD studies are the nonlinear effects, which 
have proved to be difficult to model, and which can extend to rather
large scales, making their modeling essential for using the RSD as a tool.

In recent years several studies have been performed investigating these effects 
\cite{Scoccimarro:2004tg, Taruya:2010mx, Jennings:2010uv, Tang:2011qj}.
Some of these studies included galaxies or halos,
\cite{Tinker:2006dm, Nishimichi:2011jm, Reid:2011ar, Sato:2011qr}.
Some of these methods use analysis and modeling based
on perturbation theory (PT) \cite{Bernardeau:2001qr}, but none attempt to 
rely entirely on PT to explain all of the effects. Instead, they rely on 
ansatzes with free
parameters, so that if the ansatz are accurate one can model the
effects accurately. 
Separately, there have been several approaches trying to improve perturbation
methods and to increase their ranges of validity 
\cite{Crocce:2005xy, Crocce:2005xz, Crocce:2007dt, Matsubara:2007wj,
  Matsubara:2008wx, McDonald:2006hf, Taruya:2007xy, Pietroni:2008jx,
  Valageas:2003gm, Taruya:2009ir}. 
All of these approaches adopt a single stream approximation, 
which we know breaks down on small scales inside the virialized halos and 
which is particularly problematic for modeling of RSD.

The goal of this paper is to present a systematic PT approach to all of the 
lowest order terms contributing to RSD. Our goal is to
identify which can be modeled well with PT, which can 
be modeled with extended PT methods mentioned above, and which require 
phenomenological 
additions to account for the small scale physics which cannot be modeled with 
traditional PT that does not include velocity dispersion. 
This approach is enabled by the recently developed 
distribution function approach to RSD \cite{Seljak:2011tx}, 
which decomposes RSD contributions into separate correlations between moments of distribution 
function. As such it allows us to investigate individual contributions to RSD and develop 
different PT or other approximation schemes for these terms. Whether this is ultimately 
useful for modeling RSD remains to be seen: our primary goal is to develop better physical
understanding of dominant contributions to RSD. 

The paper is organized as follows: we begin in Sec.~\ref{sec:model} by presenting a more 
detailed derivation of the angular decomposition 
of redshift space power spectra than given in \cite{Seljak:2011tx}. 
We then use in Sec.~\ref{sec:terms} the perturbative methods to model the lowest contributing terms in
this expansion, augmented by simple phenomenological models and/or beyond the 
lowest order contributions to improve the model when necessary. 
Results are also compared to numerical
simulation measurements presented in \cite{Okumura:2011pb}.
We summarize and conclude in Sec.~\ref{sec:conclusion}. 
In Appendices \ref{sec:App1}, \ref{sec:App2}, \ref{sec:App3}, \ref{sec:App4} we show some details 
of the calculations and write explicit forms of the terms contributing to the power spectra.

For this work, flat $\Lambda$CDM model is assumed 
$\Omega_{\rm m}=0.279$, 
$\Omega_{\Lambda}=0.721$, $\Omega_{\rm b}/\Omega_{\rm m}=0.165$, $h=0.701$,
$n_s=0.96$, $\sigma_8=0.807$. The primordial density field is generated using the matter transfer
function by CAMB. The positions and velocities of all the dark matter
particles are given at the redshifts $z=0,~0.509,~0.989$, and 2.070, which are for simplicity
quoted as $z=$0, 0.5, 1, and 2.

\section{Redshift-space distortions form the distribution function }
\label{sec:model}

\subsection{Generation of velocity moments}
\label{subsec:generation_of_velocity_moments}

Evolution of collisionless particles is described by the Vlasov equation ~\cite{Peebles:1994xt}
\begin{align}
    \frac{df}{d\tau}=\frac{\partial f}{\partial\tau}+\frac{\VEC{p}}{am}\cdot \VEC{$\nabla_x$}f-am\VEC{$\nabla$}\phi \cdot \VEC{$\nabla_p$}f = 0,
\label{ch1:eq1}
\end{align}
where the gravitational potential $\phi$ is given by
\begin{align}
    \nabla_x^2\phi=4\pi Ga^2 \bar{\rho}\df=\frac{3}{2}\mathcal{H}^2\Omega_{m}\df.
\end{align}
Here $f(\VEC{x},\VEC{p},\tau)$ is the particle distribution function
at a phase space point $(\VEC{x},\VEC{p})$, where $\VEC{x} $ is
the comoving position, $\VEC{p}$ is the corresponding canonical particle momentum defined by $\frac{d\VEC{p}}{d\tau}=-am\VEC{$\nabla$} \phi$.
$\tau=\int dt/a$ is the conformal time, $m$ is the particle
mass, and $ \mathcal{H}\equiv d\text{ln}a/d\tau= Ha$ is the conformal
expansion rate, where $H$ is the Hubble parameter. 

Note that in this paper we will use the canonical momentum $\VEC{p}$ rather then 
comoving $\VEC{q}=\VEC{p}/a$ defined in \cite{Seljak:2011tx}. The reason is that the comoving 
momenta $\VEC{q}$ is not the canonical coordinate to comoving position $\VEC{x}$, and this would lead to 
additional terms in the Vlasov equation (because of coordinate transformations), i.e. taking 
corresponding q-moments of usual form of Vlasov equation \ref{ch1:eq1}  would not give the standard form of continuity equation, 
Euler equation, and higher moment equations. This is not a inconvenience when the symmetries are to 
be considered, but in order to avoid this we will use the canonical momenta $\VEC{p}$. 

In the following we will drop explicitly writing the time dependence, i.e we will
write $f(\VEC{x},\VEC{p})$. The density field in real space is obtained by integrating the distribution function over the momentum space
\begin{align}
    \rho(\VEC{x})\equiv m a^{-3} \int{d^3pf(\VEC{x},\VEC{p})},
\label{eq:density}
\end{align}
and mean (bulk) velocity $\VEC{v}$ and higher moment fields can be similarly obtained by multiplying the distribution 
function by corresponding number of particle momentum $\VEC{p}=am\VEC{u}$ ($\VEC{u}$ is here a particle peculiar velocity) and then integrating over it.
The mean velocity field of a particles is then given by
\begin{align}
    \VEC{v}(\VEC{x})\equiv\frac{\int{d^3p\frac{\VEC{p}}{m a}f(\VEC{x},\VEC{p})}}{\int{d^3p f(\VEC{x},\VEC{p})}},
\label{eq:velocity}
\end{align}
and the velocity dispersion tensor is
\begin{align}
    \sigma^{ij}(\VEC{x})\equiv\frac{\int{d^3p\frac{p_i p_j}{m^2a^2}f(\VEC{x},\VEC{p})}}{\int{d^3p f(\VEC{x},\VEC{p})}}-v^iv^j,
\label{eq:dispersion}
\end{align}
i.e. $\sigma^{ij}(\VEC{x})\equiv\left\langle \partial v^i\partial v^j\right\rangle_p$ with $\partial v^i$ being the deviation of a particle's velocity
from the local mean velocity, and the average is taken over all particles at position $\VEC{x}$. 
Note the difference between the particle velocity $\VEC{u}$ and mean velocity $\VEC{v}$. 
The first one is the velocity of a single particle that corresponds to the canonical momentum $\VEC{p}$, which is one 
coordinate in the phase space. On the other hand $\VEC{v}$ is a field defined at every coordinate $\VEC{x}$ and 
is averaged over all the phase space.
In the similar way higher order moments can also be considered.

Taking a arbitrary constant unit vector $\VEC{h}$, we can construct a following
object
\begin{align}
    T^L_\VEC{h}(\VEC{x})\equiv\frac{ma^{-3}}{\bar{\rho}}\int{d^3pf(\VEC{x},\VEC{p})\left( \frac{\VEC{h}\cdot\VEC{p}}{ma}\right)^L},
\label{eq:TLa}
\end{align}
i.e. velocity moments projected on the direction of vector $\VEC{h}$, and  where $\bar{\rho}$ is
the mean mass density. If we introduce approximations in which we neglect velocity dispersion and anisotropic stress, i.e. we
neglect all the contributions from this second rank stress tensor, and similar higher rank tensors ($\sigma^{ij}=0,\ldots$) it can be shown (App.~\ref{sec:App1}) that
\ref{eq:TLa} is reduced to
\begin{align}
    T^L_\VEC{h}(\VEC{x})=\left(1+\df(\VEC{x})\right)\left( \VEC{h}\cdot\VEC{v}(\VEC{x})\right)^L,
\label{eq:TLaArox}
\end{align}
where $\df$ is a usual overdensity field ($\df\equiv\rho/\bar{\rho}-1$).

In this paper we omit the following Fourier transform ($\mathcal{F}$) 
conventions
\begin{align}
    &\tilde{f}(\VEC{k})=\mathcal{F}\left[f(\VEC{x})\right](\VEC{k})=
\int{ d^3x ~\text{exp}(i\VEC{k}\cdot\VEC{x})f(\VEC{x})},\nonumber\\
    &f(\VEC{x})=\mathcal{F}^{-1}\left[\tilde{f}(\VEC{k})\right](\VEC{x})=
\int{\frac{d^3k}{(2\pi)^3}
            ~\text{exp}(-i\VEC{k}\cdot\VEC{x})\tilde{f}(\VEC{k})}.
\end{align}

\subsection{Redshift-space distortions}
\label{subsec:redshift_space_distortions}

In redshift space the position of a particle is distorted by its peculiar velocity, thus the comoving redshift-space coordinate for 
this particle is given by
\begin{align}
   \VEC{s}=\VEC{x}+\hat{r}\frac{u_\pp}{\mathcal{H}},
\end{align}
where $\hat{r}$ is the unit vector pointing along the observer's line
of sight, $u_\pp$ is radial comoving velocity,
$a m u_\pp=p_\pp=\VEC{p}\cdot\hat{r}$. The mass density in
redshift space is then given by
\begin{align}
   \rho_s(\VEC{s})=ma^{-3}\int{d^3p~d^3x~f\left(\VEC{x},\VEC{p}\right)\df^D\left(\VEC{s}-\VEC{x}-\hat{r}\frac{u_\pp}{\mathcal{H}}\right)}=
                   ma^{-3}\int{d^3p~f\left(\VEC{s}-\hat{r}\frac{u_\parallel}{\mathcal{H}},\VEC{p}\right)}. 
\label{eq:RosR}
\end{align}
By Fourier transforming equation \ref{eq:RosR}, we get
\begin{align}
    \rho_s(\VEC{k})&=m a^{-3} \int{d^3x~d^3p~f\left(\VEC{x},\VEC{p}\right)e^{(i\VEC{k}\cdot\VEC{x}+ik_\pp u_\pp/\mathcal{H})}}\nonumber\\
                   &=m a^{-3} \int{d^3x~e^{i\VEC{k}\cdot\VEC{x}}}~\int{d^3p~f(\VEC{x},\VEC{p})e^{ik_\parallel u_\parallel/\mathcal{H}}}, 
\label{eq:RosK}
\end{align}
were $\VEC{k}$ is the wavevector in redshift space, corresponding to 
redshift-space coordinate $\VEC{s}$.

Expanding the second integral in equation \ref{eq:RosK} as a Taylor series in $k_\parallel u_\parallel/m\mathcal{H}$,
\begin{align}
    ma^{-3}\int{d^3p~f\left(\VEC{x},\VEC{p}\right)}e^{ik_\pp u_\pp/\mathcal{H}} &= ma^{-3}\int{d^3q~f\left(\VEC{x},\VEC{p}\right)}
    \sum_{L=0}\frac{1}{L!}\left(ik_\pp u_\pp/\mathcal{H}\right)^L\nonumber\\
    &=\bar{\rho}\left[\sum_{L=0}\frac{1}{L!}\left(\frac{ik_\pp}{\mathcal{H}}\right)^LT^L_\pp(\VEC{x})\right]
\end{align}
where in the last part we have used equation \ref{eq:TLa} setting the vector $\VEC{h}$ to be the unit vector pointing along the observer's
line of sight $\VEC{h}=\hat{r}$. Using that in equation \ref{eq:TLaArox} we 
have, in the case with no velocity dispersion or other second or higher rank
tensors (which we will {\it not} generally assume)
\begin{align}
    T^L_\pp(\VEC{x})=(1+\df(\VEC{x}))v^L_\pp(\VEC{x}).
\label{eq:TLpp}
\end{align}
The Fourier component of the density fluctuation in redshift space is
\begin{align}
    \df_s(\VEC{k})=\sum_{L=0}\frac{1}{L!}\left(\frac{ik_\pp}{\mathcal{H}}\right)^LT^L_\pp(\VEC{k}),
\label{eq:dsk}
\end{align}
were $T^L_\pp(\VEC{k})$ is the Fourier transform of the $T^L_\pp(\VEC{x})$. For L=0 we drop the unmeasurable $k=0$ mode,
and we are left with the density fluctuation $T^0_\pp(\VEC{k})=\df(\VEC{k})$. 

\subsection{Angular decomposition of the moments of distribution function}
\label{subsec:Angular decomposition}

In order to make the context of this paper more clear we repeat angular decomposition of the moments of distribution function from ~\cite{Seljak:2011tx},
providing more detailed derivation. The object $T^L_\VEC{h}(\VEC{x})$ introduced in equation \ref{eq:TLa} can be obtained as taking all components
of moments of distribution function in $\VEC{h}$ direction, which are the rank $L$ tensors,
\begin{align}
 T^L_{i_1,i_2,\ldots i_L}=\frac{ma^{-3}}{\bar{\rho}}\int{d^3pf(\VEC{x},\VEC{p})u_{i_1}u_{i_2}\ldots u_{i_L}}.
\label{eq:TLtensor}
\end{align}
The real-space density  field corresponds to $L = 0$, i.e. zeroth moment \ref{eq:density}, the $L = 1$ moment corresponds to the momentum density \ref{eq:velocity}, 
$L = 2$ gives the stress energy density tensor \ref{eq:dispersion} etc. These objects are symmetric under exchange of any two indices and have (L + 1)(L + 2)/2 
independent components. They can be decomposed into helicity eigenstates under rotation around $\VEC{k}$.

The full detailed derivation of this decomposition is done in (App. \ref{sec:App2}) and here we give the final result taking $\VEC{h}=\hat{r}$;
\begin{align}
 T^L_\pp(\VEC{k})=\sum_{(l=L,L-2,\ldots)}\sum^{m=l}_{m=-l}n^L_lT^{L,m}_l(k)Y_{lm}(\theta,\phi),
\label{eq:Tdecomposition}
\end{align}
where coefficients $n^L_l$ are defined in equation \ref{eq:Ilocomputation}, and spherical tensors $T^{L,m}_l$ in equation \ref{eq:TLml}, and evaluated in frame where $z\pp\VEC{k}$, so it does not contain any angular dependence.

\subsection{Redshift power spectrum}
\label{subsec:redshift_power_spectrum}
  
In our analysis we will adopt a plane-parallel approximation, were only the angle between the line of sight and the Fourier mode needs to be specified. The redshift-space
power spectrum is defined as $\left\langle\df_s(\VEC{k})|\df^*_s(\VEC{k}')\right\rangle=(2\pi)^3 P^{ss}(\VEC{k})\df^D(\VEC{k}-\VEC{k}')$. Equation \ref{eq:dsk} gives,
\begin{align}
    P^{ss}(\VEC{k})=\sum_{L=0}\sum_{L'=0}\frac{(-1)^{L'}}{L!L'!}\left(\frac{ik_\pp}{\mathcal{H}}\right)^{L+L'}P_{LL'}(\VEC{k}),
\label{eq:Pss}
\end{align}
where we define 
\begin{align}
(2\pi)^3P_{LL'}(\VEC{k})\df^D(\VEC{k}-\VEC{k}')=\left\langle T^L_\pp(\VEC{k})\right.\left|T^{*L'}_\pp(\VEC{k}')\right\rangle. 
\end{align}
Note that $P_{LL'}(\VEC{k})=P_{L'L}(\VEC{k})^*$ so that the total result is real valued 
(what is explicitly shown in PT approach in App. \ref{sec:App3}). Thus only the terms $P_{LL'}(\VEC{k})$ with $L\leq L'$ need to be considered, 
each of which comes with a factor of 2 if $L\not=L'$ and 1 if $L=L'$. If we introduce $\mu=k_\pp/k=\text{cos}\theta$, we can write,
\begin{align}
    P^{ss}(\VEC{k})=\sum_{L=0}\frac{1}{(L!)^2}\left(\frac{k\mu}{\mathcal{H}}\right)^{2L}P_{LL}(\VEC{k})
                   +2Re\sum_{L=0}\sum_{L'>L}\frac{(-1)^{L'}}{L!L'!}\left(\frac{ik\mu}{\mathcal{H}}\right)^{L+L'}P_{LL'}(\VEC{k}).
\label{eq:PssPart}
\end{align}
Next we insert the helicity decomposition of equation \ref{eq:Tdecomposition} and consider the implications
of rotational symmetry on the power spectrum. Each term $P_{LL'}(\VEC{k})$ contains products of multipole moments
\begin{align}
 T^{L,m}_l(\VEC{k})Y_{lm}(\theta,\phi)\left(T^{L',m'}_{l'}(\VEC{k})Y_{l',m'}(\theta,\phi)\right)^*\propto e^{i(m-m')\phi}.
\end{align}
Upon averaging over the azimuthal angle $\phi$ of Fourier modes all the terms with $m\neq m'$ vanish. Another way to state this is that upon rotation 
by angle $\psi$ the correlator picks up a term $e^{i(m-m')\psi}$ and
in order for the power spectrum to be rotationally invariant we
require $m=m'$. Putting all these together we find
\begin{align}
 P_{LL'}(\VEC{k})=\sum_{(l=L,L-2,\ldots)}\sum_{(l'=L',L'-2,\ldots)}\sum^l_{m=0}P^{L,L',m}_{l,l'}(k)P^m_l(\mu)P^m_{l'}(\mu),
\label{eq:PLL}
\end{align}
where $P^m_l(\mu=\cos\theta)$, are the associated Legendre polynomials, which determine the $\theta$ angular dependence of the spherical harmonics. We absorb all of the terms that 
depend on $l$ and $m$ and various constants into the definition of power spectra $P^{L,L',m}_{l,l'}(k)$. Note once again that these spectra depend only on amplitude of k. We have
\begin{align}
 P^{L,L',m}_{l,l'}(k)\propto\left<T^{L,m}_l(k)\right.\left|\left(T^{L',m'}_{l'}(k)\right)^{*}\right>.
\end{align}
We also replaced the two helicity states $\pm m$ by a single one with $m > 0$, since their $\theta$ angular dependencies are the same, and we absorbed the appropriate factors into the definition
of $P^{L,L',m}_{l,l'}(k)$.

\subsection{Perturbation theory approach}
\label{sec:ptaproach}

The parameter of the expansion in equation \ref{eq:PssPart} can roughly be defined as $k\mu v/\mathcal{H}$, where $v$ is related to a typical gravitational velocity
of the system. This velocity should be of order of a few hundred $km/s$, but note that higher and higher powers of these velocities enter the series. The expansion 
series is expected to be convergent if expansion parameter is less then unity. 

The main goal of these paper is to use perturbation theory to compute and assess contributing terms in expansion formula \ref{eq:PssPart} in next to leading order (one loop). 
There is a close, but not one to one, relation between the PT expansion and expansion in distortions function moments. Assuming that $\df$ and
$kv/\mathcal{H}$ make the same order of expansion in one loop (except $P_{04}$ where leading term is two loop quantity) regime we obtain
\begin{align}
    P^{ss}(\VEC{k})& = P_{00}(\VEC{k})+\left(\frac{k\mu}{\mathcal{H}}\right)^2P_{11}(\VEC{k})+\frac{1}{4}\left(\frac{k\mu}{\mathcal{H}}\right)^4P_{22}(\VEC{k})
    +2\text{Re}\left[\left(\frac{-ik\mu}{\mathcal{H}}P_{01}(\VEC{k})\right)\right.\nonumber\\
    &+\left(-\frac{1}{2}\left(\frac{k\mu}{\mathcal{H}}\right)^2 P_{02}(\VEC{k})\right)
    +\left(\frac{i}{6}\left(\frac{k\mu}{\mathcal{H}}\right)^3 P_{03}(\VEC{k})\right)
    +\left(-\frac{i}{2}\left(\frac{k\mu}{\mathcal{H}}\right)^3P_{12}(\VEC{k})\right)\nonumber\\
    &+\left.\left(-\frac{1}{6}\left(\frac{k\mu}{\mathcal{H}}\right)^4P_{13}(\VEC{k})\right)
    +\left(\frac{1}{24}\left(\frac{k\mu}{\mathcal{H}}\right)^4P_{04}(\VEC{k})\right)\right].
\label{eq:PssNLO}
\end{align} 
Neglecting all the velocity dispersion and anisotropic stress
contributions we can use simplified form of $T^L_\pp$ (equation \ref{eq:TLpp}).
After preforming the Fourier transformation we obtain
\begin{equation}
     T^{L}_\pp(\VEC{k})=\mathcal{F}\left[\left(1+\df(\VEC{x})\right)v^L_\pp(\VEC{x})\right](\VEC{k}).
\end{equation}
In one loop PT regime only first three momenta are needed, so we can write
\begin{align}
     &T^{1}_\pp(\VEC{k})= v_\pp(\VEC{k})+\left(v_\pp\circ\df\right)(\VEC{k}),\nonumber\\
     &T^{2}_\pp(\VEC{k})= \left(v_\pp\circ v_\pp\right)(\VEC{k})+\left(v_\pp\circ v_\pp\circ\df\right)(\VEC{k}),\nonumber\\
     &T^{3}_\pp(\VEC{k})= \left(v_\pp\circ v_\pp\circ v_\pp\right)(\VEC{k}),
\label{eq:T123con}
\end{align}
where we have used the following convention for convolution
\begin{align}
     (f\circ g)(\VEC{k})=\int{\frac{d^3q}{(2\pi)^3}f(\VEC{q})g(\VEC{k}-\VEC{q})}.
\label{eq:Convolution}
\end{align}
From the approximations we have adopted it also follows that curls of
velocity field can be neglected, 
i.e. $\nabla\times\VEC{v}(\VEC{x})=0$. Thus velocity field can be fully described 
by its divergence field $\theta(\VEC{x})=\nabla\cdot\VEC{v}(\VEC{x})$. So it follows $v_\pp(\VEC{k})=i\frac{k_\pp}{k^2}\theta(\VEC{k})$.

At this point it is useful to observe that if working in linear
perturbation regime well known Kaiser result
\cite{Kaiser:1987qv} can be easily obtained directly from equation
\ref{eq:Pss}. It follows
\begin{align}
 P^{ss}(\VEC{k})=P_{00}(\VEC{k})+2\text{Re}\left(\frac{-ik\mu}{\mathcal{H}}\right)P_{01}(\VEC{k})+\left(\frac{k\mu}{\mathcal{H}}\right)^2P_{11}(\VEC{k}),
\end{align}
and after using the facts that
$P_{\delta\theta}^{(1,1)}(\VEC{k})=-f\mathcal{H}P_{L}(\VEC{k})$ and
that
$P_{\theta\theta}^{(1,1)}(\VEC{k})=(f\mathcal{H})^2P_{L}(\VEC{k})$ we obtain
\begin{equation}
     P^{ss}(\VEC{k}) = \left(1+f\mu^2\right)^2P_{\df\df}(\VEC{k}),
\end{equation}
hence, the Kaiser formula.

\section{Perturbation theory results and comparison to the N-body simulations}
\label{sec:terms}

All of the N-body results used here have been presented in \cite{Okumura:2011pb}. Briefly, 
for all of the power spectra of the derivative expansion one needs
mass-weighted velocity moments, which can be straightforwardly
measured from simulations.  In \cite{Okumura:2011pb} a series of $N$-body simulations of
the $\Lambda$CDM cosmology seeded with Gaussian initial conditions has been used
\cite{Desjacques:2008vf}. We employ $1024^3$ particles of
mass $3.0\times 10^{11} h^{-1}M_\odot$ in a cubic box of side
$1600h^{-1}Mpc$.  We use 12 independent realizations in order to reduce the statistical
scatters.  For the details of the simulations measurements we refer to the
\cite{Desjacques:2008vf} and here we shortly repeat the basics.

\subsection{$P_{00}(\VEC{k})$: the isotropic term}
\label{sec:P00}

At the lowest order in $kv/\mathcal{H}$ expansion we have auto correlation of density field $T^0_\pp(\VEC{k})=\df(\VEC{k})$. Power spectrum, 
$P_{00}(k)\df^D(\VEC{k}-\VEC{k}')=\left<\df(\VEC{k})|\df^*(\VEC{k}')\right>$,
is well known and has been intensively studied, e.g. \cite{Carlson:2009it, Bernardeau:2001qr}.
This first term does not have any $\mu$ dependence since it is independent of red shift space distortions, it dominants for small values of $\mu$ and in the limit $\mu=0$ the transverse power spectrum 
becomes overdensity power spectrum $P_{00}(k)$. On scales smaller than $k^{-1}\sim \, 10Mpc/h$, nonlinear corrections increase the power over the linear.  

\begin{figure}[t]
    \centering
    \includegraphics[width=1.0\textwidth]{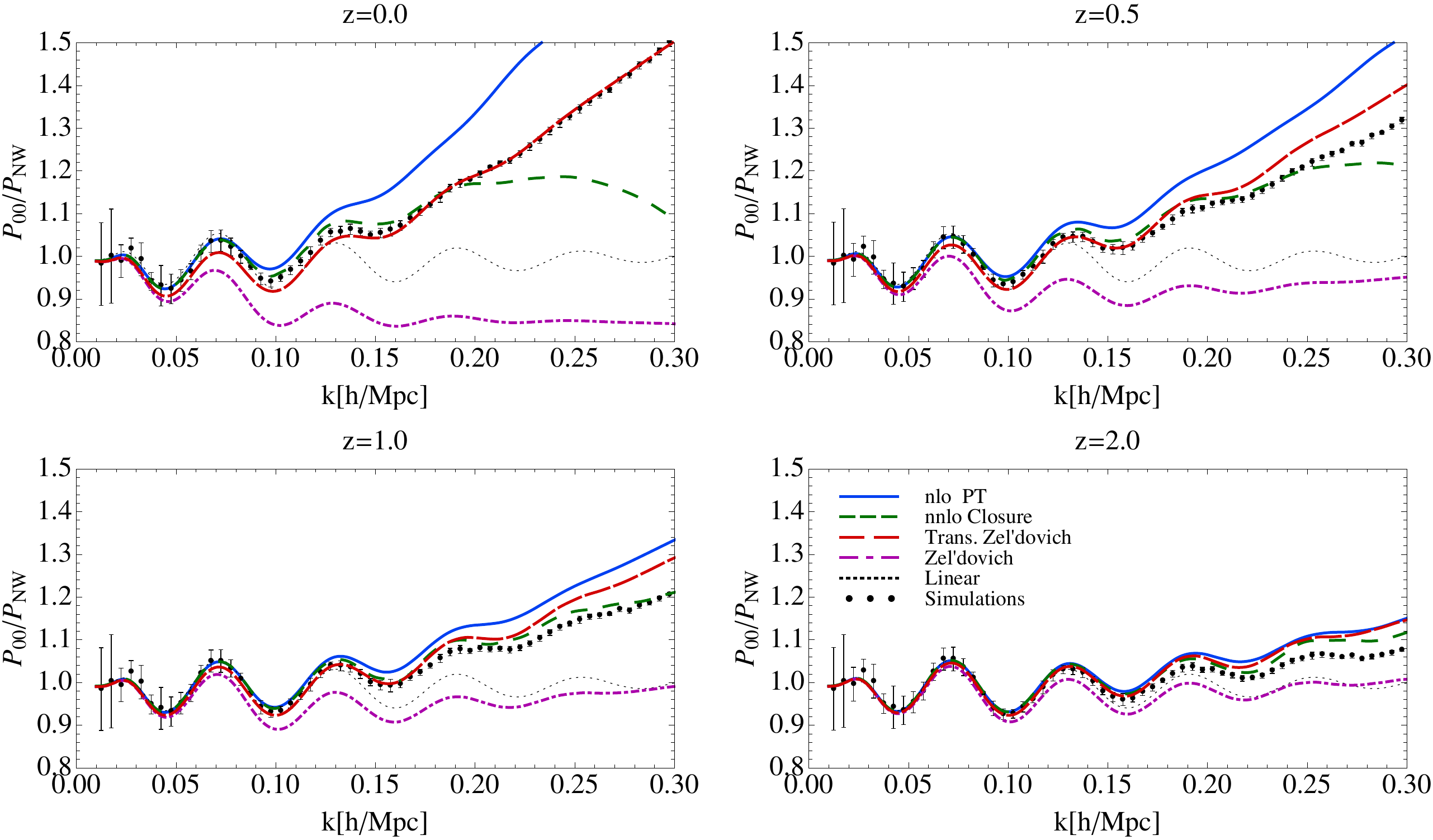}
    \caption{\small$P_{00}(k)$ power spectrum term is plotted at four
      redshifts $z=0.0,~0.5,~1.0$ and $2.0$. We show linear result
      (black, dotted), one loop PT (blue, solid), two loop closure (green,
      dashed), corrected Zel'dovich (red, long-dashed) of \cite{Tassev:2011ac} ,
      simple Zel'dovich (magenta, dot-dashed) and simulation
      measurements (black dots). The error bars show the variance among realizations in simulations. The power spectrum is divided
      by no-wiggle fitting formula from \cite{Eisenstein:1997ik}, 
      to reduce the dynamic range.}
    \label{fig:P00}
\end{figure}

Familiar one loop PT result for overdensity power spectrum is \cite{Bernardeau:2001qr}
\begin{equation}
     P_{00}(\VEC{k},\tau)=P_{\df\df}(\VEC{k},\tau)= D^2(\tau)P_{\df\df}^{(1,1)}(\VEC{k})
      +D^4(\tau)\left[P_{\df\df}^{(2,2)}(\VEC{k})+2P_{\df\df}^{(1,3)}(\VEC{k})\right],
\label{eq:P00.1}
\end{equation}
where we have restored time dependence, with $D(\tau)$ being linear cosmological growth factor. $P_{\df\df}^{(1,1)}(\VEC{k})$ is the linear power spectrum $P_{L}(\VEC{k})$, and one loop contributions are
\begin{align}
     P_{\df\df}^{(2,2)}(k)&=2\int\frac{d^3q}{(2\pi)^3}P_L(q)P_L\left(\left|\VEC{k}-\VEC{q}\right|\right)\left[F_2^{(s)}\left(\VEC{q},\VEC{k}-\VEC{q}\right)\right]^2\nonumber\\
     &=2I_{00}(k),\nonumber\\
     P_{\df\df}^{(1,3)}(k)&=3P_L(k)\int\frac{d^3q}{(2\pi)^3}P_L(q)F_3^{(s)}\left(\VEC{k},\VEC{q},-\VEC{q}\right)\nonumber\\
     &=3k^2P_L(k)J_{00}(k).
\label{eq:P00.2}
\end{align}
Explicit form of all integrals of the $I_{mn}(k)$ and $J_{mn}(k)$ type can be found in App \ref{sec:App4}. In figure \ref{fig:P00} one loop PT results for power spectrum have been presented, along with 
some of the other approaches, such as the 
closure theory approach
\cite{Taruya:2007xy} obtained from the Copter code
\cite{Carlson:2009it} and the semi-fitting method \cite{Tassev:2011ac}, 
based on power spectrum obtained from Zel'dovich approximation. Note
that if one wants to impose consistency in expansion \ref{eq:PssNLO} and
PT approach, only one loop regime PT result should be considered. All the power 
spectra on the figures are divided by the linear power spectrum fitting formula from
\cite{Eisenstein:1997ik} without BAO wiggles. We see that none of the methods give perfect agreement
across all range of scales. SPT (one loop PT) actually gives the 
best results for $k<0.05h/Mpc$, but predicts too much power at higher $k$. 

\subsection{$P_{01}(\VEC{k})$}
\label{sec:P01}

The next term to consider correlates the overdensity field $T^0_\pp(\VEC{k})=\df(\VEC{k})$ and radial component of momentum density $T^1_\pp(\VEC{k})$. This is the dominant RSD term sensitive to velocities. As we can see from equation \ref{eq:Tdecomposition}, 
momentum density can be decomposed into a scalar $(m=0)$ $T^{1,0}_1$ and two vector $(m=\pm 1)$ components $T^{1,\pm 1}_1$. Only the scalar part correlates with the density $T^{0,0}_0$, which is a
scalar field. Thus only non-vanishing contribution comes from
$P^{0,1,0}_{0,1}(k)\varpropto\left<T^{0,0}_0(k)\right.\left|\left(T^{1,0}_1(k)\right)^*\right>$,
what gives the simple angular dependence 
\begin{align}
   P_{01}(\VEC{k})=P^{0,1,0}_{0,1}(k)P^0_1(\mu)=\mu P^{0,1,0}_{0,1}(k).
\label{eq:P01.1}
\end{align}

On the other hand, correlating directly $\left<\df(\VEC{k})|T^{*1}_\pp(\VEC{k}',\tau)\right>$, from the equation \ref{eq:T123con} one gets power spectra
\begin{align}
     &P_{01}(\VEC{k})=-i\frac{\mu}{k}P_{\df\theta}(\VEC{k})-iA_{01}(\VEC{k}),
\label{eq:P01.2}
\end{align}
where the first term is also well studied correlation function of overdensity field and divergence of velocity field
\begin{align}
     &(2\pi)^3P_{\df\theta}(\VEC{k})\df^D(\VEC{k}-\VEC{k}')=\left\langle \df(\VEC{k})|\theta(\VEC{k}')\right\rangle,\nonumber\\
     &(2\pi)^3A_{01}(\VEC{k})\delta^D(\VEC{k}-\VEC{k}')=\int{\frac{d^3\VEC{q}}{(2\pi)^3}\frac{q_\pp}{q^2}}\left\langle 
     \df(\VEC{k})|\theta^*(\VEC{q})\df^*(\VEC{k}'-\VEC{q})\right\rangle.
\label{eq:P01.3}
\end{align}

 For the first term, correlation function of overdensity and divergence of velocity field, one loop PT gives
\begin{equation}
     P_{\df\theta}(k,\tau)= D^2(\tau)P_{\df\theta}^{(1,1)}(k)
      +D^4(\tau)\left[P_{\df\theta}^{(2,2)}(k)+2P_{\df\theta}^{(1,3)}(k)\right],
\label{eq:P01.4}
\end{equation}
where $P_{\delta\theta}^{(1,1)}(k)=-f\mathcal{H}P_{L}(k)$ is the contribution in the linear regime, and one loop contribution is
\begin{align}
     P_{\delta\theta}^{(2,2)}(k)&=-2f\mathcal{H}\int\frac{d^3q}{(2\pi)^3}P_L(q)P_L\left(\left|\VEC{k}-\VEC{q}\right|\right)
     F_2^{(s)}\left(\VEC{q},\VEC{k}-\VEC{q}\right)G_2^{(s)}\left(\VEC{q},\VEC{k}-\VEC{q}\right)\nonumber\\
     &=-2f\mathcal{H} I_{01}(k),\nonumber\\
     P_{\delta\theta}^{(1,3)}(k)&=-3f\mathcal{H}P_L(k)\int\frac{d^3q}{(2\pi)^3}P_L(q)
     \frac{1}{2}\left[F_3^{(s)}\left(\VEC{k},\VEC{q},-\VEC{q}\right)+
     G_3^{(s)}\left(\VEC{k},\VEC{q},-\VEC{q}\right)\right]\nonumber\\
     &=-3f\mathcal{H}k^2P_L(k)J_{01}(k)\nonumber\\
     &=-\frac{1}{2}f\mathcal{H}\left(P_{\delta\delta}^{(1,3)}(k)+\frac{P_{\theta\theta}^{(1,3)}(k)}{(f\mathcal{H})^2}\right).
\label{eq:P01.5}
\end{align}
Here we have introduced logarithmic growth rate $f=f(\tau)=d\ln{D}/d\ln{a}.$

For the second term in equation \ref{eq:P01.1}, we expand all the fields to the second order, i.e., one loop in the correlation function. Schematically, this gives
\begin{equation}
   \left\langle \delta\theta\delta\right\rangle=\left\langle \delta^{(2)}\theta^{(1)}\delta^{(1)}\right\rangle+
   \left\langle \delta^{(1)}\theta^{(2)}\delta^{(1)}\right\rangle+\left\langle \delta^{(1)}\theta^{(1)}\delta^{(2)}\right\rangle, \nonumber
\label{eq:P01.6}
\end{equation}
or in terms of power spectrum
\begin{equation}
     A_{01}(\VEC{k},\tau)=D^4(\tau)\left(A_{01}^{(211)}(\VEC{k})+A_{01}^{(112)}(\VEC{k})+A_{01}^{(112)}(\VEC{k})\right).
\label{eq:P01.7}
\end{equation}
Again, using one loop PT we obtain the contributions from each of the terms
\begin{align}
     A_{01}^{(211)}(\VEC{k})&=-2f\mathcal{H}
     \int{\frac{d^3q}{(2\pi)^3}~\frac{q_\parallel}{q^2}F_2^{(s)}(\VEC{q},\VEC{k}-\VEC{q})P_L(q)
     P_L(|\VEC{k}-\VEC{q}|)}\nonumber\\
     &=-2f\mathcal{H}\frac{\mu}{k}I_{10}(k),\nonumber\\
     A_{01}^{(121)}(\VEC{k})&=-2f\mathcal{H}~P_L(\VEC{k})
     \int{\frac{d^3q}{(2\pi)^3}~\frac{(\VEC{k}-\VEC{q})_\parallel}{(\VEC{k}-\VEC{q})^2}G_2^{(s)}(-\VEC{q},\VEC{k})
     P_L(q)}\nonumber\\
     &=-2f\mathcal{H} \mu k P_L(k)\left[3J_{10}(k)+\frac{1}{2}\left(\sigma^2_v+\frac{\sigma_0^2}{3k^2}\right)\right],\nonumber\\
     A_{01}^{(112)}(\VEC{k})&=2f\mathcal{H}~P_L(\VEC{k})
     \int{\frac{d^3q}{(2\pi)^3}~\frac{q_\parallel}{q^2}F_2^{(s)}(\VEC{q},\VEC{k})
     P_L(q)}\nonumber\\
     &= f\mathcal{H} \mu k P_L(k)\left(\sigma^2_v+\frac{\sigma_0^2}{3k^2}\right).
\label{eq:P01.8}
\end{align}
where the
$\sigma^2_v=\frac{1}{3}\int\frac{d^3q}{(2\pi)^3}\frac{P_L(q)}{q^2}$ is
the one-dimensional velocity dispersion at linear order, 
and $\sigma_0^2=\int\frac{d^3q}{(2\pi)^3}P_L(q)$.
Note that all three terms give the same angular dependence, so
$A_{01}\sim\mu$, and then follows that $P_{01}\sim\mu$, 
as was expected form the symmetry consideration on the beginning. 
Finally, collecting all the terms \ref{eq:P01.2}, \ref{eq:P01.3}, \ref{eq:P01.7},
\ref{eq:P01.8} one loop PT prediction for the $P_{01}$ follows. Now the total
contribution to the redshift power spectrum $P^{ss}$ from the
$P_{01}$  term is
\begin{align}
   P^{ss}_{01}(\VEC{k},\tau)&=2\frac{-ik\mu}{\mathcal{H}}P_{01}(\VEC{k},\tau)\nonumber\\ 
   &=2f(\tau)D^2(\tau) \mu^2\left( P_L(k) +  2D^2(\tau)\left[I_{01}(k)+I_{10}(k)+3k^2\big(J_{01}(k)+J_{10}(k)\big)P_L(k)\right]\right).
\label{eq:P01.10}
\end{align}
In this form result is naturally separated in linear and one loop contribution part. Note that linear part here is the second term of Kaiser formula.

Alternatively, the scalar mode of momentum can be obtained from the divergence of momentum and related
to $\dot{\df}$ using the continuity equation $\dot{\df}-ikp_s = 0$,
which is in terms of quantities defined previously 
\begin{align}
   \dot{T}^{0,0}_0-ikT^{1,0}_0=0.
\label{eq:P01.12} 
\end{align}
Note that the vector part of momentum field does not contribute, since it vanishes upon taking the divergence 
(i.e., vector components are orthogonal to $\VEC{k}$ and the dot product is zero).

\begin{figure}[t]
    \centering
    \includegraphics[width=1.0\textwidth]{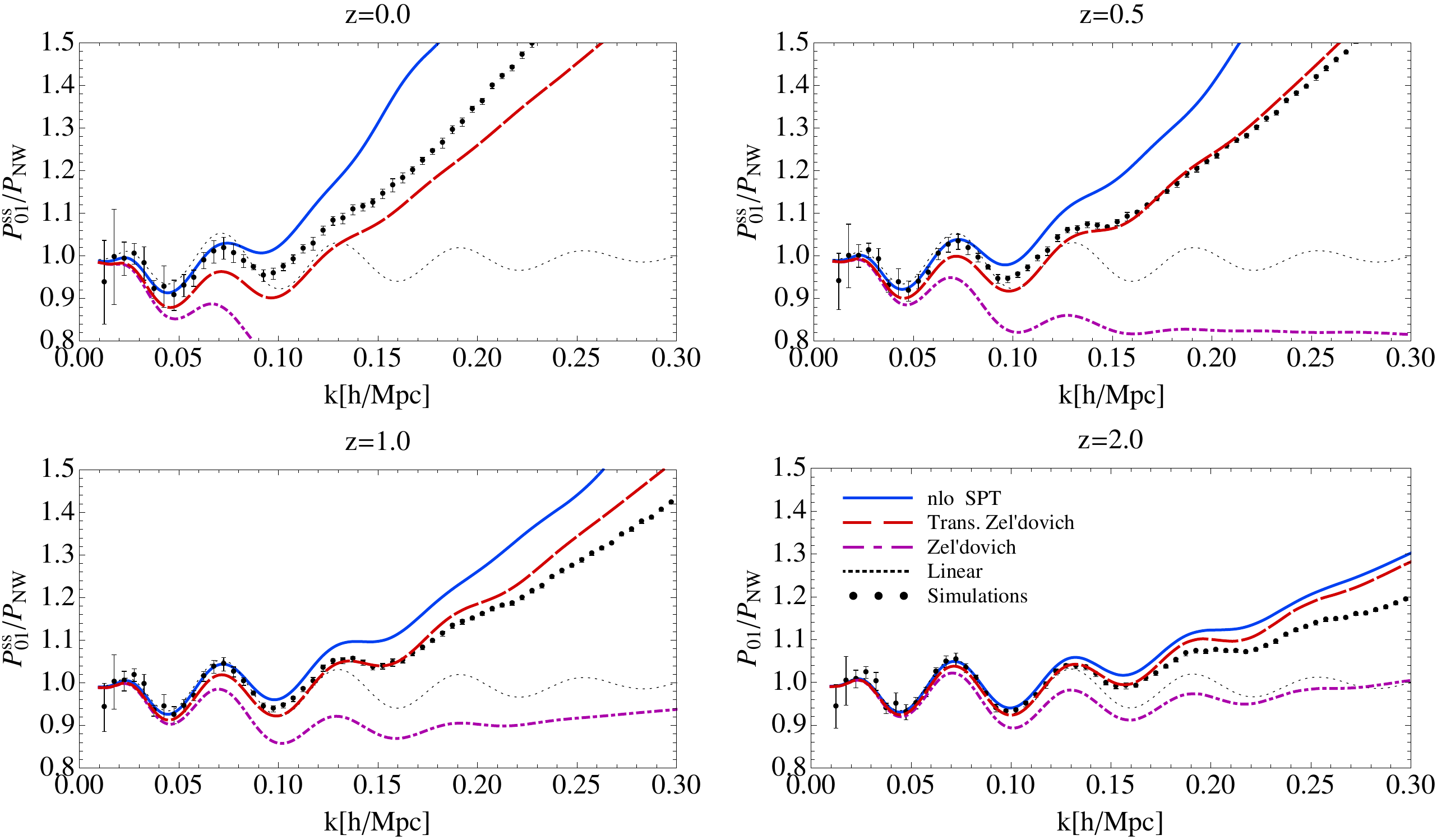}
    \caption{\small $k$-dependence of $P^{ss}_{01}$ term of redshift power spectrum is plotted at four redshifts $z=0.0,~0.5,~1.0$ and $2.0$. 
    This term has simple $\mu^2$ dependence in all nonlinear
    orders. Here we show linear Kaiser result (black,
    dotted), one loop PT (blue, solid),
    corrected Zel'dovich (red, dashed) model from \cite{Tassev:2011ac}, simple Zel'dovich (magenta,
    dot-dashed), and simulation measurements (black dots). The error bars show the variance among realizations in simulations. 
    The power spectra are divided by second, no-wiggle, term of Kaiser formula to reduce the dynamic range.}
    \label{fig:P01}
\end{figure}

It follows
\begin{align}
   P_{01}(\VEC{k},\tau)=i\frac{\mu}{k}P_{\df\dot{\df}}(k ,\tau)=i\frac{\mu}{2k}\frac{dP_{00}(k ,\tau)}{d\tau},
\label{eq:P01.13} 
\end{align}
and the total contribution to $P^{ss}(k,\tau)$ is 
\begin{align}
   P^{ss}_{01}(\VEC{k},\tau)=\mu^2\mathcal{H}^{-1}\frac{dP_{00}(k,\tau)}{d\tau}=\mu^2\frac{dP_{00}(k,a)}{d\ln a}.
\label{eq:P01.14}
\end{align}
This result, first obtained in \cite{Seljak:2011tx}, is exact for dark matter, valid also in the nonlinear regime. It shows that this term can be obtained directly from the redshift evolution of 
the dark matter power spectrum $P_{00}(k)$, so if we have an accurate PT model for $P_{00}$ then we should also have 
the same for $P_{01}$. On large scales it agrees with the linear theory predictions. If we write $P_{00}(k,\tau)=D(\tau)^2P_{L}(k,\tau)$, we find 
Kaiser part $P^{ss}_{01}=2f\mu^2P_{lin}(k)$. On smaller scales we
expect the term to deviate from the linear one, just as for
$P_{00}(k)$. Using one loop PT we simply need to calculate the derivatives of growth factor $\dot{D}(\tau)=f (\tau)\mathcal{H}D (\tau)$, and from equation \ref{eq:P00.1} we get
\begin{align}
  P_{\df\dot{\df}}(\VEC{k},\tau)&= f(\tau)\mathcal{H}D^2(\tau)\left[P_{\df\df}^{(1,1)}(k)+2D^2(\tau)\left(P_{\df\df}^{(2,2)}(k)+2P_{\df\df}^{(1,3)}(k)\right)\right].
\label{eq:P01.15}
\end{align}
Finely, plugging that in equation \ref{eq:P01.14} we get
\begin{align}
  P^{ss}_{01}(\VEC{k})=2f(\tau)D^2(\tau)\mu^2\bigg[P_L(k)+4D^2(\tau)\left(I_{00}(k) +3k^2J_{00}(k)P_L(k)\right)\bigg].
\label{eq:P01.16}
\end{align} 
After some integral transformations and calculations it can be shown that this result is equivalent to the on in equation \ref{eq:P01.10}. 
Obtained $P^{ss}_{01}$ results are presented in Figure \ref{fig:P01}. We show the one loop PT results, along with semi-fitting method
\cite{Tassev:2011ac} based on power spectrum in Zel'dovich
approximation, and simulation measurements. 
The power spectra are now divided by second term in Kaiser formula where no-wiggle linear power spectrum has been used.

\subsection{$P_{11}(\VEC{k})$}
\label{sec:P11}

The next term we are to consider is the autocorrelation of momentum
density $T^1_\pp(\VEC{k})$ field. In this case scalar $(m=0)$ $T^{1,0}_1(k)$ correlates with 
itself, and the vector $(m=\pm1)$ components $T^{1,\pm1}_1(k)$ also correlate with itself, so both components of momentum contribute,
\begin{align}
  P_{11}(k)=P^{1,1,0}_{1,1}(k)\left[P^0_1(\mu)\right]^2+P^{1,1,1}_{1,1}(k)\left[P^1_1(\mu)\right]^2.
\label{eq:P11.0}
\end{align}
Contributions to redshift space power spectrum is then given with
\begin{align}
  P^{ss}_{11}(k)=\mathcal{H}^{-2}k^2\mu^2\left[P^{1,1,0}_{1,1}(k)\mu^2+P^{1,1,1}_{1,1}(k)(1-\mu^2)\right].
\label{eq:P11.1}
\end{align}
The scalar part is the autocorrelation of the of the momentum that contributes to the continuity equation \ref{eq:P01.12}. 
In linear PT only the scalar contribution in non-zero and $P^{1,1,0}_{1,1}(k)=f^2P_L(k)$, which is the last term in Kaiser formula. 
There is another contribution to both $\mu^2$ and $\mu^4$ terms from the vector part of momentum correlator 
$P^{1,1,1}_{1,1}(k)\propto\left<|T^{1,1}_1(k)|^2\right>$, which comes
in at the second order in power spectrum, and can be computed using one loop PT. This vector part is often called the 
vorticity part of the momentum, because vorticity of momentum does not vanish, even if vorticity of velocity vanishes for a single streamed fluid \cite{McDonald:2009hs}. 
From equation \ref{eq:P11.1} can be seen that this term always adds
power to $\mu^2$ term and subtracts it in $\mu^4$ term, but is combined with a positive contribution from the scalar part in $\mu^4$ term.

Now using expressions \ref{eq:T123con} we can straightforwardly expand the correlator in density $\df$ and velocity divergence $\theta$ fields. In terms of power spectra we have
\begin{align}
     &P_{11}(\VEC{k})=\frac{\mu^2}{k^2}P_{\theta\theta}(\VEC{k})+2\frac{\mu}{k}B_{11}(\VEC{k})+C_{11}(\VEC{k}),
\label{eq:P11.3}
\end{align}
where we have introduced:
\begin{align}
     &(2\pi)^3P_{\theta\theta}(\VEC{k})\df^D(\VEC{k}-\VEC{k}')=\left\langle \theta(\VEC{k})|\theta^*(\VEC{k}')\right\rangle,\nonumber\\
     &(2\pi)^3B_{11}(\VEC{k})\df^D(\VEC{k}-\VEC{k}')=\int{\frac{d^3q}{(2\pi)^3}~\frac{q_\pp}{q^2}}\left\langle 	 
     \theta(\VEC{q})\df(\VEC{k}-\VEC{q})|\theta^*(\VEC{k}')\right\rangle,\nonumber\\
     &(2\pi)^3C_{11}(\VEC{k})\df^D(\VEC{k}-\VEC{k}')=\int{\frac{d^3q}{(2\pi)^3}\frac{d^3q'}{(2\pi)^3}
      ~\frac{q_\parallel}{q^2}}\frac{q'_\parallel}{q'^2}\left\langle \theta(\VEC{q})\df(\VEC{k}-\VEC{q})|\theta^*(\VEC{q}')
      \df^*(\VEC{k}'-\VEC{q}')\right\rangle.
\label{eq:P11.4}
\end{align}
Using one loop PT to evaluate these power spectra. First term gives familiar velocity divergence autocorrelation
\begin{equation}
      P_{\theta\theta}(\VEC{k},\tau)= D(\tau)^2P_{\theta\theta}^{(1,1)}(\VEC{k})
      +D^4(\tau)\left(P_{\theta\theta}^{(2,2)}(\VEC{k})+2P_{\theta\theta}^{(1,3)}(\VEC{k})\right),
\label{eq:P11.5}
\end{equation}
where $P_{\theta\theta}^{(1,1)}(k)$ is the linear power spectrum $(f\mathcal{H})^2P_{L}(k)$ and rest is one loop contribution to velocity divergence 
power spectrum $P_{\theta\theta}(k)$,
\begin{align}
     P_{\theta\theta}^{(2,2)}(k)&=2(f\mathcal{H})^2\int\frac{d^3q}{(2\pi)^3}P_L(q)P_L\left(\left|\VEC{k}-\VEC{q}\right|\right)
     \left[G_2^{(s)}\left(\VEC{q},\VEC{k}-\VEC{q}\right)\right]^2\nonumber\\
     &=2(f\mathcal{H})^2I_{11}(k)\nonumber\\
     P_{\theta\theta}^{(1,3)}(k)&=3(f\mathcal{H})^2P_L(k)\int\frac{d^3q}{(2\pi)^3}P_L(q)
     G_3^{(s)}\left(\VEC{k},\VEC{q},-\VEC{q}\right)\nonumber\\
     &=3(f\mathcal{H})^2k^2 P_L(k)J_{11}(k)
\label{eq:P11.6}
\end{align}

Second term can be expanded in the fields to the second order; schematically we have
\begin{equation}
    \left\langle \theta\df\theta\right\rangle=\left\langle \theta^{(2)}\df^{(1)}\theta^{(1)}\right\rangle+
    \left\langle \theta^{(1)}\df^{(2)}\theta^{(1)}\right\rangle+\left\langle \theta^{(1)}\df^{(1)}\theta^{(2)}\right\rangle. \nonumber
\label{eq:P11.7}
\end{equation}
This gives in terms of the power spectrum $B_{11}(\VEC{k})$
\begin{equation}
     B_{11}(\VEC{k},\tau)=D^4(\tau)\left(B_{11}^{(211)}(\VEC{k})+B_{11}^{(112)}(\VEC{k})+B_{11}^{(112)}(\VEC{k})\right).
\label{eq:P11.8}
\end{equation}
where contributing terms are
\begin{align}
     B_{11}^{(211)}(\VEC{k})&=2(f\mathcal{H})^2P_L(\VEC{k})\int{\frac{d^3q}{(2\pi)^3}~\frac{(\VEC{k}-\VEC{q})_\parallel}
     {(\VEC{k}-\VEC{q})^2}G_2^{(s)}(\VEC{k},-\VEC{q})P_L(\VEC{q})} \nonumber\\
     &=2(f\mathcal{H})^2\mu k P_L(k) \left[3J_{10}(k)+\frac{1}{2}\left(\sigma^2_v+\frac{\sigma_0^2}{3k^2}\right)\right], \nonumber\\
     B_{11}^{(121)}(\VEC{k})&=2(f\mathcal{H})^2P_L(\VEC{k})\int{\frac{d^3q}{(2\pi)^3}~\frac{q_\parallel}{q^2}
     F_2^{(s)}(\VEC{k},-\VEC{q})P_L(\VEC{q})} \nonumber\\
     &=-(f\mathcal{H})^2\mu k P_L(k)\left(\sigma^2_v+\frac{\sigma_0^2}{3k^2}\right), \nonumber\\
     B_{11}^{(112)}(\VEC{k})&=2(f\mathcal{H})^2\int{\frac{d^3q}{(2\pi)^3}~\frac{q_\parallel}{q^2}
     G_2^{(s)}(\VEC{k}-\VEC{q},\VEC{q})P_L(\VEC{k}-\VEC{q})P_L(\VEC{q})} \nonumber\\
     &=2(f\mathcal{H})^2\frac{\mu}{k}I_{22}(k).
\label{eq:P11.9}
\end{align}
Similarly, for the last term in equation \ref{eq:P11.3}, we have
\begin{align}
     C_{11}^{(1111)}(\VEC{k})&=(f\mathcal{H})^2\int{\frac{d^3q}{(2\pi)^3}~\frac{q_\parallel}{q^2}\left(\frac{q_\parallel}{q^2}+
     \frac{(\VEC{k}-\VEC{q})_\parallel}{(\VEC{k}-\VEC{q})^2}\right)P_L(\VEC{k}-\VEC{q})P_L(\VEC{q})}\nonumber\\
     &=(f\mathcal{H})^2 k^{-2}\left(I_{31}(k)+\mu^2I_{13}(k)\right).
\label{eq:P11.10}
\end{align}

Combining all that, we can write the contribution to redshift space power spectrum $P^{ss}$ from $P_{11}$ term
\begin{align}
    P^{ss}_{11}(\VEC{k})&=\left(\frac{k \mu}{\mathcal{H}}\right)^2P_{11}(\VEC{k})=f^2(\tau)D^2(\tau)\mu^2\left(\mu^2P_L(k)+D^2(\tau)I_{31}(k)\right)\nonumber\\
    &+f^2(\tau)D^4(\tau)\mu^4\left[2I_{11}(k)+4I_{22}(k)+I_{13}(k)+6k^2\big(J_{11}(k)+2J_{10}(k)\big)P_L(k)\right].  
\label{eq:P11.12}
\end{align}
As can be seen we obtained $\mu^2$ and $\mu^4$ angular dependence from this term, 
as was argued from symmetry consideration in \cite{Seljak:2011tx}. Vector contribution can be identified as
the part multiplying $\mu^2$ \cite{Seljak:2011tx}.

On the other hand, we could have started directly from equation \ref{eq:P11.0}. If we chose to work in the frame where $z\pp\VEC{k}$ one can write the 
decomposition \ref{eq:Tdecomposition} of momentum density $T^1_\pp(\VEC{k})=p_\pp(\VEC{k})=\hat{r}\cdot\VEC{p}(\VEC{k})=p_s \cos \theta + 
p_v \sin \theta \cos \phi$, where we have chosen, without loss of generality, for $\hat{r}$ to be in $x-z$ plane, and $p_s$ and $p_v$ represent scalar 
and vector part of decomposition, respectively. After averaging over $\phi$ angle, this enables us to write $P^{1,1,0}_{1,1}=P_{p_s,p_s}$ and
$P^{1,1,1}_{1,1}=P_{p_v,p_v}$. Just as before, scalar part can be determined directly from continuity equation \ref{eq:P01.12}.
We can again use one loop PT to evaluate scalar and vector contributions
\begin{align}
    P_{p_s,p_s}=k^{-2}P_{\dot{\df},\dot{\df}}&=(f\mathcal{H})^2D(\tau)^2k^{-2}\left(P_{\df\df}^{(1,1)}(\VEC{k})+
    D^2(\tau)\left[4P_{\df\df}^{(2,2)}(\VEC{k})+6P_{\df\df}^{(1,3)}(\VEC{k})\right]\right)\nonumber\\
    &=(f\mathcal{H})^2D(\tau)^2 k^{-2}\left(P_L(k)+D(\tau)^2\left[8I_{00}(k)+18k^2J_{00}(k)P_L(k)\right]\right),\nonumber\\
    P_{p_v,p_v}&=(f\mathcal{H})^2\int{\frac{d^3q}{(2\pi)^3}\frac{\left|\hat{k}\times\VEC{q}\right|^2}{q^4}}
    \frac{k^2-2\VEC{k}\cdot\VEC{q}}{(\VEC{k}-\VEC{q})^2}P_L(k)P_L(\left|\VEC{k}-\VEC{q}\right|)\nonumber\\
    &=(f\mathcal{H})^2D^4(\tau)k^{-2}I_{31}(k).
\label{eq:P11.13}
\end{align}
Thus, the contribution to the total red shift power spectrum from $P_{11}$ term is
\begin{align}
    P^{ss}_{11}=f^2(\tau)D^2(\tau)\mu^2\bigg[\mu^2P_L(k)+\mu^2D(\tau)^2\big(8I_{00}(k)+18J_{00}(k)P_L(k)\big)
    +(1-\mu^2)D(\tau)^2I_{31}(k)\bigg].
\label{eq:P11.14}
\end{align}
Again, after some coordinate transformations and algebra
it can be shown that this result is equivalent to the one we
obtained earlier in equation \ref{eq:P11.12}. 

In order to improve our prediction for the vector part we can take
into consideration the most relevant higher order loop contributions. Starting
from definition of $C_{11}$ (equation \ref{eq:P11.4}), which gives
raise to the vector part of $P_{11}$, and generalizing our one loop prediction in equation 
\ref{eq:P11.10} we can write 
\begin{align}
     C_{11}(\VEC{k})&=\int{\frac{d^3q}{(2\pi)^3}~\frac{q^2_\parallel}{q^4}P_{\tf\tf}(\VEC{q})P_{\df\df}(\VEC{k}-\VEC{q})}+
     \int{\frac{d^3q}{(2\pi)^3}~\frac{q_\parallel}{q^2}\frac{(\VEC{k}-\VEC{q})_\parallel}{(\VEC{k}-\VEC{q})^2}P_{\df\tf}(\VEC{q})P_{\df\tf}(\VEC{k}-\VEC{q})}.
\label{eq:P11.15}
\end{align}
In low $k$ limit this gives back the previous result from equation
\ref{eq:P11.10}, and in high $k$ limit the first term dominates what is
giving $(f\mathcal{H}D)^2P_{00}(k)\sigma_v^2$ for the vector part.

\begin{figure}[t]
    \centering
    \includegraphics[width=1.0\textwidth]{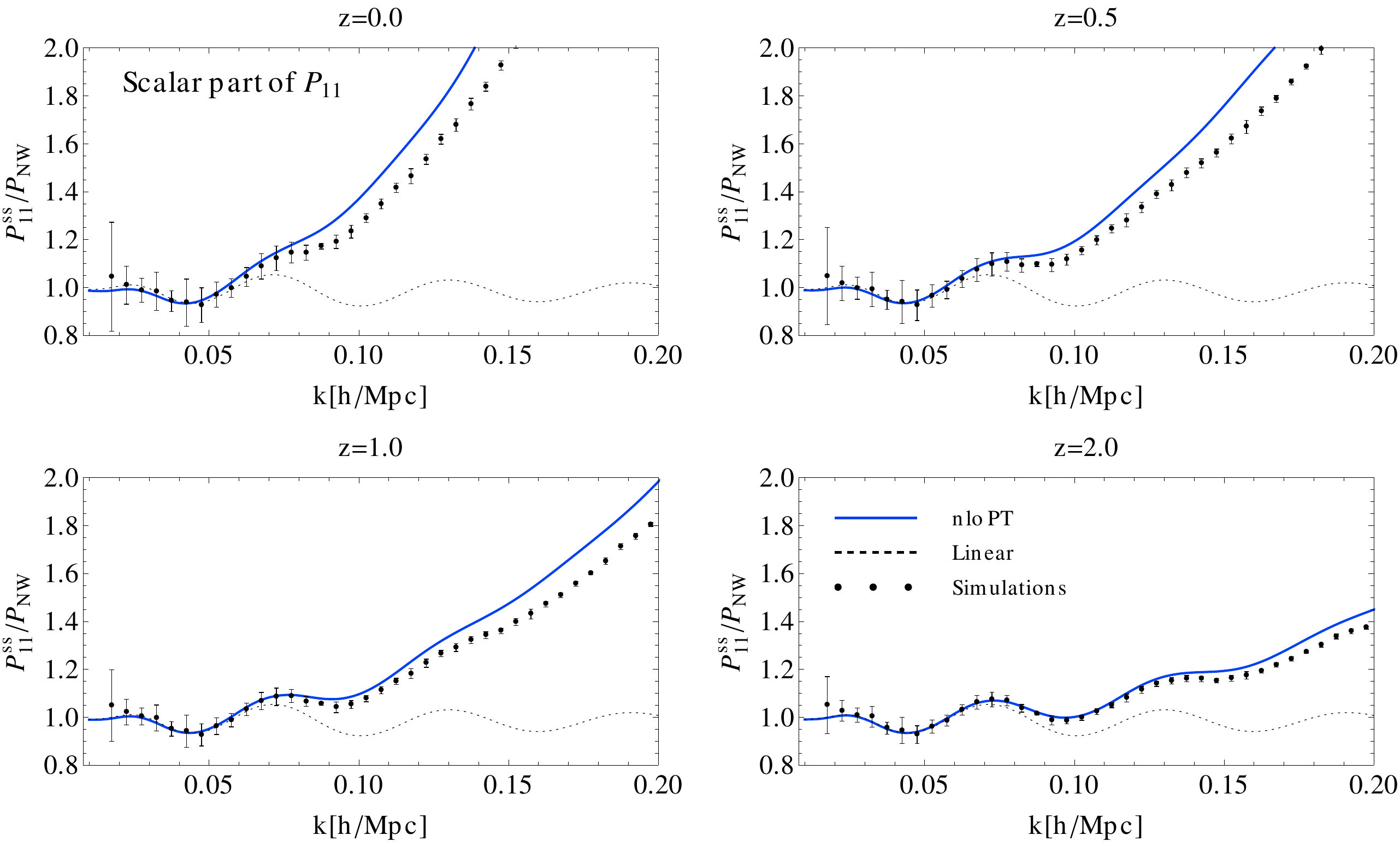}
    \caption{\small $k$-dependence of the scalar part of $P^{ss}_{11}$ term. 
Power spectrum is plotted at four redshifts $z=0.0,~0.5,~1.0$ and $2.0$. 
    This term has a simple $\mu^4$ dependence. Here we show linear Kaiser (black,
    dotted) and one loop PT (blue, solid) result, and compare it to
    simulation measurements (black dots). The error bars show the variance among realizations in simulations.
    The power spectra are divided by the no-wiggle linear term. }
    \label{fig:P11.1}
\end{figure}

In figure \ref{fig:P11.1} we show scalar part of $P^{ss}_{11}$ which comes from scalar contributions. It has a simple $\mu^4$ angular dependence, and corresponds to the third  
Kaiser term. We divide the plots by this Kaiser limit, using the no-wiggle linear power spectrum. One loop PT results are compared to the 
simulation measurements. We see that PT is quite successful in reproducing the nonlinear evolution of this term. 

\begin{figure}[t]
    \centering
    \includegraphics[width=1.0\textwidth]{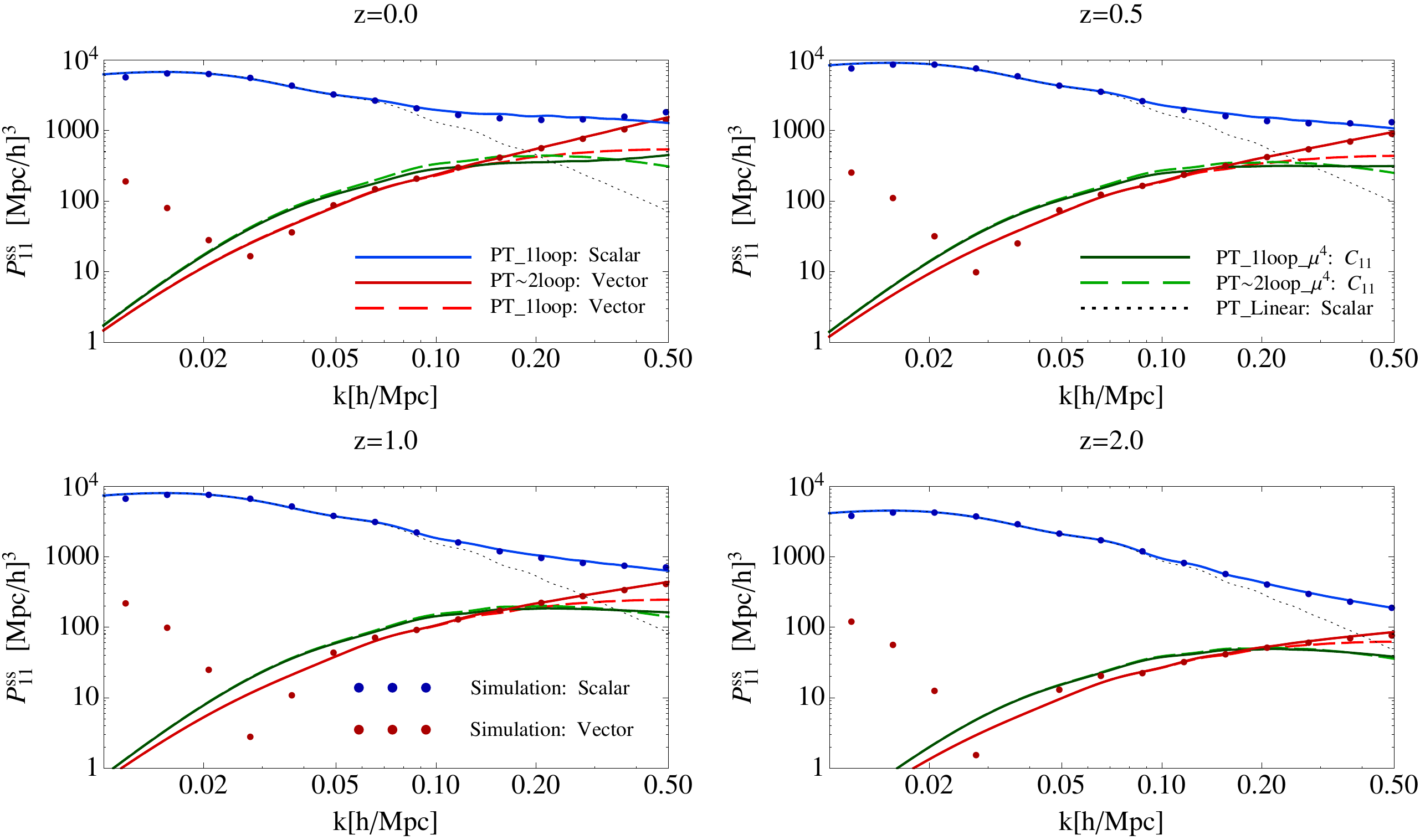}
    \caption{\small $k$-dependence of scalar and vector part of $P^{ss}_{11}$ 
term of the redshift power spectrum is plotted at four redshifts 
$z=0.0,~0.5,~1.0$ and $2.0$, assuming $\mu=1$. 
    Scalar part has simple $\mu^4$ angular dependence while the vector part has
 $~\mu^2(1-\mu^2)$ angular dependence at all (nonlinear) orders. 
    We show linear/Kaiser result (black, dotted), one loop PT
    result for scalar part (blue, solid), one loop PT
    result for vector part (lighter red, dashed), relevant part of two loop PT for
    vector part (red, solid) and simulations for scalar (blue points)
    and vector (red points) part. We also show scalar contributions of
    $C_{11}$ term
    at one (lighter green, dashed) and two (green, solid) loop order.}
    \label{fig:P11.2}
\end{figure}

In figure \ref{fig:P11.2} we show the vector $\mu^2(1-\mu^2)$ part. We
see that one loop PT is successful in 
reproducing the simulations for $k<0.2h/Mpc$ (the disagreement for
$k<0.03h/Mpc$ is likely numerical), and adding two
loop corrections increases these rage to larger $k$.
We also see that this vector contribution is considerably smaller
than the scalar part for $\mu=1$ for most of the $k$-range shown here, becoming comparable only at $k \sim 0.5h/Mpc$.  
However, because this vector term scales as $\mu^2$ while the linear scalar term scales as $\mu^4$, the vector 
terms always dominates for sufficiently small $\mu$. So for $\mu=0.1$ the nonlinear vector part exceeds linear 
scalar part already at $k \sim 0.05h/Mpc$.  

\subsection{$P_{02}(\VEC{k})$}
\label{sec:P02}

At orders higher than $P_{11}(k)$ there are no linear contributions, hence these terms are usually not of interest for extracting 
the cosmological information. However, these terms are known to be important on 
surprisingly large scales. 
These terms have usually been modeled phenomenologically in terms of adopting a simple functional form for $k$ and $\mu$ dependence 
and are related to the so called Fingers-of-God (FoG) effect. 
We begin with the dominant $P_{02}$ term, which, as we will see,  is the last term to contribute to $\mu^2$ dependence.
 
We correlate the scalar density filed $T^{0,0}_0=\df$ with
the tensor field $T^{2,m}_l$. Since scalars only correlate with scalars,
there are only two different terms that contribute \cite{Seljak:2011tx},
\begin{align}
     P_{02}(\VEC{k})=P^{0,2,0}_{0,0}(k)\left[P^0_0(\mu)\right]^2+P^{0,2,0}_{0,2}(k)P^0_0(\mu)P^0_2(\mu).
\label{eq:P02.0}
\end{align}
In terms of the contribution to the redshift space power spectrum this gives
\begin{align}
     P^{ss}_{02}=-\left(\frac{k\mu}{\mathcal{H}}\right)^2\left[P^{0,2,0}_{0,0}(k)+\frac{1}{2}P^{0,2,0}_{0,2}(k)(3\mu^2-1)\right].
\label{eq:P02.1}
\end{align}
The first term is the correlation between the isotropic part of the mass weighted square of velocity, i.e. the energy density $T^{2,0}_0=(1+\df)v^2$, 
and the density field $T^{0,0}_0=\df$, and the second term comes from the scalar part of the anisotropic stress $T^{2,0}_2$, correlated with the density $T^{
0,0}_0 =\df$.

Before using PT to model these terms let us consider what we can expect from 
physical grounds. As argued in ~\cite{Seljak:2011tx}, in systems with a large 
rms velocity, the first,
isotropic part $P^{0,2,0}_{0,0}$ should scale as $~P_{00}(k)\sigma^2$, where $\sigma^2$ has units of velocity square and 
includes the small scale velocity dispersion generated inside nonlinear halos. Some of 
this contribution cannot be modeled with simple fluid based PT, since not all of velocity
dispersion is captured in this approach. As a result, we should not even hope that PT can be reliable for this term: we will 
need to add an extra contribution to account for the small scale velocity dispersion. 

Expanding the fields we can write the contributing terms as following
\begin{equation}
     P_{02}(\VEC{k},\tau)=-D^4(\tau)(A_{02}(\VEC{k})+B_{02}(\VEC{k})),
\label{eq:P02.3}
\end{equation}
where we have
\begin{align}
     (2\pi)^3A_{02}(\VEC{k})\df^D(\VEC{k}-\VEC{k}')&=\int{\frac{d^3q}{(2\pi)^3}~\frac{q_\pp}{q^2}}
     \frac{(\VEC{k}'-\VEC{q})_\pp}{(\VEC{k}'-\VEC{q})^2}\left\langle \df(\VEC{k})|\theta^*
     (\VEC{q})\theta^*(\VEC{k}'-\VEC{q})\right\rangle,\nonumber\\
     (2\pi)^3B_{02}(\VEC{k})\df^D(\VEC{k}-\VEC{k}')&=\int{\frac{d^3q}{(2\pi)^3}\frac{d^3q'}{(2\pi)^3}~\frac{q_\pp}
     {q^2}\frac{q'_\pp}{q'^2}}\left\langle \df(\VEC{k})|\theta^*(\VEC{q})\theta^*(\VEC{q}')
     \df^*(\VEC{k}'-\VEC{q}-\VEC{q}')\right\rangle.
\label{eq:P02.4}
\end{align}
Using the one loop PT to evaluate this terms we expand these terms in the following way
\begin{align}
     &A_{02}(\VEC{k})=A_{02}^{(211)}(\VEC{k})+A_{02}^{(121)}(\VEC{k})+A_{02}^{(112)}(\VEC{k}),\nonumber\\
     &B_{02}(\VEC{k})=B_{02}^{(1111)}(\VEC{k}),
\label{eq:P02.5}
\end{align}
which after some computation give
\begin{align}
     A_{02}^{(211)}(\VEC{k})&=2(f\mathcal{H})^2
     \int{\frac{d^3q}{(2\pi)^3}~\frac{q_\pp}{q^2}\frac{(\VEC{k}-\VEC{q})_\pp}{(\VEC{k}-\VEC{q})^2}
     F_2^{(s)}(\VEC{q},\VEC{k}-\VEC{q})P_L(\VEC{q})P_L(\VEC{k}-\VEC{q})}\nonumber\\
     &=(f\mathcal{H})^2 k^{-2} \left(I_{02}(k)+\mu^2I_{20}(k)\right),\nonumber\\
     A_{02}^{(121)}(\VEC{k})&=A_{20}^{(112)}(\VEC{k})=2(f\mathcal{H})^2P_L({k})
     \int{\frac{d^3q}{(2\pi)^3}~\frac{q_\pp}{q^2}\frac{(\VEC{k}-\VEC{q})_\pp}{(\VEC{k}-\VEC{q})^2}
     G_2^{(s)}(\VEC{q},-\VEC{k})P_L(\VEC{q})}\nonumber\\
     &=(f\mathcal{H})^2 P_L({k}) \left(J_{02}(k)+\mu^2J_{20}(k)\right),\nonumber\\
     B_{02}^{(1111)}(\VEC{k})&=-(f\mathcal{H})^2 P_L(k)
     \int{\frac{d^3q}{(2\pi)^3}~\frac{q_\pp^2}{q^4}P_L(\VEC{q})}\nonumber\\
     &=-(f\mathcal{H})^2 P_L(k) \sigma_v^2.
\label{eq:P02.6}
\end{align}
Putting together all of the above we obtain for the $P_{02}$ contribution to the total redshift power spectrum
\begin{align}
   P^{ss}_{02}(\VEC{k},\tau)&=-\left(\frac{k\mu}{\mathcal{H}}\right)^2 P_{02}(\VEC{k})&\nonumber\\
   &=f^2(\tau)D^4(\tau)\mu^2\bigg[I_{02}(k)+k^2\Big(2J_{02}(k)-\sigma_v^2\Big)P_L(k)+\mu^2\Big(I_{20}(k)+2k^2J_{20}(k)P_L(k)\Big)\bigg].
\label{eq:P02.7}
\end{align}
As we mentioned above, we have the contribution of form $-\mu^2k^2\sigma^2P_L(k)$,
which suppresses the linear power spectrum with a $k^2$ like effect, increasing towards higher $k$. 
This is the lowest order FoG term, which we see contributes as $(k\mu)^2$ and so effects the $\mu^2$ term of total $P_{ss}$.  
Because small scale velocity dispersion effects cannot be modeled by 
PT, which is restricted to the weakly non-linear regime, we will consider
a model where we add to the PT predicted value for velocity dispersion
$\sigma_v^2$ the contributions coming from
small scales. In the equations above
we can then replace $\sigma_v^2\to\sigma_v^2+\sigma_{02}^2/(f\mathcal{H}D)^2$, where
$\sigma_{02}^2$ is the small scale addition to the velocity
dispersion, and which we treat here as a free parameter. List of
values used here for these parameters (depending on redshift), is given in the 
table
\ref{tb:SigPar}, in section \ref{sec:hm}. In these section we also
consider the explanation of these values using the halo model, see for
example \cite{Seljak:2000}. 
In addition to small scale velocity dispersion model, we also include the most relevant higher order PT terms.
If we consider higher order contributions to $\left\langle \df |\df v_\pp^2\right\rangle$ term we see that it has subsets of diagrams 
where  $\left\langle \df \right|$ is not connected to any of the velocity fields, so we can write $\left\langle \df |\df v_\pp^2\right\rangle=\
\left\langle \df|\df\right\rangle \left\langle v_\pp^2\right\rangle$. Formally, in one loop computation only the leading 
term of this subset contributes in equation \ref{eq:P02.6}.
Collecting these we see that we can model $B_{02}$ term by replacement
\begin{align}
  D^4(\tau) B_{02}({k})=-\left(f\mathcal{H}D^2\right)^2 \sigma_{v}^2P_L(k)~\to-(f\mathcal{H}D)^2\left(\sigma_{v}^2+\sigma_{02}^2/(f\mathcal{H}D)^2\right)P_{00}(k,\tau).
\end{align}
 
\begin{figure}[t]
    \centering
    \includegraphics[width=1.0\textwidth]{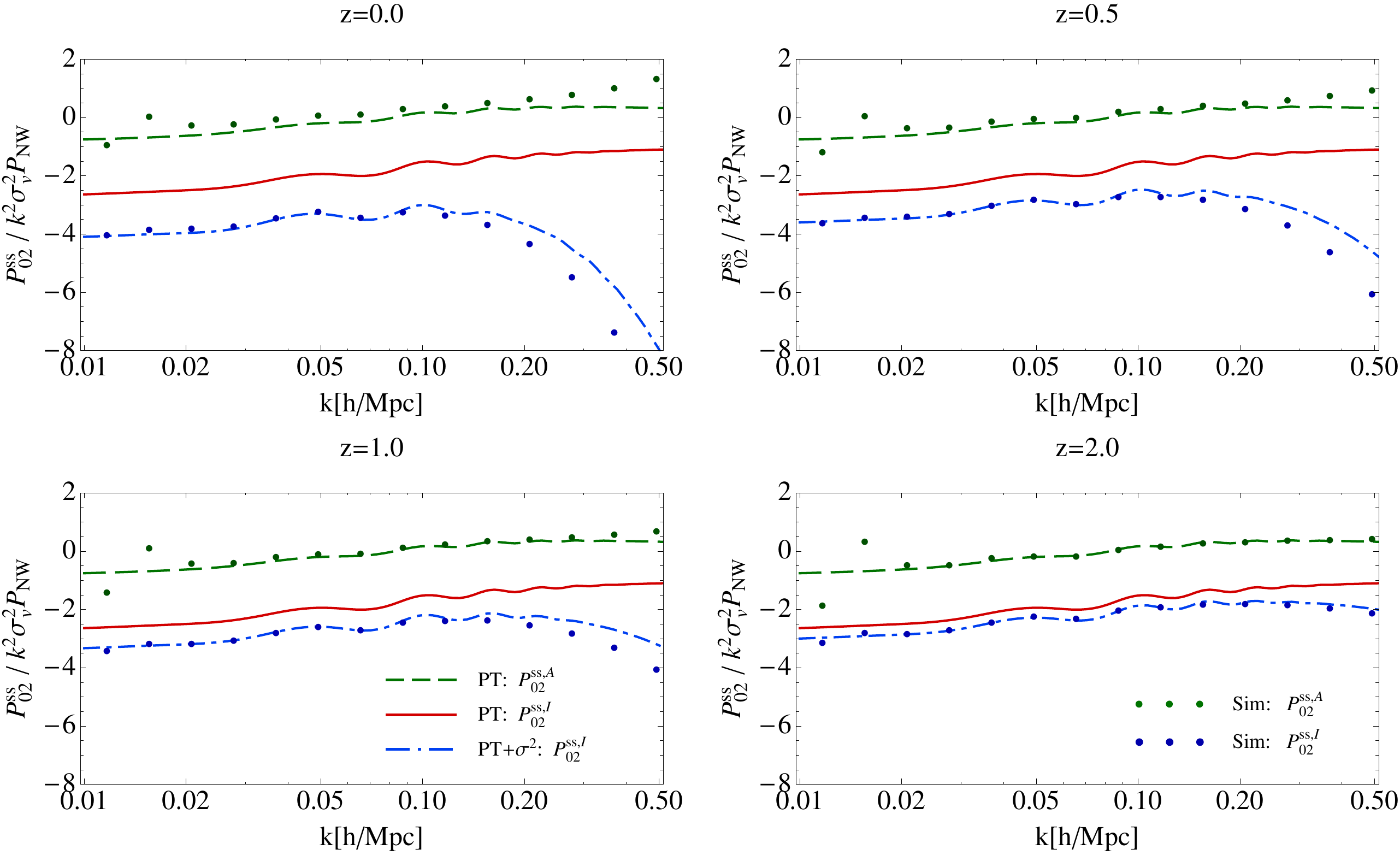}
    \caption{\small $k$-dependence of isotropic and anisotropic part of $P^{ss}_{02}$ term of redshift power spectrum is plotted at four redshifts $z=0.0,~0.5,~1.0$ and $2.0$. 
    Isotropic part $P^{ss,I}_{02}$, computed in one loop PT (red, solid) is plotted, as well as using the model presented above (blue, dot-dashed). Isotropic part has simple $~\mu^2$ angular dependence while the 
    anisotropic part $P^{ss,A}_{02}$ (green, dashed) has $~\mu^2(3\mu^2-1)/2$ angular dependence. Simulation measurements (dots) for the corresponding terms are also presented.
    The power spectra are divided by $k^2\sigma_v^2P^{\text{nw}}_L$
    without the wiggles.}
    \label{fig:P02}
\end{figure}

In order to discuss the results let us first rewrite equation
\ref{eq:P02.7} in  form of isotropic and anisotropic part as for
$P_{02}$. We have
$P^{ss}_{02}=\mu^2\left(P^{ss,I}_{02}+\frac{1}{2}(3\mu^2-1)P^{ss,A}_{02}\right)$,
where
\begin{align}
   P^{ss,I}_{02}(\VEC{k},\tau)=&\frac{f^2(\tau)D^4(\tau)}{3}  \bigg[
   3I_{02}(k)+I_{20}(k)+2k^2\big(3J_{02}(k)+J_{20}(k)\big)P_L(k)
   \bigg]\nonumber\\
   &~-f^2(\tau)D^2(\tau)k^2\left(\sigma_{v}^2+\sigma_{02}^2/(f\mathcal{H}D)^2\right)P_{00}(k,\tau),\nonumber\\
   P^{ss,A}_{02}(\VEC{k},\tau)=&\frac{2f^2(\tau)D^4(\tau)}{3}\bigg[I_{20}(k)+2k^2J_{20}(k)P_L(k)\bigg].
\label{eq:P02.8}
\end{align}
In figure \ref{fig:P02} we show isotropic and anisotropic part of the $P_{02}$ contribution to the total redshift power spectrum. All power spectrum contributions are divided 
by the $(fDk)^2\sigma_v^2P^{\text{nw}}_{L}(k)$, where we again used
the no-wiggle power spectrum. We can see that the contribution to
$\mu^2$ is always negative, while the corresponding vector term
from $P_{11}$ always adds power and the two partially cancel out \cite{Seljak:2011tx}. As we see the scalar anisotropic
stress-density correlator $P^{0,2,0}_{0,2}$ contributes to the 
$\mu^2$ angular term, as well as to the $\mu^4$
term. The anisotropic term is reasonably well
modeled with PT and has smaller magnitude than the isotropic term, as expected, since the velocity dispersion in virialized objects is
essentially isotropic. The isotropic term is poorly modeled with just PT: we need a significant contribution from the 
small scale velocity dispersion, which can be seen to essentially double the amplitude of this term at low $k$, and 
far more than that at high $k$. 
In figure \ref{fig:P02}  we can see that this term helps the model
considerably, but of course it has one free parameter. 

We can also write this result in powers of $\mu$,
\begin{align}
   P^{ss}_{02}[\mu^2]&=f^2(\tau)D^4(\tau)\Big( I_{02}(k)+2k^2J_{02}(k)P_L(k) \Big)-f^2(\tau)D^2(\tau)k^2\left(\sigma_{v}^2+\sigma_{02}^2/(f\mathcal{H}D)^2\right)P_{00}(k,\tau),\nonumber\\
   &=\bar{P}^{ss}_{02}[\mu^2]-f^2(\tau)D^2(\tau)k^2\left(\sigma_{v}^2+\sigma_{02}^2/(f\mathcal{H}D)^2\right)P_{00}(k,\tau),\nonumber\\
   P^{ss}_{02}[\mu^4]&=\bar{P}^{ss}_{02}[\mu^4]=f^2(\tau)D^4(\tau)\Big(I_{20}(k)+2k^2J_{20}(k)P_L(k)\Big),
\label{eq:P02.9}
\end{align}
where we have implicitly defined $\bar{P}^{ss}_{02}$ by omitting the
velocity dispersion part from $P^{ss}_{02}$.

\subsection{$P_{12}(\VEC{k})$}
\label{sec:P12}

As we can see the lowest order in $\mu$ with which correlators contribute is the $L+L'$ or $L+L'+1$. So contributions to $\mu^2$ comes only 
from terms $P_{01}$, $P_{11}$ and $P_{02}$. The next order in powers of $\mu^2$ will be $\mu^4$ terms. As we have seen $P_{11}$ and $P_{02}$
also have contributions to $\mu^4$, with $P_{11}$ having the linear
order term which dominates on large scales. 

Here we correlate the momentum filed $T^{1,m}_l$ with the tensor field
$T^{2,m}_l$. Because of rotational invariance we can correlate only
scalar to scalar field and vector to vector field
\begin{align}
     P_{12}(\VEC{k})=P^{1,2,0}_{1,0}(k)\left[P_0^0(\mu)P_1^0(\mu)\right]+P^{1,2,0}_{1,2}(k)\left[P_1^0(\mu)P_2^0(\mu)\right]+P^{1,2,1}_{1,2}(k)\left[P_1^1(\mu)P_2^1(\mu)\right]
\label{eq:P12.0}
\end{align}
In terms of the contribution to the redshift space power spectrum this gives
\begin{align}
     P^{ss}_{12}(\VEC{k})=-i\left(\frac{k\mu}{\mathcal{H}}\right)^3\left[P^{1,2,0}_{1,0}(k)\mu+\frac{1}{2}P^{1,2,0}_{1,2}(k)\mu(3\mu^2-1)+3P^{1,2,1}_{1,2}(k)\mu(1-\mu^2)\right].
\label{eq:P12.1}
\end{align}

Using the one loop PT we get both $\mu^4$ and $\mu^6$ angular terms, 
but since there are 3 terms we cannot 
distinguish between them in equation \ref{eq:P12.1}.
Using equation \ref{eq:T123con} and one loop PT we get
\begin{equation}
     P_{12}(\VEC{k})=-D^4(\tau)\frac{i}{k^2}\left(k_\pp A_{12}(\VEC{k})+k^2 B_{12}(\VEC{k})+k_\pp C_{12}(\VEC{k})\right),
\label{eq:P12.2}
\end{equation}
where the contributing terms are
\begin{align}
     &(2\pi)^3A_{12}(\VEC{k})\df^D(\VEC{k}-\VEC{k}')=\int{\frac{d^3q}{(2\pi)^3}~\frac{q_\pp}{q^2}}
     \frac{(\VEC{k}'-\VEC{q})_\pp}{(\VEC{k}'-\VEC{q})^2}\left\langle \theta(\VEC{k})|\theta^*
     (\VEC{q})\theta^*(\VEC{k}'-\VEC{q})\right\rangle,\nonumber\\
     &(2\pi)^3B_{12}(\VEC{k})\df^D(\VEC{k}-\VEC{k}')=\int{\frac{d^3q~d^3q'}{(2\pi)^6}~\frac{q_\pp}{q^2}
     \frac{q'_\pp}{q'^2}}\frac{(\VEC{k}'-\VEC{q}')_\pp}{(\VEC{k}'-\VEC{q}')^2}\left\langle
     \theta(\VEC{q})\df(\VEC{k}-\VEC{q})|\theta^*(\VEC{q}')\theta^*(\VEC{k}'-\VEC{q}')\right\rangle,\nonumber\\
     &(2\pi)^3C_{12}(\VEC{k})\df^D(\VEC{k}-\VEC{k}')=\int{\frac{d^3q~d^3q'}{(2\pi)^6}~\frac{q_\pp}{q^2}
     \frac{q'_\pp}{q'^2}}\left\langle \theta(\VEC{k})|\theta^*(\VEC{q})\theta^*(\VEC{q}')\df^*(\VEC{k}'-\VEC{q}-\VEC{q}')\right\rangle.
\label{eq:P12.3}
\end{align}

The first of these terms we can be expanded further 
\begin{align}
     &A_{12}(\VEC{k})=A_{12}^{(211)}(\VEC{k})+A_{1}^{(121)}(\VEC{k})+A_{1}^{(112)}(\VEC{k}),
\label{eq:P12.4}
\end{align}
and computing these terms gives;
\begin{align}
     A_{12}^{(211)}(\VEC{k})&=-2(f\mathcal{H})^3
     \int{\frac{d^3q}{(2\pi)^3}~\frac{q_\pp}{q^2}\frac{(\VEC{k}-\VEC{q})_\pp}{(\VEC{k}-\VEC{q})^2}
     G_2^{(s)}(\VEC{q},\VEC{k}-\VEC{q})P_L(\VEC{q})P_L(\VEC{k}-\VEC{q})}\nonumber\\
     &=-(f\mathcal{H})^3 k^{-2} \left(I_{12}(k)+\mu^2I_{21}(k)\right),\nonumber\\
     A_{12}^{(121)}(\VEC{k})&=A_{12}^{(112)}(\VEC{k})=-2(f\mathcal{H})^3P_L(\VEC{k})
     \int{\frac{d^3q}{(2\pi)^3}~\frac{q_\pp}{q^2}\frac{(\VEC{k}-\VEC{q})_\pp}{(\VEC{k}-\VEC{q})^2}
     G_2^{(s)}(\VEC{q},-\VEC{k})P_L(\VEC{q})}\nonumber\\
     &=-(f\mathcal{H})^3 P_L(\VEC{k}) \left(J_{02}(k)+\mu^2J_{20}(k)\right),\nonumber\\
     B_{12}(\VEC{k})&=-2(f\mathcal{H})^3\int{\frac{d^3q}{(2\pi)^3}~\frac{q_\pp^2}{q^4}
     \frac{(\VEC{k}-\VEC{q})_\pp}{(\VEC{k}-\VEC{q})^2}P_L(\VEC{q})P_L(\VEC{k}-\VEC{q})}\nonumber\\
     &=(f\mathcal{H})^3 \mu k^{-3}\left(I_{03}(k)+\mu^2 I_{30}(k)\right),\nonumber\\
     C_{12}(\VEC{k})&=(f\mathcal{H})^3 P_L(\VEC{k})
     \int{\frac{d^3q}{(2\pi)^3}~\frac{q_\pp^2}{q^4}P_L(\VEC{q})},\nonumber\\
     &=(f\mathcal{H})^3\sigma_v^2P_L(\VEC{k}).
\label{eq:P12.5}
\end{align}
All this gives us the contribution to total redshift space power spectrum
\begin{align}
   P^{ss}_{12}(\VEC{k},\tau)=-i\left(\frac{k\mu}{\mathcal{H}}\right)^3 P_{12}(\VEC{k},\tau)=&f(\tau)^3D(\tau)^4\mu^4
   \bigg[I_{12}(k)-I_{03}(k)+2k^2J_{02}(k)P_L(k)-k^2\sigma_v^2P_L(k) \nonumber\\
   & +\mu^2\Big(I_{21}(k)-I_{30}(k)+2k^2J_{20}P_L(k)\Big)\bigg].
\label{eq:P12.6}
\end{align}
 We can again add the 
small scale velocity dispersion in by hand, as was done and explained
in case of $P_{02}$.
Considering the relevant higher order contributions we see that the isotropic part of
$\left\langle T_\pp^1|\df v_\pp^2\right\rangle$ can be modeled by
\begin{align}
 -iD^4(\tau)\frac{\mu}{k}C_{12}(k)=-iD^4(\tau)\frac{\mu}{k}(f\mathcal{H})^3\sigma_v^2 P_L(k)\to-(f\mathcal{H}D)^2\left(\sigma_{v}^2+\sigma_{12}^2/(f\mathcal{H}D)^2\right)P_{01}(\VEC{k},\tau),\nonumber
\end{align}
where we again treat small scale velocity dispersion $\sigma_{12}^2$
as a free parameter with values for different redshifts given in table
\ref{tb:SigPar}. These values are the same as for $P_{02}$ case, and
the reasons and explanation in term of halo model is given in section \ref{sec:hm}.

As mentioned earlier, since we have only $\mu^4$ and $\mu^6$ angular
dependence we can not determine all three terms in equation
\ref{eq:P12.1} separately. Let us instead separate the angular dependences itself and collect the terms
\begin{align}
   P^{ss}_{12}\left[\mu^4\right]
   &=\left[P^{ss}_{12}\right]_{1,0}^{1,2,0}-\frac{1}{2}\left(\left[P^{ss}_{12}\right]_{1,2}^{1,2,0}-6\left[P^{ss}_{12}\right]_{1,2}^{1,2,1}\right) \nonumber\\
   & =f(\tau)^3D(\tau)^4\Big[I_{12}(k)-I_{03}(k)+2k^2J_{02}(k)P_L(k) \Big]\nonumber\\
   &\qquad-\frac{1}{2}f(\tau)^2D(\tau)^2k^2\left(\sigma_{v}^2+\sigma_{12}^2/(f\mathcal{H}D)^2\right)P^{ss}_{01}(k,\tau),\nonumber\\
                               &=\bar{P}^{ss}_{12}\left[\mu^4\right]-\frac{1}{2}f(\tau)^2D(\tau)^2k^2\left(\sigma_{v}^2+\sigma_{12}^2/(f\mathcal{H}D)^2\right)P^{ss}_{01}(k,\tau) \nonumber\\
   P^{ss}_{12}\left[\mu^6\right]&=\bar{P}^{ss}_{12}\left[\mu^6\right]=\frac{3}{2}\left(\left[P^{ss}_{12}\right]_{1,2}^{1,2,0}-2\left[P^{ss}_{12}\right]_{1,2}^{1,2,1}\right) \nonumber\\
                                & =f(\tau)^3D(\tau)^4\Big[I_{21}(k)-I_{30}(k)+2k^2J_{20}P_L(k)\Big],
\label{eq:P12.6}
\end{align}
where we again implicitly define $\bar{P}^{ss}_{12}$, by omitting the
velocity dispersion part. 

\begin{figure}[t]
    \centering
    \includegraphics[width=1.0\textwidth]{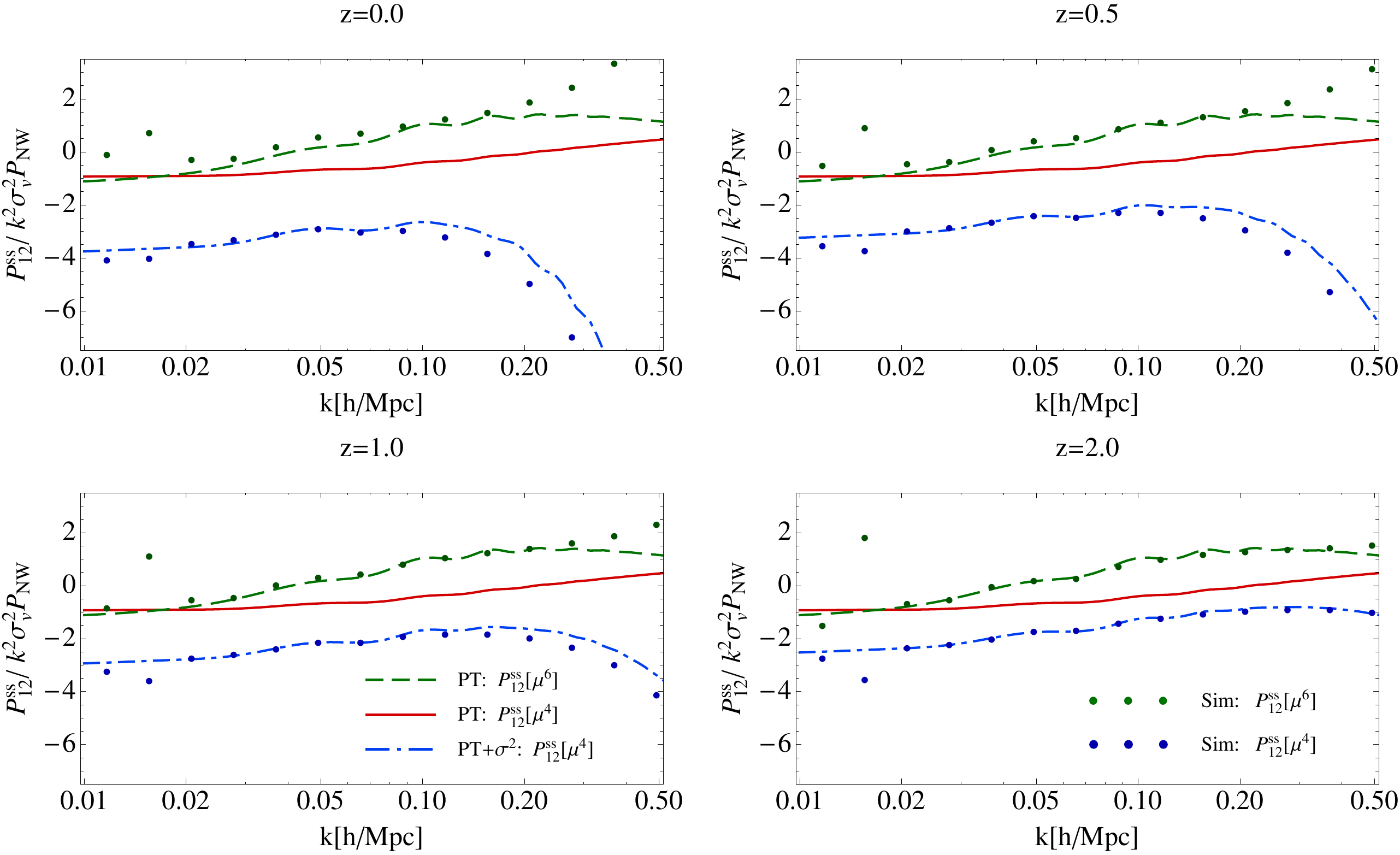}
    \caption{\small $k$-dependence of $\mu^4$ and $\mu^6$ part of $P^{ss}_{12}$ term of redshift power spectrum is plotted at four redshifts, $z=0.0,~0.5,~1.0$ and $2.0$. 
    $\mu^4$ part $P^{ss}_{02}[\mu^4]$ is computed in one loop PT (red, solid) regime, and using model presented above (blue, dot-dashed). We show $\mu^6$ part $P^{ss}_{02}[\mu^6]$
    computed in one loop PT (green, dashed), and simulation measurements (dots) for the corresponding terms.
    All power spectra are divided by $k^2\sigma_v^2P^{\text{nw}}_L$ without the wiggles. }
    \label{fig:P12}
\end{figure}

In figure \ref{fig:P12} we show $P^{ss}_{12}\left[\mu^4\right]$ and $P^{ss}_{12}\left[\mu^6\right]$ parts to the total redshift power spectrum. Power spectrum contributions are divided 
by the $(fDk)^2\sigma^2P^{\text{nw}}_{L}(k)$, where we again used no-wiggle power spectrum. 

\subsection{$P_{22}(\VEC{k})$}
\label{sec:P22}

\begin{figure}[t]
    \centering
    \includegraphics[width=1.0\textwidth]{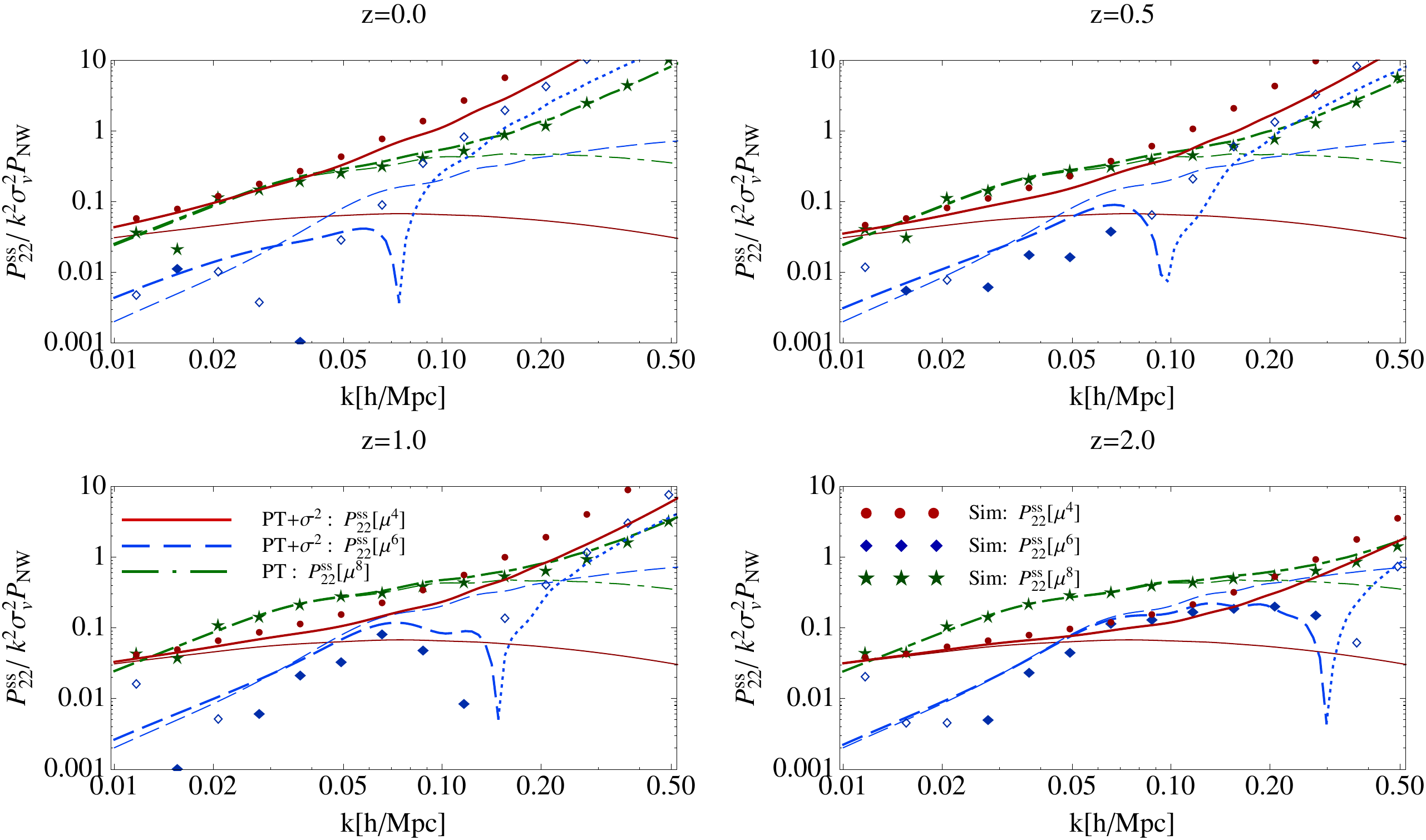}
    \caption{\small $k$-dependence of $\mu^4$, $\mu^6$ and $\mu^8$
      parts of $P^{ss}_{22}$ term of redshift power spectrum is
      plotted at four redshifts, $z=0.0,~0.5,~1.0$ and $2.0$. 
    $\mu^4$ part $P^{ss}_{22}[\mu^4]$ is shown for one loop PT (red, solid,
    thin) regime, and for improved two loop PT model with small scale velocity dispersion  (red,
    solid, thick), as well as for simulations (red
    dots). Similarly, $P^{ss}_{22}[\mu^6]$ part is shown
    using one loop PT (blue, dashed, thin) result, improved two loop
    PT model with small scale velocity dispersion 
    (blue, dashed/dotted, thick), and simulations (blue,
    empty/full, diamonds). Dashed/full results present positive values
    of $\mu^6$ dependence and dotted/empty negative values. $\mu^8$
    contribution is shown for one (thin, green, dot-dashed) and two
    loop PT (thick, green, dot-dashed), with the 
corresponding simulations (green stars). 
    All power spectra are divided by $k^2\sigma_v^2P^{\text{nw}}_L$
    without the wiggles.}
    \label{fig:P22}
\end{figure}

Next we consider correlator of tensor $T^{2,m}_l$ field  with itself. This term
will give $\mu^4$, $\mu^6$ and $\mu^8$ contributions. One loop PT gives first order contributions to all of these angular terms. 
From the expansion of power spectrum \ref{eq:PLL} we can see that the constant contribution to $P_{22}$ is coming from the scalar term,
$P^{2,2,0}_{0,0}$ and partially from $P^{2,2,0}_{0,2}$ and $P^{2,2,0}_{2,2}$. This will give $\mu^4$ as the lowest order contribution to 
the total $P^{ss}_{22}$, and all of the other terms will come as $\mu^6$, $\mu^8$. Let us now assess these contributions using
one loop PT
\begin{align}
     \bar{P}_{22}(\VEC{k},\tau)&=2(f\mathcal{H}D)^4\int{\frac{d^3q}{(2\pi)^3}~\left[\frac{q_\pp}{q^2}
     \frac{(\VEC{k}-\VEC{q})_\pp}{(\VEC{k}-\VEC{q})^2}\right]^2P_L(\VEC{q})P_L(\VEC{k}-\VEC{q})} \nonumber\\
     &=(f\mathcal{H}D)^4 \frac{1}{4}k^{-4}\bigg(I_{23}(k)+2\mu^2I_{32}(k)+\mu^4I_{33}(k)\bigg),
\label{eq:P22.1}
\end{align}
which gives rise to the total red shift power spectrum contribution
\begin{equation}
     P^{ss}_{22}(\VEC{k},\tau)=\frac{1}{4}\left(\frac{k \mu}{\mathcal{H}}\right)^4\bar{P}_{22}(\VEC{k},\tau)=\frac{1}{16}f^4(\tau)D^4(\tau)\mu^4
     \bigg(I_{23}(k)+2\mu^2I_{32}(k)+\mu^4I_{33}(k)\bigg).
\label{eq:P22.2}
\end{equation}
These are the leading order contributions to the angular dependence of this term. Now let us also investigate the
most important contributions from the higher orders. For that purpose let us write the full correlator in terms of the 
density and velocity fields
\begin{align}
 \left\langle T^{2}_\pp\right.\left| T^2_\pp\right\rangle = 
 \left\langle v^{2}_\pp\right.\left|v^2_\pp\right\rangle+2\left\langle v^{2}_\pp\right.\left| \df v^2_\pp\right\rangle
 +\left\langle \df v^{2}_\pp\right.\left|\df v^2_\pp\right\rangle.
\label{eq:P22.3}
\end{align}
In equation \ref{eq:P22.2} we have considered only the first of these
three terms, but we should also include some of the most important
contributions from the remaining terms. 
From two loop considerations first we improve the first
term \ref{eq:P22.1} by exchanging linear power spectrum $P_L$ with
one loop $P_{\tf\tf}$.  The most important contributions of the other two
terms can be modeled as 
\begin{align}
\left\langle v^{2}_\pp\right.\left| \df v^2_\pp\right\rangle &\sim (f\mathcal{H}D)^2\sigma_v^2 \bar{P}_{02},\nonumber\\
\left\langle \df v^{2}_\pp\right.\left|\df v^2_\pp\right\rangle &\sim (f\mathcal{H}D)^4 (\sigma^2_v)^2  P_{00}+\bar{P}_{22}\circ P_{00}.\nonumber
\end{align}
These are of course not the only higher order term, but after a
detailed analysis these terms turn out to be the most relevant and the 
rest can be neglected.  We can again include the
small scale velocity dispersion extending $\sigma_v^2\to\sigma_{v}^2+\sigma_{22}^2/(f\mathcal{H}D)^2$
as we did previously for $P_{02}$ and $P_{12}$. Combining all we obtain a model
\begin{align}
 P_{22}(k,\mu)=&\bar{P}_{22}(k,\mu)-2
 (f\mathcal{H}D)^2\left(\sigma_{v}^2+\sigma_{22}^2/(f\mathcal{H}D)^2\right)\bar{P}_{02}(k,\mu)\nonumber\\
&+(f\mathcal{H}D)^4\left(\sigma_{v}^2+\sigma_{22}^2/(f\mathcal{H}D)^2\right)^2P_{00}(k)+(\bar{P}_{22}\circ P_{00})(k).
\end{align}
In high $k$ limit last (convolution) term corresponds to $2(f\mathcal{H}D)^4\sigma_v^4P_{00}(k)$.
In figure \ref{fig:P22} we show the individual angular contributions for
one loop PT calculus and for the improved model suggested above, and compare them to
simulation measurements. We see that using the proposed model 
improves results in comparison to the one loop PT
contributions, but still only qualitatively agrees with the simulations. One would find much better agreement if 
not imposing $\sigma_{22}=\sigma_{02}$, i.e. with more free parameters. We mention that most of the correction to the $\mu^4$ term
comes from the isotropic modeling of the last $\left\langle \df v^{2}_\pp\right.\left|\df
  v^2_\pp\right\rangle$ terms. Term $\left\langle
  v^{2}_\pp\right.\left| \df v^2_\pp\right\rangle$ also contributes to 
$\mu^4$ but less than the previous term. Corrections to
the $\mu^6$ come from the angular dependency of $A_{02}$ term and we see
that it can explain the change of sign and scale growth trends. The additional terms do not affect the $\mu^8$ term, 
which is well predicted (at least relative to $\mu^4$ and $\mu^6$)
with two loop PT model of first term. 

\subsection{$P_{03}(\VEC{k})$,  $P_{13}(\VEC{k})$ and  $P_{04}(\VEC{k})$}
\label{sec:P03}

Our goal is to consider all terms at the $\mu^4$ order. There are 3 left. 
First we consider terms $P_{03}$ and $P_{13}$.  We correlate overdensity field or momentum field with rank three tensor field $T^3_\pp(\VEC{x})$.
Angular decomposition for $P_{03}$ is relatively simple since it has
only scalar contributions, but $P_{13}$ has scalar and vector contributions.
Using the angular expansion we get the following angular dependence
\begin{align}
     &P_{03}(\VEC{k})=P^{0,3,0}_{0,1}(k)\mu+P^{0,3,0}_{0,3}(k)\frac{1}{2} \mu\left(5 \mu^2-3\right),\nonumber\\
     &P_{13}(\VEC{k})=P^{1,3,0}_{1,1}(k)\mu^2+P^{1,3,1}_{1,1}(k)\left(1-\mu^2\right)+P^{1,3,0}_{1,3}(k)\frac{1}{2} \mu^2 \left(5 \mu^2-3\right)
     -P^{1,3,1}_{1,3}(k)\frac{3}{2} \left(5 \mu^4-6 \mu^2+1\right).
\label{eq:P03P13.1}
\end{align}
In one loop PT these terms are 
\begin{align}
     P_{03}(\VEC{k})&=i3(f\mathcal{H})^3D^4(\tau) 
     \mu k^{-1} P_L({k})\int{\frac{d^3q}{(2\pi)^3}~\left(\frac{q_\pp}{q^2}\right)^2P_L(q)} \nonumber\\
     &=i3(f\mathcal{H})^3D^4(\tau) \mu k^{-1} P_L(k)\sigma_v^2,\nonumber\\
     P_{13}(\VEC{k})&=3(f\mathcal{H}D(\tau))^4 \mu^2 k^{-2} P_L({k})
     \int{\frac{d^3q}{(2\pi)^3}~\left(\frac{q_\pp}{q^2}\right)^2P_L(q)} \nonumber\\
     &=3(f\mathcal{H})^4D^4(\tau) \mu^2 k^{-2}P_L(k)\sigma_v^2.
\label{eq:P03P13.2}
\end{align}
From angular decomposition of $P_{03}$ we have scalar terms,
$P^{0,3,0}_{0,1}$ and $P^{0,3,0}_{0,3}$, contributing with angular
dependence $\mu^4$ and $\mu^6$.  
One could evaluate these terms in PT, but at least two loop order is required for $\mu^6$, 
since in one loop order gives just $\mu^4$ dependence. For $P_{13}$ we see that the lowest 
angular dependence comes from the vector contribution and not the scalar,
although the scalar has lower perturbative order.
Similar case was
discussed for $P_{11}$,  where the vector part, which
is of one loop order, contributes to $\mu^2$, while the leading linear order scalar part has $\mu^4$ dependence.

\begin{figure}[htp]
    \centering
    \includegraphics[width=1.0\textwidth]{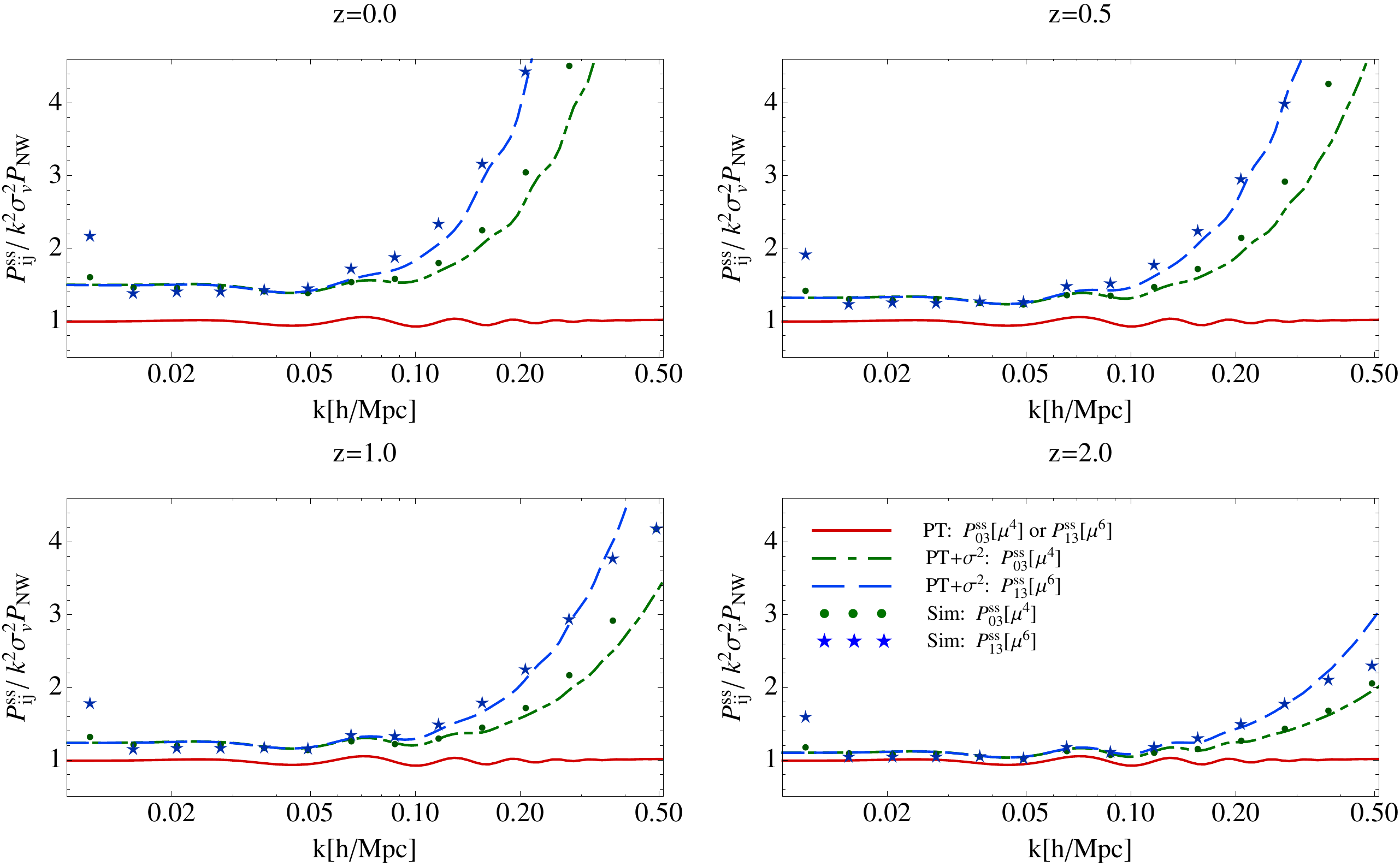}
    \caption{\small $\mu^4$ dependence of $P^{ss}_{03}$
      and $\mu^6$ dependence of $P^{ss}_{13}$ is plotted at four
      redshifts, $z=0.0,~0.5,~1.0$ and $2.0$. 
      One loop PT result is plotted (red, solid), as well as results
      of improved model discussed in the text for $P^{ss}_{03}[\mu^4]$ (blue,
      dashed) and $P^{ss}_{13}[\mu^6]$ (green, dot-dashed). Results are
      compared to the simulation measurements; $P^{ss}_{03}[\mu^4]$ (blue,
      stars) and $P^{ss}_{13}[\mu^6]$ (green,
      dots).  All the plots are
      divided by no-wiggle $-f^3(D\mu)^4k^2\sigma_v^2P^{\text{nw}}_L$ for
      $P^{ss}_{03}$ and $f\mu^2$ times this for the $P^{ss}_{13}$ term.}
    \label{fig:P0313}
\end{figure}

\begin{figure}[htp]
    \centering
    \includegraphics[width=1.0\textwidth]{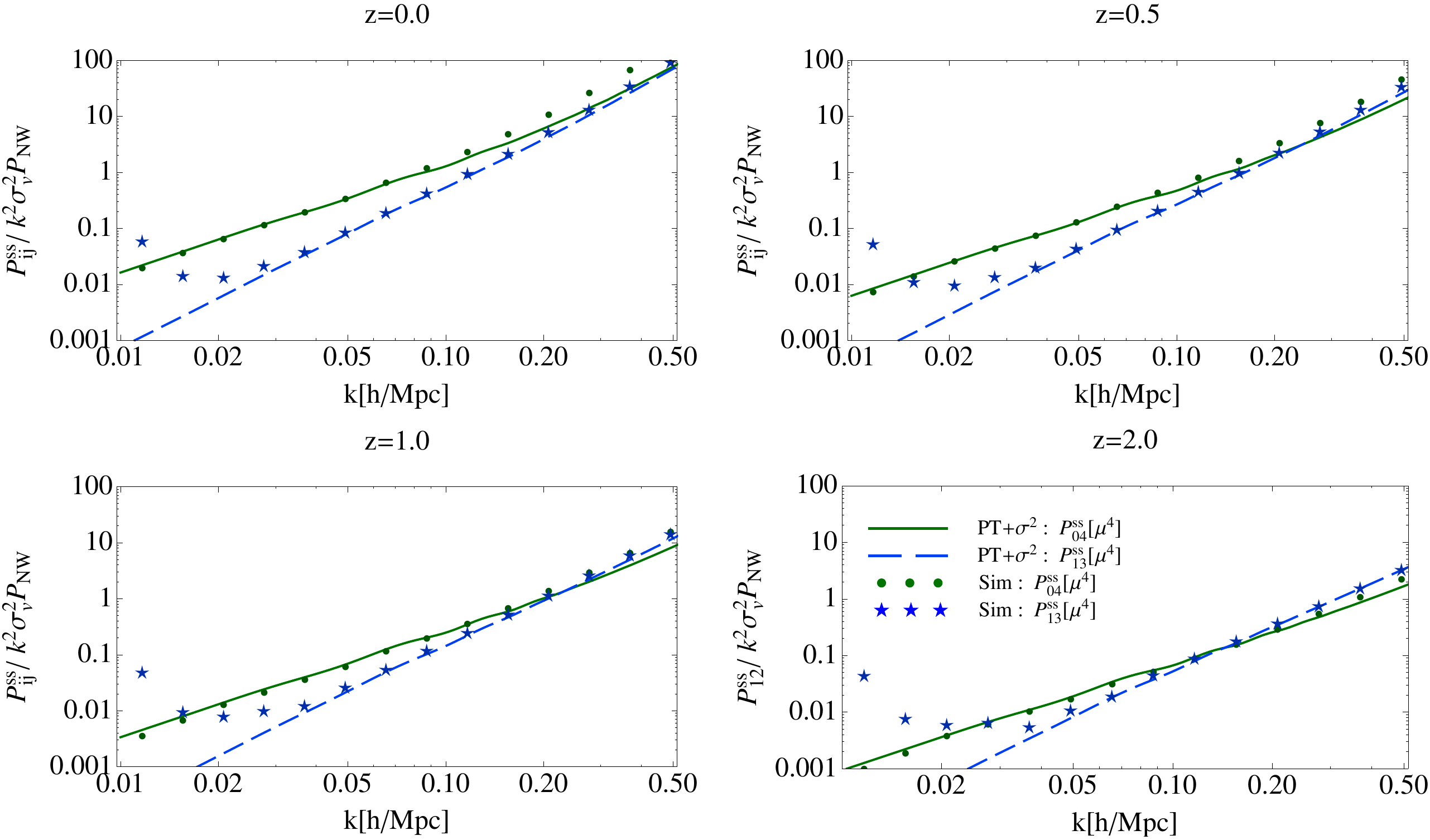}
    \caption{\small $\mu^4$ dependence of $P^{ss}_{13}$
      and $P^{ss}_{04}$ term is plotted at four
      redshifts, $z=0.0,~0.5,~1.0$ and $2.0$. 
      Simple modeled results for $P^{ss}_{13}$ (blue, dashed), and
      $P^{ss}_{04}$ (green, solid) are shown and compared to
      corresponding simulation measurements; $P^{ss}_{13}$ (blue, stars), and
      $P^{ss}_{04}$ (green, dots). Both, simulations and
      model results for $P^{ss}_{13}$ have negative values. All the plots are
      divided by no-wiggle $(fD\mu)^4k^2\sigma_v^2P^{\text{nw}}_L$.}
    \label{fig:P0413}
\end{figure}

One loop PT contribution to total $P^{ss}$ give
\begin{align}
     P^{ss}_{03}(\VEC{k},\tau)&=\frac{i}{3}\left(\frac{k\mu}{\mathcal{H}}\right)^3P_{03}(\VEC{k})=-f^3(\tau)D^4(\tau)\mu^4k^2\sigma_v^2P_L({k}), \nonumber\\
     P^{ss}_{13}(\VEC{k},\tau)&=-\frac{1}{3}\left(\frac{k\mu}{\mathcal{H}}\right)^4 P_{13}(\VEC{k})=-f^4(\tau)D^4(\tau)\mu^6k^2\sigma_v^2P_L({k}).
\label{eq:P03.2}
\end{align}
We can again include some higher order terms based on small scale velocity dispersion type arguments. 
For example, let us asses contributions to each of the terms above as if fully coming from
\begin{align}
  \left\langle T^0_\pp\right.\left| T^3_\pp\right\rangle& = 3\sigma^2_v\left\langle T^0_\pp\right.\left| T^1_\pp\right\rangle,\nonumber\\
  \left\langle T^1_\pp\right.\left| T^3_\pp\right\rangle& = 3\sigma^2_v\left\langle T^1_\pp\right.\left| T^1_\pp\right\rangle,\nonumber
\end{align}
and neglecting other two loop contribution. 
Taking this into account we can model terms
above by replacing $P_{03}\to 3(f\mathcal{H}D)^2\sigma_v^2P_{01}$ and $P_{13}\to 3(f\mathcal{H}D)^2\sigma_v^2P_{11}$. In figure \ref{fig:P0313} we show results for both 
$\mu^4$ part of $P_{03}$ and for $\mu^6$ part of $P_{13}$. On the same plot we show one loop PT prediction for
both terms (keeping  in mind that in the overall contribution they differ relative to each other by the factor of $-\mu^2f(\tau)$. 
We compare model results presented above to simulations. The specific shape in simulations is explained
by the proposed model, while it is not in one loop PT result. This effect arises
from substitution of $P_L$ with
$P_{01}$ or $P_{11}$. We can again add the small scale velocity
dispersion $\sigma^2_v\to\sigma^2_{v}+\sigma_{03}^2/(f\mathcal{H}D)^2$ (or
equivalently $\sigma^2_{13}$) which is not included in PT analysis. In the model we
suggest for $P_{13}$ we get, in addition to the $\mu^6$ dependence, also
$\mu^4$ dependence, which comes from the vector part of $P_{11}$. In
figure \ref{fig:P0413} we show that this can explain the trends and amplitude seen
in simulations for this term. In the case of the leading contributions $P_{03}[\mu^4]$ and $P_{13}[\mu^6]$ we found that lower 
value for small scale velocity dispersion $\sigma_{03}$ and
$\sigma_{13}$ is needed (table \ref{tb:SigPar}). This can be describer
using the halo model and we return to that in section \ref{sec:hm}. 
Note that this value only affects the total amplitude, i.e. translates whole result up and down, but does not affect the shape.

In a similar fashion we can estimate contribution of $P_{04}$
term, which is the last term we need to consider at $\mu^4$ order. 
Formally this term does not even contribute at the one loop order
in PT, but we can do two loop considerations as we did before.
Considering the most relevant two loop contributions (from partially disconnected
diagrams) this term can be modeled as 
\begin{align}
\left\langle T^0_\pp\right.\left|\  T^4_\pp\right\rangle= 6(f\mathcal{H}D)^2\sigma^2_v\Big<\df\left|  v_\pp^2\right\rangle+ 3(f\mathcal{H}D)^4\sigma^4_v\left\langle
\df\right.\left|\df\right\rangle + \left\langle\df\right.\left|\df\right\rangle\left\langle v^2_\pp \right.\left| v^2_\pp\right\rangle _c,\nonumber
\end{align}
where we used subscript $c$ to label the connected part of the correlator.
Here we again include the small scale velocity dispersion using $\sigma_v^2\to\sigma_{v}^2+\sigma_{04}^2/(f\mathcal{H}D)^2$,
just as in $P_{02}$ case. We can write the $P^{ss}_{04}$ contribution
\begin{align}
   P^{ss}_{04}\left[\mu^4\right]=&-\frac{1}{2}f(\tau)^2D(\tau)^2
   k^2\left(\sigma_{v}^2+\sigma_{04}^2/(f\mathcal{H}D)^2\right)\bar{P}^{ss}_{02}\left[\mu^2\right]\nonumber\\
   &+\frac{1}{4}f(\tau)^4D(\tau)^4k^4\left(\sigma_{v}^2+\sigma_{04}^2/(f\mathcal{H}D)^2\right)^2P^{ss}_{00}(k,\tau)
   +\frac{1}{12} P^{ss}_{00}(k,\tau)\int{\frac{d^3q}{(2\pi)^3}\bar{P}_{22}(\VEC{q},\tau)} \nonumber\\
   P^{ss}_{04}\left[\mu^6\right]=&-\frac{1}{2}f(\tau)^2D(\tau)^2 k^2\left(\sigma_{v}^2+\sigma_{04}^2/(f\mathcal{H}D)^2\right)\bar{P}^{ss}_{02}\left[\mu^4\right].
\label{eq:P04.2} 
\end{align}

\subsection{Halo model and small scale velocity dispersion}
\label{sec:hm}

In our analysis of correlators that contribute in expansion of the
RSD power spectrum we find that some of them have terms proportional
to velocity dispersion of dark matter particles. Particles moving in
the gravitational potential can have large velocities even on the small
scales, so they have significant contribution to the total velocity
dispersion. Using PT we can evaluate velocity dispersion
\begin{equation}
\left(f\mathcal{H}D\sigma_{v}\right)^2=\frac{1}{3}\int{\frac{d^3q}{(2\pi)^3}\frac{P_{\tf\tf}(q,\tau)}{q^2}}.
\end{equation}
Linear theory gives $\sigma_v\simeq600km/s$, but this does not properly take into account small scale
contributions, which come from within virialized halos where PT cannot be used. 
Thus to take into account all nonlinear contributions we have
to add this to our model in order to match the simulation
predictions. In table \ref{tb:SigPar}
we show the values used for small scale velocity dispersion for
modeled terms. These values are obtained by fitting our PT model
using the free parameter for small scale dispersion
($\sigma_{02},~\sigma_{12},~\ldots$) in order to match simulation predictions.
We see that we can classify these into a few groups which have approximately 
the same value. 

\begin{table}[ht]
\caption{Small scale velocity dispersions as described in the paper (in km/s).} 
\centering 
\setlength{\tabcolsep}{8pt}
\renewcommand{\arraystretch}{1.0}
\begin{tabular}{c|cccc|c|cccc}
\hline\hline
z & 0.0 & 0.5 & 1.0 & 2.0 &z & 0.0 & 0.5 & 1.0 & 2.\\ [0.5ex] 
\hline
$\sigma_{02}$, $\sigma_{12}$, $\sigma_{13}$ (vector),
$\sigma_{22}$ & 375 & 356 & 282 & 144& $\sigma_{bv^2}$  & 377 & 267 & 190 & 105\\
$\sigma_{03}$, $\sigma_{13}$ (scalar) & 209 & 198 & 159 & 80  &$\sigma_{v^2}$  & 221 & 154 & 106 & 56\\
$\sigma_{04}$ & 432 & 382 & 315 & 144& $\sigma_{bv^4}$  & 510 & 371 & 270 & 153 \\
\hline
$\sigma_{FoG}$ & 346 & 322 & 249 & 133& $(\bar\sigma^2)^{1/2}$  & 387 & 278 & 200 & 111 \\ [1ex]
\hline
\end{tabular}
\label{tb:SigPar}
\end{table}

We can understand the fact that some terms have equal velocity dispersion to others, and some 
do not, using the halo model \cite{Peacock:2000qk,Ma:2000yt, Seljak:2000,Berlind:2002rn,Cooray:2002dia}.  
We can distinguish between three types of
contributions to velocity dispersion. For terms
$P_{02},~P_{12},~P_{13}(vector)$ and $P_{22}$  we find that the same
value is needed. In these terms $v_\pp^2$ always comes weighted by $1+\delta$. 
In a halo model we divide all the mass into halos, such that the integral over the 
halo mass function times mass gives the mean density of the universe, 
\begin{equation}
\int{dM\frac{dn}{dM} M}=\bar{\rho},
\end{equation}
where $dn/dM$ is halo mas function.
Each halo has a bias $b(M)$, which describes how strongly the halo clusters relative to 
the mean, since 
\begin{equation}
\df(\VEC{k})=\frac{1}{\bar{\rho}}\int{dM\frac{dn}{dM} Mb(M)\df(\VEC{k})}.
\label{bint}
\end{equation}
Each halo also has a small scale 1-d velocity dispersion $v_{\pp}^2(M) \propto M^{2/3}$, 
where the latter relation is only approximate and does not take into account effects 
such as halo profile dependence on the halo mass etc. 

We now decompose the terms into halos of different mass, accounting for small 
scale velocity dispersion $v_{\pp}^2(M)$, and accounting for biasing whenever this is 
multiplied by density $\delta$. 
For example for term $P_{02}$ schematically we can
write
\begin{align}
P_{02}=\left<\df|(1+\df) v_\pp^2
\right> \sim \left<\df|\df v_\pp^2
\right>=\left<\df|\df\right>\frac{1}{\bar{\rho}}\int{dM\frac{dn}{dM} M
  b(M) v_{\pp}^2(M)} \equiv P_{00}\sigma^2_{bv^2},
\end{align}
i.e. we find that the velocity dispersion is weighted by bias. Note that we should have written the 
term $\df$ in halo model as well, but since the bias integrates to unity (equation \ref{bint})
we do not have a contribution from the left hand side. Note also that we only include the small 
scale velocity dispersion effects here that come on top of the PT calculations above. 
Same quantity enters also in $~P_{12},~P_{13}(vector)$ and $P_{22}$. 

For terms $P_{03}$ and $P_{13}(scalar)$
we have a different contribution to small scale velocity dispersion because one of the
velocity field in $v_\pp^3$ correlates with the density field and we can approximate $1+\delta$ with 
1 at the lowest order. As a result $v_\pp^2$ 
is not density weighted.  For example for $P_{03}$ we have
contributions from term
\begin{align}
P_{03}=\left<\df|(1+\df)v_\pp^3\right> \sim \left<\df|v_\pp^3\right>=
3\left<\df|v_{\pp}\right>\frac{1}{\bar{\rho}}\int{dM\frac{dn}{dM} M
 v_{\pp}^2(M)}\equiv 3P_{\df v_\pp}\sigma^2_{v^2}.
\end{align}
Since there is no biasing and since $b(M)>1$ at high mass halos which dominate the 
velocity dispersion these terms have a smaller value of velocity dispersion than we had
in the first case. This is precisely what we find when fitting to the simulations.

Finally, for the term $P_{04}$ we find contribution
\begin{align} P_{04}=\left<\df|(1+\df)
  v_\pp^4\right> \sim
\left<\df|\df
  v_\pp^4\right>=\left<\df|\df\right>\frac{1}{\bar{\rho}}\int{dM\frac{dn}{dM}
  M b(M) v_{\pp,s}^4(M)}\equiv P_{00}\sigma^2_{bv^4}.
\end{align}
This term gives a value bigger then previous two because higher mass halos give a larger
weight and they are more biased,
which is also consistent with what we observe in simulations, and is
presented in table \ref{tb:SigPar}. To convert $v_{\pp}$ into velocity
dispersion we use the relation
\begin{align}
 v_{\pp,s}^2(M)=(235\text{km/s})^2\left(\frac{M}{h10^{13} M_\odot}\right)^{2/3},
\end{align}
see for example \cite{Seljak:2002zr}. 
We use standard Sheth-Tormen model for halo mass function and halo bias \cite{Sheth:1999mn}. 
We see that predictions from
the halo model presented in \ref{tb:SigPar} agree qualitatively but 
not quantitatively.  This could be a consequence of the simplifying assumptions, such 
as ignoring the internal structure of the halo and its mass dependence.
Note also that there are no errors in the analysis: it is possible that the sampling 
variance errors are large, specially for $\sigma^2_{bv^4}$, which receives dominant 
contributions from the very high mass halos which 
may or may not be present in our simulations, depending on the realization. 
We do not go into a more detailed modeling here, but 
it is possible that with a more detailed model the agreement would improve. 
Even at 
this level the halo model gives an insight in hierarchy of the
contributions $\sigma^2_{v^2}<\sigma^2_{bv^2}<\sigma^2_{bv^4}$,
and offers a qualitative picture why different
terms in expansion need different values for velocity dispersion. 

\subsection{Putting it all together: $\mu^{2j}$ terms, finger of god
  resummation and Legendre moments}
\label{sec:mu2mu4}

\begin{figure}[t]
    \centering
    \includegraphics[width=1.0\textwidth]{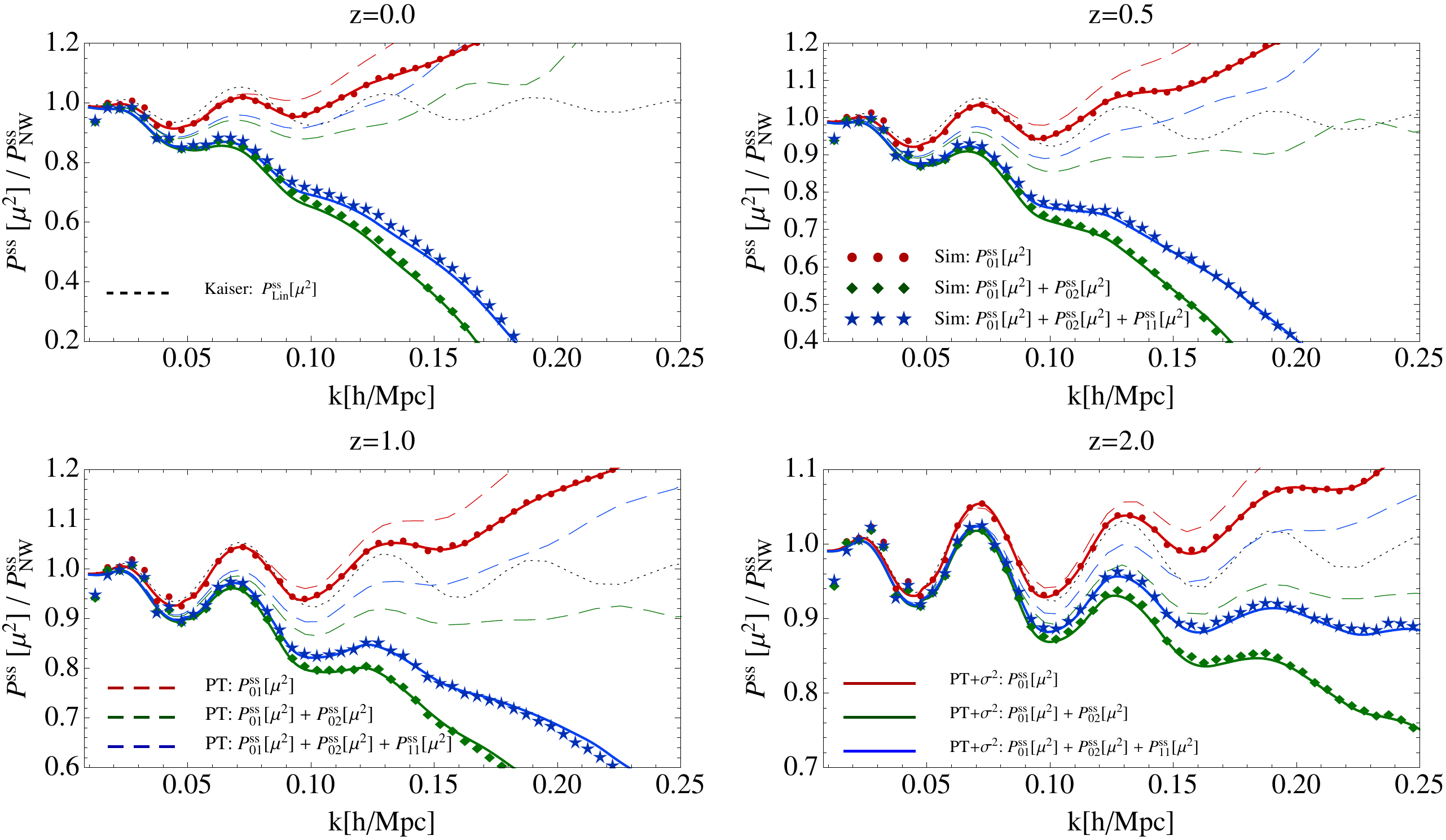}
    \caption{\small $\mu^2$ dependence of $P^{ss}$ at four
      redshifts, $z=0.0,~0.5,~1.0$ and $2.0$. 
      We show separately contributions of the PT (dashed lies), the improved velocity dispersion
 model (solid lines), and simulation measurements (points). 
      The leading term is $P^{ss}_{01}$ (red), to which we add $P^{ss}_{02}$ (green), and to which we add $P^{ss}_{11}$ to get the total (blue). Kaiser $\mu^2$ term
      (black, dotted) is also shown. All the lines are divided by no-wiggle $\mu^2$ Kaiser term.}
    \label{fig:mu2}
\end{figure}

\begin{figure}[t]
    \centering
    \includegraphics[width=1.0\textwidth]{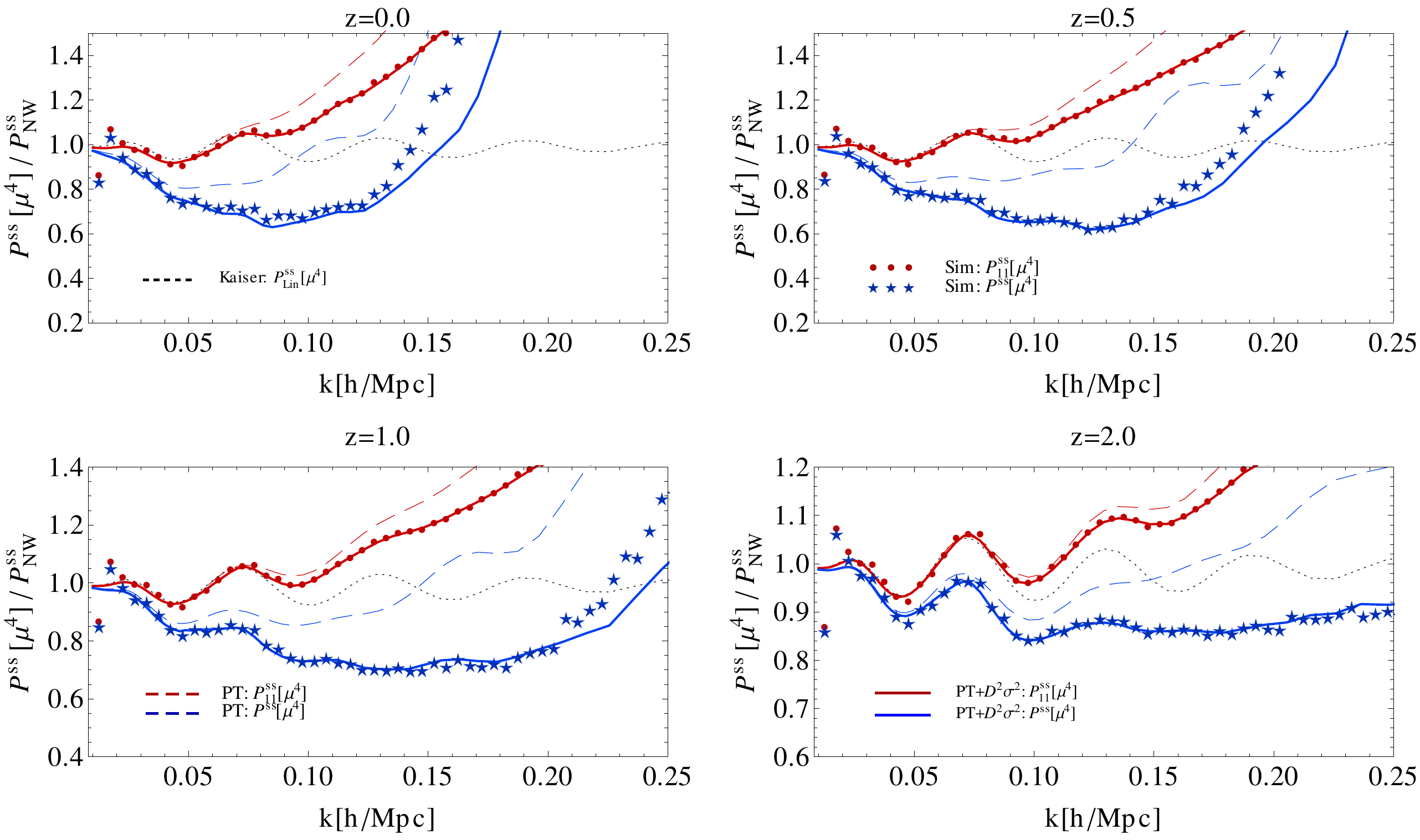}
    \caption{\small $\mu^4$ dependence of $P^{ss}$ at four
      redshifts, $z=0.0,~0.5,~1.0$ and $2.0$. 
      We show the PT (blue, dashed line) results, the improved velocity dispersion model
      presented in this paper (blue, solid
      line), and simulation measurements (blue, stars). 
      Also, the leading $P^{ss}_{11}$ term  is shown  in red.
	Kaiser $\mu^4$ term
      (black, dotted) is also shown.
      All the lines are divided by no-wiggle $\mu^4$ Kaiser term.}
    \label{fig:mu4}
\end{figure}

 There are a finite number of velocity moment terms at each order of $\mu^{2j}$, in contrast to
the Legendre multipoles expansion (monopole, quadrupole, hexadecapole etc),
which receive contributions from all orders in moments of distribution function.
We will thus investigate $\mu^{2j}$ expansion, with the lowest 3 orders 
containing cosmological information, while the rest can be treated as nuisance parameters to be 
marginalized over. Even in that case a good prior for these higher order angular terms would be 
very useful, although given the large number of terms that contribute to it it seems easier to 
be guided by the simulations rather than the PT. 
In this section we collect all the previous terms with $\mu^2$ and $\mu^4$ dependence. 
At $\mu^2$ level the 
contributions come from $P_{01}$, $P_{02}$ and $P_{11}$ terms, and for 
$\mu^4$ from $P_{11}$, $P_{02}$, $P_{12}$, $P_{22}$, $P_{03}$, $P_{13}$ and $P_{04}$ terms.
In figures \ref{fig:mu2} and \ref{fig:mu4} we show $\mu^2$ and $\mu^4$ dependence of these terms divided by the corresponding 
no-wiggle Kaiser term. We show both the simplest PT model and the improved model that includes velocity dispersion effects. 
For modeling some of the terms we have been using the model for
velocity dispersion $\sigma^2_v\to\sigma^2_v+\sigma_{ij}^2$, where the
added value $\sigma_{ij}^2$ for term $P_{ij}$ is given by the table \ref{tb:SigPar}.
These model was optimized to fit corresponding terms primarily on large scales, where the dominant contributions comes from $P_{01}$
for $\mu^2$ and $P_{11}$ form $\mu^4$ terms. To improve the model
further for $P_{01}$ and scalar part of $P_{11}$, instead of PT
predictions, we use exact values obtained from the simulations. We expect that ongoing 
activities in the modeling of nonlinear power spectrum will result in a successful model of these terms (note that $P_{01}$ 
is given by the time derivative of the nonlinear power spectrum $P_{00}$). Although we have introduced a free parameters in our model note that 
$P_{01}$ and $P_{11}$ terms do not contain any free parameters, so we can use simulation results as well as any other method to predict these terms.

The leading order in RSD is the $\mu^2$ term. On large scales it is given by the Kaiser expression, but note that the deviations 
from the linear theory are of the order of 10\% at $z=0$ already at $k \sim 0.05{\rm h/Mpc}$. These nonlinear effects are dominated 
by the small scale velocity dispersion effects, which cannot be modeled by PT (a smaller effect, of the order of 2\% at these scales, 
is caused by nonlinear effects in $P_{01}$ which are modeled in PT). This is a serious challenge for the RSD models 
and the ability to extract cosmological information from RSD: 
any additional free parameter that needs to be determined from the data will reduce the statistical power of the data set. 
Note however that we do not observe dark matter, but galaxies, so to address this concern in a proper way one will need 
to repeat this study with galaxies. We plan to pursue this in the near future.  
At higher redshifts these nonlinear effects are smaller: at $z=1$ the 10\% nonlinear suppression happens at $k \sim 0.1{\rm h/Mpc}$. 
In all cases the dominant nonlinear effect is to suppress the small scale power, as expected by the phenomenological 
models like \cite{Scoccimarro:2004tg}, where a Gaussian smoothing is
added to the extension of Kaiser formula. 

The $\mu^4$ terms show considerably more structure in the nonlinear effects: the overall power is initially suppressed relative to the 
linear term, stays flat for a while and then increases again (above $k \sim 0.1{\rm h/Mpc}$ at $z=0$). The effects are large: 
20\% suppression of power at $k \sim 0.05{\rm h/Mpc}$ for $z=0$ relative to linear. The model has some success in 
predicting some of these details, but is far from perfect and again it relies on the free parameters. 
The nonlinear effects are smaller at higher redshift, but remain significant. These $\mu^4$ terms have an important contribution 
to RSD. For example, at higher redshift (where $f \sim 1$) they contribute about 30\% to the quadrupole on large scales, with the 
dominant 70\% contribution coming from $\mu^2$ term. As for the $\mu^2$ term, it remains to be seen how well we can model these 
terms such that we can extract the maximal information from the data, but the fact that the nonlinear effects are so large 
already on very large scales is a cause for concern. 

 Our result for the RSD power spectrum $P^{ss}$ can be compactified in so called finger of good resummation. 
Following the ideas presented in \cite{Okumura:2011pb} we can show explicitly that up to $\mu^4$ order our
result in equation \ref{eq:PssNLO} can be written in the following way
\begin{equation}
  P^{ss}_{\text{FoG}}(k,\mu)=\exp\big[-k^2\mu^2\sigma_{\text{FoG}}^2\big]\Big(A(k)+\mu^2B(k)+\mu^4C(k)+\ldots\Big),
\label{eq:FoG_1}
\end{equation}
where we have defined
\begin{align}
 A(k)&=P_{00},\nonumber\\
 B(k)&=P^{ss}[\mu^2]+k^2\sigma_{\text{FoG}}^2P_{00},\nonumber\\
 C(k)&=P^{ss}[\mu^4]+k^2\sigma_{\text{FoG}}^2P^{ss}[\mu^2]+\frac{1}{2}\big(k^2\sigma_{\text{FoG}}^2\big)^2P_{00},
\end{align}
i.e. 
there is no change to $\mu^2$, and $\mu^4$ terms, since
the terms $P^{ss}[\mu^2]$ and $P^{ss}[\mu^4]$ contain all the terms
discussed previously 
and the FoG terms cancel by construction of $B(k)$ and $C(k)$. 

If we set the value of $\sigma_{\text{FoG}}^2=(fD)^2\sigma_{v}^2$ this reduces to simple form where $\sigma^2_{v}$
is now present only in the exponent in equation \ref{eq:FoG_1} and not in the brackets, i.e.
\begin{align}
 A(k)&=P_{00},\nonumber\\
 B(k)&=P^{ss}_{01}+\bar{P}^{ss}_{02}[\mu^2]+P^{ss}_{01}[\mu^2],\nonumber\\
 C(k)&=P^{ss}_{11}[\mu^4]+\bar{P}^{ss}_{02}[\mu^4]+\bar{P}^{ss}_{12}[\mu^4]+\bar{P}^{ss}_{22}[\mu^4],
\end{align}
where all the $P^{ss}$ and $\bar{P}^{ss}$ terms 
here, as defined in previous sections, do not contain velocity dispersion
contributions. This argument also generalizes to the 
case where we replace $\sigma_{\text{FoG}}^2=(fD)^2\sigma_{v}^2+\sigma^2$, where $\sigma^2$ is the 
small scale velocity dispersion. This 
is the basic justification for using the FoG model. 

Unfortunately, the fact that $\sigma_v^2$ cancels out is of limited use, since in practice the 
velocity dispersion is not dominated by linear 
$\sigma_{v}^2$, but by small scale velocity dispersions, and as argued above, there is 
no single $\sigma^2$, but instead there are several 
different velocity dispersions entering in the detailed RSD analysis
at $\mu^2$ and $\mu^4$ order, $\sigma_{bv^2}$, $\sigma_{v^2}$ and $\sigma_{bv^4}$. 
In fact, in our analysis we include these terms already so one can argue that it is the 
next term that we do not include that should enter in $\sigma_{\text{FoG}}^2$. At $\mu^6$ order 
there are again several velocity dispersions that can be defined and that have a wide 
range of values, so we cannot simply write down a value without explicitly evaluating 
all the terms at this order. It is however likely that their values will be of the same 
order as $\sigma_{bv^2}$, $\sigma_{v^2}$ and $\sigma_{bv^4}$. In table \ref{tb:SigPar} we compare the 
root mean square average of these velocity dispersion values to the best fit value for $\sigma_{\text{FoG}}$, 
showing that indeed the value of $\sigma_{\text{FoG}}^2$ is indeed related to these other values. 

It is customary to expand the redshift-space power spectrum in terms of Legendre multipole moments. The motivation for this
is that when using the full angular information Legendre moments are uncorrelated on scales small relative to the survey size.  
Using ordinary Legendre polynomials ${\cal  P}_l(\mu)$, we have
\begin{equation}
  P^{ss}(k,\mu)=\sum_{l=0,2,4,\cdots}P^{ss}_l(k){\cal P}_l(\mu),
\end{equation}
where multipole moments, $P^{ss}_l$, are given by 
\begin{equation}
  P^{ss}_l(k)=(2l+1)\int^{1}_{0}P^{ss}(k,\mu){\cal P}_l(\mu)d\mu ~. 
\end{equation}
where ${\cal P}_l(\mu)$ are the ordinary Legendre polynomials, ${\cal P}_0(\mu)=1$, 
${\cal P}_2(\mu)=(3\mu^2-1)/2$ and ${\cal P}_4(\mu)=(35\mu^4-30\mu^2+3)/8$.
In the RSD analyses we are usually limited to modeling the monopole ($l=0$) and quadrupole ($l=2$) terms,
although some information is also contained in hexadecapole term ($l=4$).

\begin{figure}[t]
    \centering
    \includegraphics[width=1.0\textwidth]{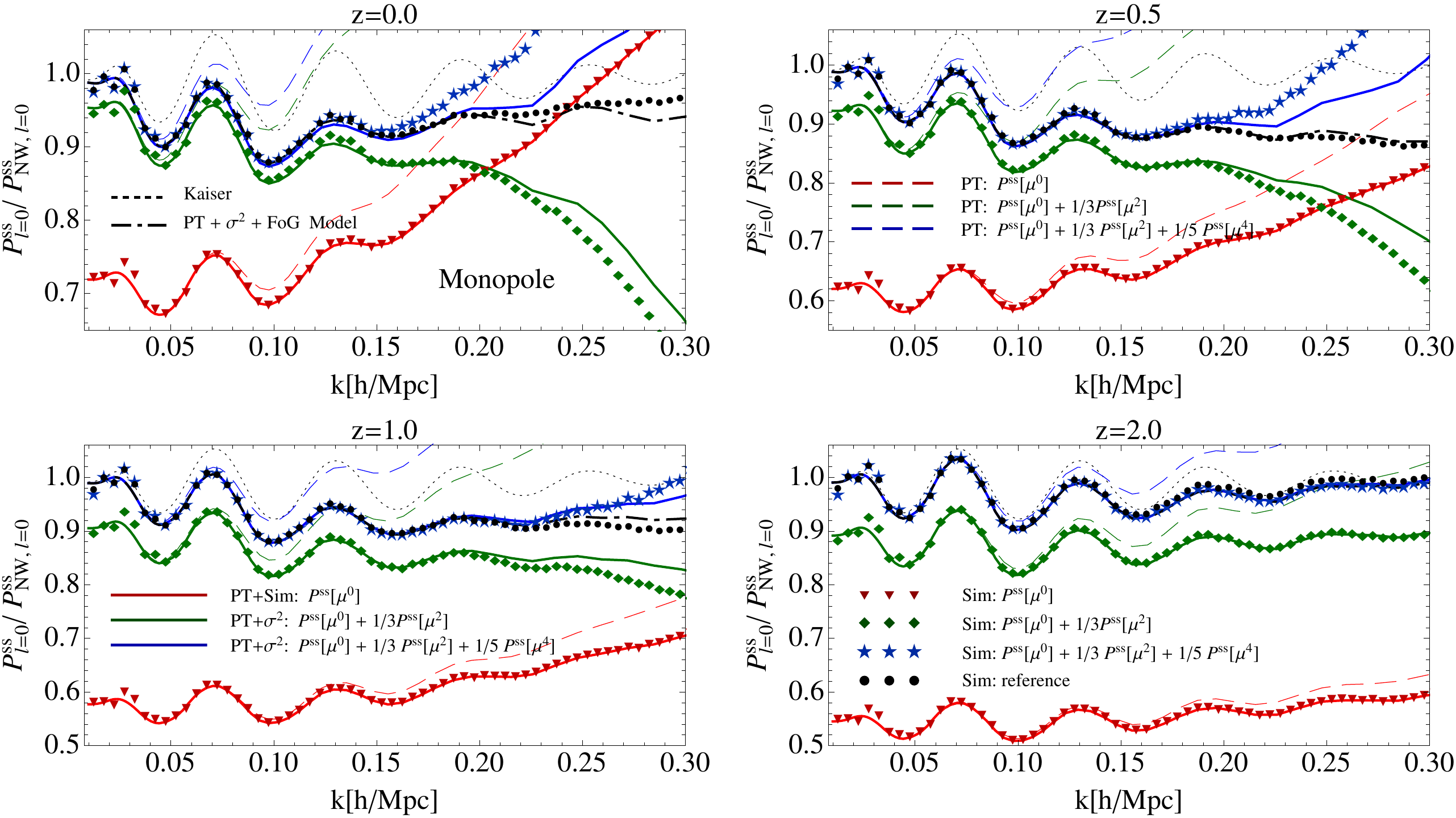}
    \caption{\small Monopole moment $P_0^{ss}$ is plotted at four
      redshifts, $z=0.0,~0.5,~1.0$ and $2.0$. To the first isotropic term
      (red) of $P^{ss}$ expansion we first add $\mu^2$ (green)
      term and then also $\mu^4$ term (blue).  We show PT (dashed lines) results, improved velocity dispersion model
      presented in this paper (solid lines), simulations up to $\mu^4$ contributions
      (triangles, diamonds and stars), and reference
      simulation results (points). 
      Resummed FoG model from equation \ref{eq:FoG_1}  (black dot-dashed line) and linear Kaiser model (black dotted line) are also shown.
      All the results shown are divided by monopole contributions
      of the no-wiggle Kaiser model.}
    \label{fig:multi1}
\end{figure}

\begin{figure}[t]
    \centering
    \includegraphics[width=1.0\textwidth]{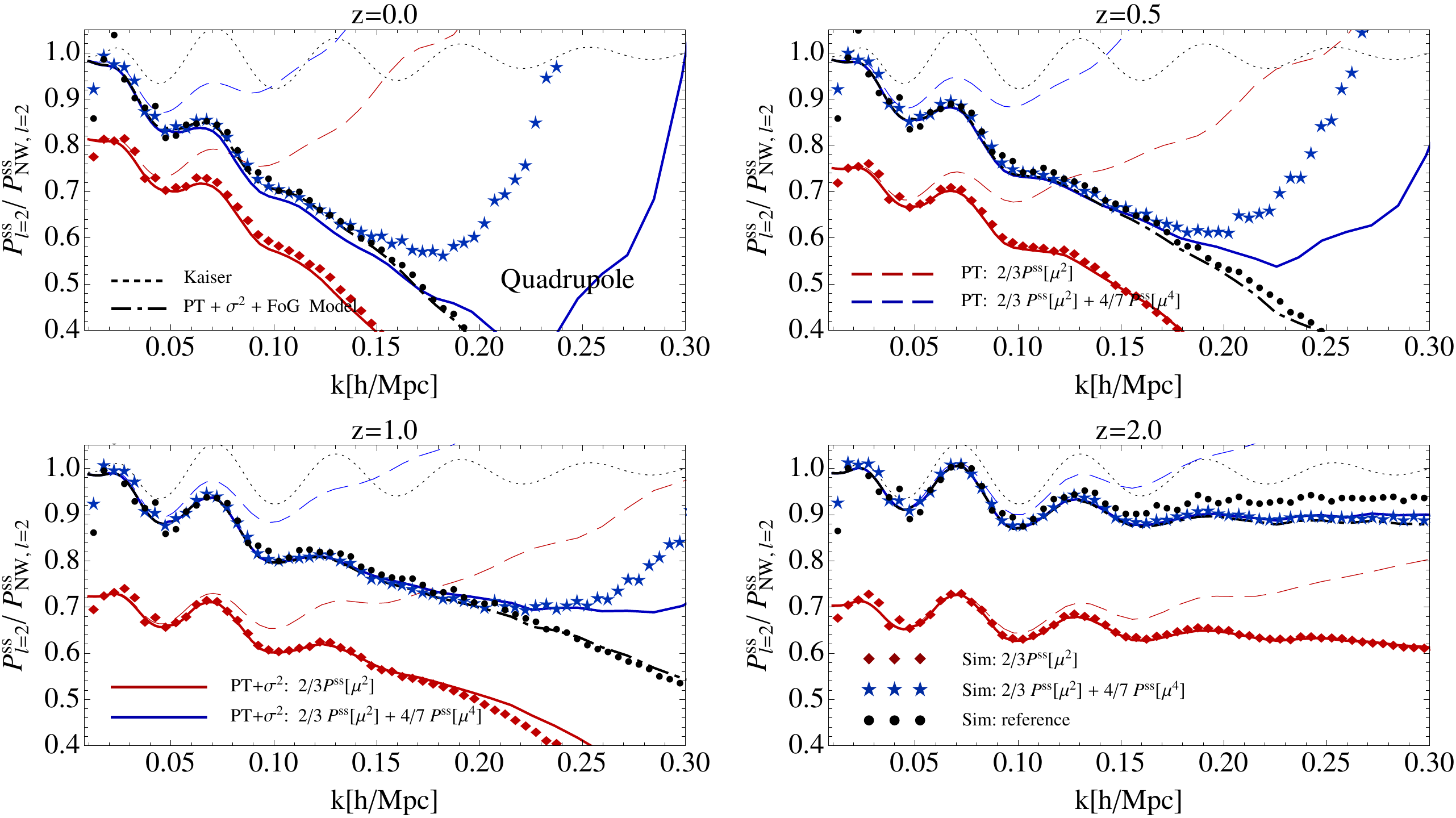}
    \caption{\small Quadrupole moment $P_2^{ss}$ is plotted at four
      redshifts, $z=0.0,~0.5,~1.0$ and $2.0$. To $\mu^2$ (red) term of $P^{ss}$ expansion
      we add $\mu^4$ term (blue).  We show PT (dashed lines) results, improved velocity dispersion model
      presented in this paper (solid lines), simulations up to $\mu^4$ contributions
      (diamonds and stars), and reference
      simulation results (points). 
      Resummed FoG model from equation \ref{eq:FoG_1}  (black dot-dashed line) and linear Kaiser model (black dotted line) are also shown.
      All the results shown are divided by quadrupole contributions
      of the no-wiggle Kaiser model.}

    \label{fig:multi2}
\end{figure}

In figures \ref{fig:multi1} and \ref{fig:multi2} we show monopole and quadrupole power spectra predictions of improved velocity dispersion model presented in the paper,
as well as one loop PT result. We also show resummed FoG result
choosing for $\sigma_{\text{FoG}}$ values given in the last line in table
\ref{tb:SigPar}. We compare this to the reference multipole results
obtained from full simulation redshift space power spectra. 
We also show simulation results where only terms up to $\mu^4$ are considered. In case of monopole we see that these two simulation results 
agree on scales larger then $k\sim (0.15-0.20)$h/Mpc (depending on redshift) but then start to deviate one from an other. In the case of the quadrupole these 
deviations start to be more then 1\% for $k> 0.15$h/Mpc. This trend is due to the $\mu^6$ term
which is weighted by 1/7 for the monopole but 11/21 for the quadrupole, which is almost the same weight (4/7) as for $\mu^4$ term.
At higher $k$ higher $\mu$ terms ($\mu^6$, $\mu^8$, ...) start to be relevant and contribute significantly to the total redshift 
power spectrum. These higher $\mu$ contributions have large amplitudes with differing signs \cite{Okumura:2011pb}, 
which would suggest that 
we might not be able to rely on our expansion in low-$k$ any longer, although FoG resummation can still 
help here.
From figures we can also see relative contributions to the total monopole
and quadrupole power from $\mu^2$ and $\mu^4$ terms. We see that at
scales larger than $k\sim0.15$h/Mpc $\mu^4$ term contributes with 5-10\%
(depending on the redshift) to the total power of monopole and with 15-30\% for
the quadrupole. For the quadrupole all the remaining power comes form the
$\mu^2$ term while for the monopole the $\mu^2$ term constitutes 
25-35\%  of power and the rest comes from isotropic $P_{00}$ term. 
To reduce the dynamical range we again divide monopole results by the no-wiggle monopole Kaiser term
and the quadrupole results by the no-wiggle quadrupole Kaiser term. 

\section{Conclusions}
\label{sec:conclusion}

In this paper we use the distribution function approach to redshift space distortions (RSD)
that decomposes RSD into moments of distribution function. Our goal is 
to model the terms that contribute to the redshift space distortions using perturbation theory. 
We first repeat the derivations presented 
in \cite{Seljak:2011tx}, explicitly deriving the
decomposition of the moments into helicity eigenstates, based on 
their transformation properties under rotation around the direction of the Fourier mode. 
We give the explicit forms of correlators of the moments of distribution function 
and their angular dependencies. 

It is worth comparing the phase space approach to the redshift space power spectrum to the alternative perturbative 
derivations 
that can be found in the literature, e.g. \cite{Matsubara:2007wj,Taruya:2010mx}. 
The advantage of phase space approach as 
presented here lies in the decomposition 
into the hierarchy of terms that contribute to the redshift space power spectrum with an explicit dependence on the expansion parameters.
In this way a systematic expansion approach is possible and a physical meaning of each term is revealed that enables effective 
modeling of each of the contributing correlators in the relation \ref{eq:PssPart}. This also allows
a detailed term-by-term comparison of the simulation results, since one can compare each term to the 
simulations rather than just the final RSD power spectrum.
In this paper we focus on the PT modeling and comparing the results to the simulations 
we are able to clearly show where 
the PT modeling preforms well, even at one loop, and where it does less so. 
The approach also allows us to identify physical reasons for failure of PT in individual terms, which 
are mostly related to various small scale velocity dispersion effects, and for which  
other modeling methods may be required.
This also enables us to argue that it is more physical to try to model some of the terms in PT going beyond one loop,
while remaining at the one loop level for the other terms. We should also mention that if just PT is used to evaluate all the terms 
at one loop, the results of phase space approach should correspond to \cite{Matsubara:2007wj}, although only 
monopole predictions are presented there (compare e.g. 
figure 10 in \cite{Matsubara:2007wj} to SPT predictions for monopole in figure \ref{fig:multi1}). 

The leading order contributions to RSD
can be classified in terms of their angular dependence, with the lowest order being $\mu^2$ and $\mu^4$, where $\mu$ 
is the angle between the Fourier mode and the line of sight. 
There are three terms contributing to $\mu^2$ and seven terms contributing to $\mu^4$. We evaluate all of these 
terms using the lowest order PT (one loop) and compare them to simulation results. 
For some terms adding two loop contributions proves to be important and we extend
our models to include the relevant contributions. 
Also, for some of these terms standard PT is not sufficient and 
we propose physically motivated ansatz that goes beyond the loop
analysis. These are based on the small scale induced velocity dispersion effects 
which multiply the long range correlations, such as density-density or density-velocity correlations. 
Such ansatz has a free parameter, small scale velocity dispersion,
which cannot be modeled using PT. 
We found that a number of these terms have the same value, but also that not all velocity 
dispersions should be equal. We developed a halo model to describe the hierarchy of these 
terms and shown that the model can qualitatively explain the simulation results. 
Our analysis systematically accounts 
for all of the PT terms at one loop order and the small scale dispersion parameters, while 
necessary for a good description of RSD, have physically motivated values. 
In this sense our model goes beyond previous analyses \cite{Scoccimarro:2004tg, Taruya:2010mx}, 
which include some, but not 
all of the PT terms and which often treat FoG parameters as fitting parameters without a physical 
meaning. 

The dominant term to RSD is the $\mu^2$ term and its dominant contribution is 
the momentum density correlated with the density. This term can be written in terms of a time derivative of the power spectrum 
\cite{Seljak:2011tx} and so can be modeled using dark matter power spectrum emulators. 
Two other terms contribute to $\mu^2$, the vector part of the momentum density-momentum density correlation, and
the scalar part of energy density-density correlation. 
We find that they affect RSD at a 10\% level already at $k \sim 0.05 {\rm h/Mpc}$.  
The energy density-density correlation term is the dominant nonlinear effect, is negative for all scales
and thus reduces the total $\mu^2$ power. It is related to the Fingers-of-God effect. This term contains velocity
dispersion term which cannot be modeled in PT and requires a free parameter in the model. 

The next angular term has $\mu^4$ dependence and there are 
seven terms that contribute to the total power spectrum, of which one, 
scalar part of $P_{11}$, contains a linear contribution that does not
vanish on large scales. We evaluate all of these terms in PT. 
Some of these terms are well modeled by PT, while 
others also require velocity dispersion type parameters. 
With these the modeling achieves some level of precision compared to the simulations, but is still
limited in the dynamic range, with an error of about 5\% at $k \sim 0.2 {\rm h/Mpc}$ at $z=1$. 

Our ultimate goal is to develop accurate models of RSD that can be applied to observations. 
We observe galaxies, not dark matter, 
and understanding the physical processes that lead to RSD in dark matter is just the first step
towards the goal of understanding the RSD in galaxies. 
The results presented here are only a rough guide for 
the challenges awaiting us when applying these techniques to the data, but there are some 
lessons learned that are likely to be valid also for galaxies. 
One is the importance of velocity dispersion effects, which dominate our model uncertainties on 
large scales. 
The good news may be that the velocity dispersion effects, which are the main source of the 
modeling difficulties in this paper, may be smaller for galaxies than for the dark matter, 
specially if a sample of central galaxies can be selected. 
We also expect that the halo model for computing velocity dispersion should be applicable to 
galaxies as well. 
However, galaxies also have additional 
challenges not present for dark matter: galaxy biasing 
will introduce additional scale dependent effects in redshift space that will need to be modeled, even if 
there are no such scale dependent biases in real space ~\cite{Seljak:2011tx, Okumura:2012xh}. 
The success of modeling the RSD and extracting 
the cosmological information from it depends on our ability to model these 
galaxy biasing and velocity dispersion terms. 
We plan to address some of these issues in the future work. 

\acknowledgments

We would like to thank Nico Hamaus, Darren Reed and Lucas Lombriser for useful discussions and comments. 
ZV would like to thank the Berkeley Center for Cosmological Physics and the Lawrence Berkeley Laboratory for their hospitality.
This work is supported by the DOE, the Swiss National Foundation under contract 200021-116696/1 and WCU grant R32-10130.
The simulations were performed on the ZBOX3 supercomputer of the Institute for Theoretical Physics
at the University of Z\"{u}rich.

\bigskip

\appendix

\section{Components of moments of distribution function}\label{sec:App1}

Starting from equation \ref{eq:TLa} we can chose some basis, for example Cartesian, to express scalar product $\VEC{h}\cdot\VEC{q}$. It follows
\begin{align}
    T^L_\VEC{h}(\VEC{x})=\frac{m a^{-3}}{\bar{\rho}}\int{d^3p f(\VEC{x},\VEC{p})\left(h_ip_i/ma\right)^L},
\label{eq:TLaCartesian}
\end{align}
where summation over $i=1,2,3$ is implied. Using the multinomial theorem:
\begin{align}
 (x_{1}+x_{2}+\ldots+x_{m})^n=\sum_{k_1+k_2+\ldots+k_m=n}\frac{n!}{k_1!k_2!\cdots k_m!}x_1^{k_1}x_2^{k_2}\cdots x_m^{k_m},
\label{eq:MultiTM}
\end{align}
it follows that
\begin{align}
 T^L_\VEC{h}(\VEC{x})=\frac{ma^{-3}}{\bar{\rho}}\sum_{k_1+k_2+k_3=L}\frac{L!}{k_1!k_2!k_3!}h_1^{k_1}h_2^{k_2}h_3^{k_3}\int{d^3pf(\VEC{x},\VEC{p})p_1^{k_1}p_2^{k_2}p_3^{k_3}}/(ma)^L.
 \label{eq:TLaMulti}
\end{align}
Neglecting velocity dispersion and anisotropic stress second rank tensor, and similar higher rank tensors contributions ($\sigma^{ij}=0,\ldots$) we are left with 
\begin{align}
 \frac{ma^{-3}}{\bar{\rho}}\int{d^3pf(\VEC{x},\VEC{q})p_1^{k_1}p_2^{k_2}p_3^{k_3}}/(ma)^L=(1+\df(\VEC{x}))v_1^{k_1}v_2^{k_2}v_3^{k_3},
\end{align}
where $v_i$ is given in equation \ref{eq:velocity}. Returning this back into equation \ref{eq:TLaMulti} and using multinomial theorem again we get
\begin{align}
 T^L_\VEC{h}(\VEC{x})=(1+\df(\VEC{x}))\sum_{k_1+k_2+k_3=L}\frac{L!}{k_1!k_2!k_3!}(h_1v_1)^{k_1}(h_2v_2)^{k_2}(h_3v_3)^{k_3}=(1+\df(\VEC{x}))(\VEC{h}\cdot\VEC{v}(\VEC{x}))^L.
\end{align}
Thus we have retrieved result of equation \ref{eq:TLaArox}, and choosing 
$\VEC{h}=\hat{r}$ we get equation \ref{eq:TLpp}.

\section{Decomposition of $T^L_\VEC{h}$ in spherical tensors}
\label{sec:App2}

In this section we want to retrieve, starting from equation \ref{eq:TLtensor}, equation \ref{eq:Tdecomposition}. Let us consider the object 
$T^L_\VEC{h}(\VEC{x})$ as defined in \ref{eq:TLa}, which can actually be constructed by contracting all the components of rank $L$ of tensor $T^L_{i_1,i_2,\ldots i_L}$ (equation \ref{eq:TLtensor})
with unit $\VEC{h}$ vectors. Fourier transforming this object gives us simply
\begin{align}
 T^L_\VEC{h}(\VEC{k})=\frac{m a^{-3}}{\bar{\rho}}\int{d^3p f(\VEC{k},\VEC{p})\left( \frac{\VEC{h}\cdot\VEC{p}}{am}\right)^L}.
 \label{eq:TLak}
\end{align}
Since we have translation symmetry it follows 
\begin{align}
 \left<\VEC{x}\right.\left|\VEC{x}+\VEC{r}\right>=\xi(\VEC{r})\qquad\Rightarrow\qquad\left<\VEC{k}\right.\left|\VEC{k}'\right>=(2\pi)^3P(\VEC{k})\df^D(\VEC{k}-\VEC{k}'),\nonumber
\end{align}
where it is implied that we take correlation of a general function of similar form like equation \ref{eq:TLa}. This enables us to work with each Fourier mode separately, 
and add them appropriately in the end when we discuss the power spectrum. By symmetry of the problem we may choose a reference frame where $z$-axis is along $\VEC{h}$ vector.
Since spherical harmonics form a complete set of orthonormal functions and thus form an orthonormal basis of the Hilbert space of square-integrable functions, we can expanded $f(\VEC{k},\VEC{p})$
in that frame as a linear combination,
\begin{align}
 f(\VEC{k},p,\theta,\phi)=\sum_{l=0}^\infty\sum_{m=-l}^{m=l}f^m_l(\VEC{k},p)Y_{lm}(\theta,\phi),
\label{eq:fkseries}
\end{align}
where $p$ is the amplitude of momentum. Let us now consider the transformation properties of $f^m_l(\VEC{k},p)$ under rotation around the $z$-axis. We can think of 
rotation by some angle $\psi$ (i.e. $\phi'=\phi+\psi$) in two ways
\begin{align}
 f(\VEC{k},p,\theta,\phi')=f(\VEC{k},p,\theta,\phi+\psi)&=\sum_{l=0}^\infty\sum_{m=-l}^{m=l}f^m_l(\VEC{k},p)'Y_{lm}(\theta,\phi)\nonumber\\
                                                        &=\sum_{l=0}^\infty\sum_{m=-l}^{m=l}f^m_l(\VEC{k},p)Y_{lm}(\theta,\phi')\nonumber.
\end{align}
From rotation properties of spherical harmonics it follows that $f^m_l(\VEC{k},p)$ transform as
\begin{align}
 f^m_l(\VEC{k},p)'=e^{im\psi}f^m_l(\VEC{k},p),
\end{align}
so it is an eigenstate of an helicity operator $i\partial/\partial\phi$, with helicity eigenvalue, or simply helicity, $m$. A quantity with helicity 0 is
called a scalar, that with helicity $m = ±1$ is called a vector and that with $m = ±2$ a tensor, but the
expansion goes to arbitrary values of $m$. It is possible to do similar considerations for arbitrary rotation and so it can be shown that $f^m_l$ transform as spherical tensors.

In the chosen reference frame, using $(\VEC{h}\cdot\VEC{p})=p_z=p~\text{cos}\theta$, and inserting equation \ref{eq:fkseries} into equation 
\ref{eq:TLak} we obtain
\begin{align}
 T^L_\VEC{h}(\VEC{k})&=\frac{m a^{-3}}{\bar{\rho}}\sum_{l,m}\int{dp~p^2f^m_l(\VEC{k},p)(p/am)^L}\int{d\Omega~Y_{lm}(\theta,\phi)\text{cos}^L\theta}\nonumber\\
                     &=\sum_{l,m}T^{L,m}_l(\VEC{k})I_{lm}\df_{m0},
\label{eq:TLaizvodza}
\end{align}
where we have defined helicity eigenstates of moments of the distribution function
\begin{align}
 T^{L,m}_l(\VEC{k})=4\pi\frac{m a^{-3}}{\bar{\rho}}\int{dp~p^2(p/am)^Lf^m_l(\VEC{k},p)},
\label{eq:TLml}
\end{align}
and used abbreviation for the integral
\begin{align}
 I_{lm}=\frac{1}{2}\sqrt{\frac{2l+1}{4\pi}\frac{(l-m)!}{(l+m)!}}\int{dx~x^LP^m_l(x)}.
 \label{eq:Ilm}
\end{align}
We have used definition of spherical harmonics $Y_{lm}(\theta,\phi)=\sqrt{\frac{2l+1}{4\pi}\frac{(l-m)!}{(l+m)!}}P^m_l(\text{cos}\theta)e^{im\phi}$, and $x=\text{cos} \theta$ abbreviation.
Since in equation \ref{eq:TLaizvodza} we have Kronecker delta $\df_{m0}$ it is easy to evaluate integral
\begin{align}
 I_{l0}&=\frac{1}{2}\sqrt{\frac{2l+1}{4\pi}}\int^1_{-1}{dx~x^LP_l(x)}=\frac{(-1)^l L!}{2^{l+1}(l!)^2}\sqrt{\frac{2l+1}{4\pi}}\int^1_{-1}{dx~x^{L-l}(x^2-1)^l}\nonumber\\
 &=\begin{cases}
        \sqrt{\frac{2l+1}{4\pi}}\frac{n^L_l}{2}(1+(-1)^{L-l}) & \text{if } l\leq L,\\
        0 & \text{if }  l > L,
   \end{cases}
\label{eq:Ilocomputation}
\end{align}
and here we have used $n^L_l=\frac{1}{2^{l+1}}\dbinom{L}{l}\frac{\Gamma(l+1)\Gamma\left(\frac{1}{2}(L-l+1)\right)}
{\Gamma\left(\frac{1}{2}(L+l+3)\right)}$. Collecting all that, equation \ref{eq:TLaizvodza} becomes
\begin{align}
 T^L_\VEC{h}(\VEC{k})=\sum_{(l=L,L-2,\ldots)}\sqrt{\frac{2l+1}{4\pi}}n^L_lT^{L,0}_l(\VEC{k}).
\label{eq:TLaza}
\end{align}
Since we are working in $\hat{z}\pp \VEC{h}$ frame it is apparent that for some arbitrary $\VEC{k}$, the angular dependence is contained in $T^{L,0}_l(\VEC{k})$ spherical tensors.
The goal now is to disentangle the angular dependence from the radial. The procedure depends on whether one is using active or passive interpretation of rotation transformation. 
Let us first look at active interpretation. Then the completely contracted tensors, like the one we are dealing with equation \ref{eq:TLaza}, can be obtained from the same one evaluated in 
$\VEC{k}_z=k\hat{z}\pp\VEC{h}$ direction by rotating it in general $\VEC{k}$ direction. Because $T^{L,m}_l$ are spherical tensors it follows
\begin{align}
 T^L_\VEC{h}(\VEC{k})=\mathcal{D}(R)T^L_\VEC{h}(R^{-1}\VEC{k})
 &=\sum_{(l=L,L-2,\ldots)}\sqrt{\frac{2l+1}{4\pi}}n^L_l\mathcal{D}(R)T^{L,0}_l(k\hat{z})\nonumber\\
 &=\sum_{(l=L,L-2,\ldots)}\sum^{m=l}_{m=-l}\sqrt{\frac{2l+1}{4\pi}}n^L_l\mathcal{D}^{(l)}_{0m}(R)T^{L,m}_l(k\hat{z}),
 \label{eq:TaLrotation}
\end{align}
where $\mathcal{D}(R)$ is the Wigner rotation matrix and $\mathcal{D}^{(l)}_{m'm}(R)$ its matrix elements. Using the well known relations, $\mathcal{D}^{(l)}_{m'm}(R^{-1})=\mathcal{D}^{(l)*}_{mm'}(R)$ and
$\mathcal{D}^{(l)}_{m0}(\phi,\theta,0)=\sqrt{\frac{4\pi}{2l+1}}Y^*_{lm}(\theta,\phi)$, we get
\begin{align}
 T^L_\VEC{h}(\VEC{k})=\sum_{(l=L,L-2,\ldots)}\sum^{m=l}_{m=-l}n^L_lT^{L,m}_l(k)Y_{lm}(\theta,\phi),
 \label{eq:TLazk}
\end{align}
where spherical harmonics now describe rotation form direction $\VEC{k}$ back to $\VEC{h}$. On the other hand, using the passive interpretation we argue that $T^L_\VEC{h}(\VEC{k})$ in frame $z\pp\VEC{h}$ can
be obtained by rotating it from $z'\pp\VEC{k}$ frame, which is described by the same equations as before \ref{eq:TaLrotation}. Note now that we were able to express result in terms of $T^{L,m}_l(k)$, 
just a function of amplitude $k$, and all angular dependence is given with spherical harmonics which now describe the angular dependence in $\VEC{h}$ direction seen from $z'\pp\VEC{k}$ frame. 
Now setting simply $\VEC{h}=\hat{r}$ along a line of sight direction we get result \ref{eq:Tdecomposition}, where $\text{cos}\theta=\hat{r}\cdot\VEC{k}/k=\mu$.

Finally we show that in decomposition form \ref{eq:TLazk} we retrieve the same number of independent components $(L+1)(L+2)/2$ as inferred from symmetries of equation \ref{eq:TLtensor},
since for even $L$, i.e. $l=2n$ we have
\begin{align}
 \sum^{L/2}_{n=0}(2l+1)=\sum^{L/2}_{n=0}(4n+1)=1+\frac{L}{2}+4\frac{\frac{L}{2}(\frac{L}{2}+1)}{2}=\frac{(L+1)(L+2)}{2},\nonumber
\end{align}
and for odd $L$, i.e. $l=2n+1$, we have
\begin{align}
 \sum^{(L-1)/2}_{n=0}(2l+1)=\sum^{(L-1)/2}_{n=0}(4n+3)=3+3\frac{L-1}{2}+4\frac{\frac{L-1}{2}(\frac{L-1}{2}+1)}{2}=\frac{(L+1)(L+2)}{2}.\nonumber
\end{align}

\section{Conjugation properties of $P_{LL'}(\VEC{k})$}\label{sec:App3}

In this section we investigate the conjugation properties of $P_{LL'}(\VEC{k})$ functions. Starting from the condition
that overdensity field $\df(\VEC{x})$ and velocity field $\VEC{v}(\VEC{x})$ are real valued fields, it 
follows that for Fourier space fields $\df(\VEC{k})$, $\theta(\VEC{k})$ and $v_\pp(\VEC{k})$ we have $f^*(\VEC{k})=f(-\VEC{k})$.
This is valid also for more complex fields like 
\begin{align}
  p_n(\VEC{k})=\int{\frac{d^3q_1d^3q_2\ldots d^3q_n}{(2\pi)^{3n}}f_1(\VEC{q}_1)f_2(\VEC{q}_2)\ldots f_n(\VEC{q}_n)\df^D(\VEC{k}-\VEC{q}_1-\VEC{q}_1-\ldots-\VEC{q}_n)}.
\label{eq:App3.0}
\end{align}
If we compute conjugated field we get
\begin{align}
  p^*_n(\VEC{k})&=\int{\frac{d^3q_1d^3q_2\ldots d^3q_n}{(2\pi)^{3n}}f^*_1(\VEC{q}_1)f^*_2(\VEC{q}_2)\ldots f^*_n(\VEC{q}_n)\df^D(\VEC{k}-\VEC{q}_1-\VEC{q}_1-\ldots-\VEC{q}_n)}\nonumber\\
                &=\int{\frac{d^3q_1d^3q_2\ldots d^3q_n}{(2\pi)^{3n}}f_1(-\VEC{q}_1)f_2(-\VEC{q}_2)\ldots f_n(-\VEC{q}_n)\df^D(\VEC{k}-\VEC{q}_1-\VEC{q}_1-\ldots-\VEC{q}_n)}\nonumber\\
                &=p_n(-\VEC{k}).
 \label{eq:App3.1}
\end{align}
From equation \ref{eq:TLaArox} it follows that $T^{*L}_\pp(\VEC{k})=T^{L}_\pp(-\VEC{k})$,  so for correlator we have
\begin{align}
 \left<T^{L}_\pp(\VEC{k})\right.\left|T^{*L'}_\pp(\VEC{k}')\right>=\left<T^{L'}_\pp(-\VEC{k})\right.\left|T^{*L}_\pp(-\VEC{k}')\right>
 =\left<T^{L'}_\pp(\VEC{k})\right.\left|T^{*L}_\pp(\VEC{k}')\right>^*,
 \label{eq:App3.2}
\end{align}
thus we have $P_{LL'}(\VEC{k})=P^*_{L'L}(\VEC{k})$. So, for sum of two correlator we can write
\begin{align}
 \left<T^{L}_\pp(\VEC{k})\right.\left| T^{*L'}_\pp(\VEC{k}')\right>+\left<T^{L'}_\pp(\VEC{k})\right.\left|T^{*L}_\pp(\VEC{k}')\right>
 =2\text{Re}\left<T^{L}_\pp(\VEC{k})\right.\left|T^{*L'}_\pp(\VEC{k}')\right>.
 \label{eq:App3.4}
\end{align}

\section{Integrals $I(k)$ and $J(k)$}
\label{sec:App4}

Here we define integrals $I_{nm}(k)$ and $J_{nm}(k)$ used in previous chapters:
\begin{align}
     I_{nm}(k)=\int {\frac{d^3q}{(2\pi)^3}~f_{nm}(\VEC{k},\VEC{q}) P_L(q)P_L(|\VEC{k}-\VEC{q}|)}
    \qquad\text{and}\qquad
     J_{nm}(k)=\int {\frac{dq^3}{(2\pi)^3}~g_{nm}\left(\frac{q}{k}\right)\frac{P_L(q)}{q^2}},
\end{align}
where we define kernels $f_{nm}(\VEC{k},\VEC{q})$, and use $r=q/k$ and $x=\VEC{k}\cdot\VEC{q}/(kq)$:

\begin{center}
   \begin{tabular}{p{8cm} p{8cm} }

$f_{00}(\VEC{k},\VEC{q})=\left(\frac{7x+3r-10rx^2}{14r(1+r^2-2rx)}\right)^2$,   &   $f_{01}(\VEC{k},\VEC{q})=\frac{\left(7x+3r-10rx^2\right)\left(7x-r-6rx^2\right)}{(14r(1+r^2-2rx))^2}$,\\ 
$f_{10}(\VEC{k},\VEC{q})=\frac{x\left(7x+3r-10rx^2\right)}{14r^2(1+r^2-2rx)}$,  &  $f_{11}(\VEC{k},\VEC{q})=\left(\frac{7x-r-6rx^2}{14r(1+r^2-2rx)}\right)^2$,\\ 
$f_{02}(\VEC{k},\VEC{q})=\frac{(x^2-1)(7x+3r-10rx^2)}{14r(1+r^2-2rx)^2}$,       &  $f_{20}(\VEC{k},\VEC{q})=\frac{(2x+r-3rx^2)(7x+3r-10rx^2)}{14r^2(1+r^2-2rx)^2}$,\\ 
$f_{12}(\VEC{k},\VEC{q})=\frac{(x^2-1)(7x-r-6rx^2)}{14r(1+r^2-2rx)^2}$,         &  $f_{21}(\VEC{k},\VEC{q})=\frac{\left(2x+r-3rx^2\right)\left(7x-r-6rx^2\right)}{14r^2(1+r^2-2rx)^2}$,\\ 
$f_{22}(\VEC{k},\VEC{q})=\frac{x\left(7x- r- 6rx^2\right)}{14r^2(1+r^2-2rx)}$,  &  $f_{03}(\VEC{k},\VEC{q})=\frac{(1-x^2)(3rx-1)}{r^2(1+r^2-2rx)}$,\\ 
$f_{30}(\VEC{k},\VEC{q})=\frac{1-3x^2-3rx+5rx^3}{r^2(1+r^2-2rx)}$,              &  $f_{31}(\VEC{k},\VEC{q})=\frac{(1-2rx)(1-x^2)}{2r^2(1+r^2-2rx)}$,\\ 
$f_{13}(\VEC{k},\VEC{q})=\frac{4rx+3x^2-6rx^3-1}{2r^2(1+r^2-2rx)}$,             &  $f_{23}(\VEC{k},\VEC{q})=\frac{3(1-x^2)^2}{(1+r^2-2rx)^2}$,\\ 
$f_{32}(\VEC{k},\VEC{q})=\frac{(1-x^2)(2-12rx-3r^2+15r^2x^2)}{r^2(1+r^2-2rx)^2}$,  &  $f_{33}(\VEC{k},\VEC{q})=\frac{-4+12x^2+24rx-40rx^3+3r^2-30r^2x^2+35r^2x^4}{r^2(1+r^2-2rx)^2}$.\\ 
   \end{tabular}
\end{center}

Also we have kernels $g_{nm}(r)$:
\begin{align}
    &g_{00}(r)=\frac{1}{3024}\left(\frac{ 12}{r^2}-158+100r^2-42r^4+\frac{3}{r^3}\left(r^2-1\right)^3\left(7r^2+2\right)\ln{\left[\frac{r+1}{\left|r-1\right|}\right]}\right),\nonumber\\
    &g_{01}(r)=\frac{1}{3024}\left(\frac{ 24}{r^2}-202+ 56r^2-30r^4+\frac{3}{r^3}\left(r^2-1\right)^3\left(5r^2+4\right)\ln{\left[\frac{r+1}{\left|r-1\right|}\right]}\right),\nonumber\\
    &g_{10}(r)=\frac{1}{1008}\left(               -38+  48r^2-18r^4+\frac{9}{r}(r^2-1)^3\ln\left[\frac{r+1}{\left|r-1\right|}\right]\right),\nonumber\\
    &g_{11}(r)=\frac{1}{1008}\left(\frac{ 12}{r^2}- 82  +4r^2- 6r^4+\frac{3}{r^3}\left(r^2-1\right)^3\left(  r^2+2\right)\ln{\left[\frac{r+1}{\left|r-1\right|}\right]}\right),\nonumber\\
    &g_{02}(r)=\frac{1}{ 224}\left(\frac{2}{r^2} \left(r^2+1\right)\left(3 r^4-14 r^2+3\right)-\frac{3}{r^3} \left(r^2-1\right)^4 \ln \left[\frac{r+1}{|r-1|}\right]\right),\nonumber\\
    &g_{20}(r)=\frac{1}{ 672}\left(\frac{2}{r^2}  \left(9-109 r^2+63 r^4-27 r^6\right)+\frac{9}{r^3} \left(r^2-1\right)^3 \left(3 r^2+1\right)\ln \left[\frac{r+1}{|r-1|}\right]\right),
\end{align}
and all the rest vanish in next to leading order regime.

\bigskip

\bibliographystyle{JHEP}
\bibliography{Bib}

\providecommand{\href}[2]{#2}\begingroup\raggedright\begin{thebibliography}{10}

\bibitem{Kaiser:1987qv}
N.~Kaiser, {\it {Clustering in real space and in redshift space}},  {\em Mon.
  Not. Roy. Astron. Soc.} {\bf 227} (1987) 1--27.

\bibitem{Hamilton:1997zq}
A.~Hamilton, {\it {Linear redshift distortions: A Review}},
  \href{http://xxx.lanl.gov/abs/astro-ph/9708102}{{\tt astro-ph/9708102}}.
  Published in The Evolving Universe. Edited by D. Hamilton, Kluwer Academic,
  1998, p. 185-275.

\bibitem{Cole:1993kh}
S.~Cole, K.~B. Fisher, and D.~H. Weinberg, {\it {Fourier analysis of redshift
  space distortions and the determination of Omega}},  {\em
  Mon.Not.Roy.Astron.Soc.} {\bf 267} (1994) 785,
  [\href{http://xxx.lanl.gov/abs/astro-ph/9308003}{{\tt astro-ph/9308003}}].

\bibitem{White:2008jy}
M.~White, Y.-S. Song, and W.~J. Percival, {\it {Forecasting Cosmological
  Constraints from Redshift Surveys}},  {\em Mon.Not.Roy.Astron.Soc.} {\bf 397}
  (2008) 1348--1354, [\href{http://xxx.lanl.gov/abs/0810.1518}{{\tt
  arXiv:0810.1518}}].

\bibitem{McDonald:2008sh}
P.~McDonald and U.~Seljak, {\it {How to measure redshift-space distortions
  without sample variance}},  {\em JCAP} {\bf 0910} (2009) 007,
  [\href{http://xxx.lanl.gov/abs/0810.0323}{{\tt arXiv:0810.0323}}]. * Brief
  entry *.

\bibitem{Bernstein:2011ju}
G.~M. Bernstein and Y.-C. Cai, {\it {Cosmology without cosmic variance}},
  \href{http://xxx.lanl.gov/abs/1104.3862}{{\tt arXiv:1104.3862}}.

\bibitem{Amara:2006kp}
A.~Amara and A.~Refregier, {\it {Optimal Surveys for Weak Lensing Tomography}},
   {\em Mon.Not.Roy.Astron.Soc.} {\bf 381} (2007) 1018--1026,
  [\href{http://xxx.lanl.gov/abs/astro-ph/0610127}{{\tt astro-ph/0610127}}].

\bibitem{Casarini:2012qj}
L.~Casarini, S.~A. Bonometto, S.~Borgani, K.~Dolag, G.~Murante, {\em et.~al.},
  {\it {Tomographic weak lensing shear spectra from large N-body and
  hydrodynamical simulations}},  \href{http://xxx.lanl.gov/abs/1203.5251}{{\tt
  arXiv:1203.5251}}.

\bibitem{Scoccimarro:2004tg}
R.~Scoccimarro, {\it {Redshift-space distortions, pairwise velocities and
  nonlinearities}},  {\em Phys.Rev.} {\bf D70} (2004) 083007,
  [\href{http://xxx.lanl.gov/abs/astro-ph/0407214}{{\tt astro-ph/0407214}}].

\bibitem{Taruya:2010mx}
A.~Taruya, T.~Nishimichi, and S.~Saito, {\it {Baryon Acoustic Oscillations in
  2D: Modeling Redshift-space Power Spectrum from Perturbation Theory}},  {\em
  Phys.Rev.} {\bf D82} (2010) 063522,
  [\href{http://xxx.lanl.gov/abs/1006.0699}{{\tt arXiv:1006.0699}}].

\bibitem{Jennings:2010uv}
E.~Jennings, C.~M. Baugh, and S.~Pascoli, {\it {Modelling redshift space
  distortions in hierarchical cosmologies}},  {\em Mon.Not.Roy.Astron.Soc.}
  {\bf 410} (2011) 2081, [\href{http://xxx.lanl.gov/abs/1003.4282}{{\tt
  arXiv:1003.4282}}].

\bibitem{Tang:2011qj}
J.~Tang, I.~Kayo, and M.~Takada, {\it {Likelihood reconstruction method of
  real-space density and velocity power spectra from a redshift galaxy
  survey}},  \href{http://xxx.lanl.gov/abs/1103.3614}{{\tt arXiv:1103.3614}}.

\bibitem{Tinker:2006dm}
J.~L. Tinker, {\it {Redshift-Space Distortions with the Halo Occupation
  Distribution II: Analytic Model}},  {\em Mon.Not.Roy.Astron.Soc.} {\bf 374}
  (2007) 477--492, [\href{http://xxx.lanl.gov/abs/astro-ph/0604217}{{\tt
  astro-ph/0604217}}].

\bibitem{Nishimichi:2011jm}
T.~Nishimichi and A.~Taruya, {\it {Baryon Acoustic Oscillations in 2D II:
  Redshift-space halo clustering in N-body simulations}},  {\em Phys.Rev.} {\bf
  D84} (2011) 043526, [\href{http://xxx.lanl.gov/abs/1106.4562}{{\tt
  arXiv:1106.4562}}].

\bibitem{Reid:2011ar}
B.~A. Reid and M.~White, {\it {Towards an accurate model of the redshift space
  clustering of halos in the quasilinear regime}},
  \href{http://xxx.lanl.gov/abs/1105.4165}{{\tt arXiv:1105.4165}}.

\bibitem{Sato:2011qr}
M.~Sato and T.~Matsubara, {\it {Nonlinear Biasing and Redshift-Space
  Distortions in Lagrangian Resummation Theory and N-body Simulations}},  {\em
  Phys.Rev.} {\bf D84} (2011) 043501,
  [\href{http://xxx.lanl.gov/abs/1105.5007}{{\tt arXiv:1105.5007}}].

\bibitem{Bernardeau:2001qr}
F.~Bernardeau, S.~Colombi, E.~Gaztanaga, and R.~Scoccimarro, {\it {Large scale
  structure of the universe and cosmological perturbation theory}},  {\em
  Phys.Rept.} {\bf 367} (2002) 1--248,
  [\href{http://xxx.lanl.gov/abs/astro-ph/0112551}{{\tt astro-ph/0112551}}].

\bibitem{Crocce:2005xy}
M.~Crocce and R.~Scoccimarro, {\it {Renormalized cosmological perturbation
  theory}},  {\em Phys.Rev.} {\bf D73} (2006) 063519,
  [\href{http://xxx.lanl.gov/abs/astro-ph/0509418}{{\tt astro-ph/0509418}}].

\bibitem{Crocce:2005xz}
M.~Crocce and R.~Scoccimarro, {\it {Memory of initial conditions in
  gravitational clustering}},  {\em Phys.Rev.} {\bf D73} (2006) 063520,
  [\href{http://xxx.lanl.gov/abs/astro-ph/0509419}{{\tt astro-ph/0509419}}].

\bibitem{Crocce:2007dt}
M.~Crocce and R.~Scoccimarro, {\it {Nonlinear Evolution of Baryon Acoustic
  Oscillations}},  {\em Phys.Rev.} {\bf D77} (2008) 023533,
  [\href{http://xxx.lanl.gov/abs/0704.2783}{{\tt arXiv:0704.2783}}].

\bibitem{Matsubara:2007wj}
T.~Matsubara, {\it {Resumming Cosmological Perturbations via the Lagrangian
  Picture: One-loop Results in Real Space and in Redshift Space}},  {\em
  Phys.Rev.} {\bf D77} (2008) 063530,
  [\href{http://xxx.lanl.gov/abs/0711.2521}{{\tt arXiv:0711.2521}}].

\bibitem{Matsubara:2008wx}
T.~Matsubara, {\it {Nonlinear perturbation theory with halo bias and
  redshift-space distortions via the Lagrangian picture}},  {\em Phys.Rev.}
  {\bf D78} (2008) 083519, [\href{http://xxx.lanl.gov/abs/0807.1733}{{\tt
  arXiv:0807.1733}}].

\bibitem{McDonald:2006hf}
P.~McDonald, {\it {Dark matter clustering: a simple renormalization group
  approach}},  {\em Phys.Rev.} {\bf D75} (2007) 043514,
  [\href{http://xxx.lanl.gov/abs/astro-ph/0606028}{{\tt astro-ph/0606028}}].

\bibitem{Taruya:2007xy}
A.~Taruya and T.~Hiramatsu, {\it {A Closure Theory for Non-linear Evolution of
  Cosmological Power Spectra}},  \href{http://xxx.lanl.gov/abs/0708.1367}{{\tt
  arXiv:0708.1367}}.

\bibitem{Pietroni:2008jx}
M.~Pietroni, {\it {Flowing with Time: a New Approach to Nonlinear Cosmological
  Perturbations}},  {\em JCAP} {\bf 0810} (2008) 036,
  [\href{http://xxx.lanl.gov/abs/0806.0971}{{\tt arXiv:0806.0971}}].

\bibitem{Valageas:2003gm}
P.~Valageas, {\it {A new approach to gravitational clustering: a path-integral
  formalism and large-n expansions}},  {\em Astron.Astrophys.} {\bf 421} (2004)
  23--40, [\href{http://xxx.lanl.gov/abs/astro-ph/0307008}{{\tt
  astro-ph/0307008}}].

\bibitem{Taruya:2009ir}
A.~Taruya, T.~Nishimichi, S.~Saito, and T.~Hiramatsu, {\it {Non-linear
  Evolution of Baryon Acoustic Oscillations from Improved Perturbation Theory
  in Real and Redshift Spaces}},  {\em Phys.Rev.} {\bf D80} (2009) 123503,
  [\href{http://xxx.lanl.gov/abs/0906.0507}{{\tt arXiv:0906.0507}}].

\bibitem{Seljak:2011tx}
U.~Seljak and P.~McDonald, {\it {Distribution function approach to redshift
  space distortions}},  \href{http://xxx.lanl.gov/abs/1109.1888}{{\tt
  arXiv:1109.1888}}.

\bibitem{Okumura:2011pb}
T.~Okumura, U.~Seljak, P.~McDonald, and V.~Desjacques, {\it {Distribution
  function approach to redshift space distortions: N-body simulations}},
  \href{http://xxx.lanl.gov/abs/1109.1609}{{\tt arXiv:1109.1609}}.

\bibitem{Peebles:1994xt}
P.~J.~E. Peebles, {\it {Principles of physical cosmology}}, . Princeton, USA:
  Univ. Pr. (1993) 718 p.

\bibitem{Desjacques:2008vf}
V.~Desjacques, U.~Seljak, and I.~Iliev, {\it {Scale-dependent bias induced by
  local non-Gaussianity: A comparison to N-body simulations}},
  \href{http://xxx.lanl.gov/abs/0811.2748}{{\tt arXiv:0811.2748}}.

\bibitem{Carlson:2009it}
J.~Carlson, M.~White, and N.~Padmanabhan, {\it {A critical look at cosmological
  perturbation theory techniques}},  {\em Phys. Rev.} {\bf D80} (2009) 043531,
  [\href{http://xxx.lanl.gov/abs/0905.0479}{{\tt arXiv:0905.0479}}].

\bibitem{Tassev:2011ac}
S.~Tassev and M.~Zaldarriaga, {\it {The Mildly Non-Linear Regime of Structure
  Formation}},  \href{http://xxx.lanl.gov/abs/1109.4939}{{\tt
  arXiv:1109.4939}}.

\bibitem{Eisenstein:1997ik}
D.~J. Eisenstein and W.~Hu, {\it {Baryonic Features in the Matter Transfer
  Function}},  {\em Astrophys. J.} {\bf 496} (1998) 605,
  [\href{http://xxx.lanl.gov/abs/astro-ph/9709112}{{\tt astro-ph/9709112}}].

\bibitem{McDonald:2009hs}
P.~McDonald, {\it {How to generate a significant effective temperature for cold
  dark matter, from first principles}},  {\em JCAP} {\bf 1104} (2011) 032,
  [\href{http://xxx.lanl.gov/abs/0910.1002}{{\tt arXiv:0910.1002}}].

\bibitem{Seljak:2000}
U.~Seljak, {\it {Analytic model for galaxy and dark matter clustering}},  {\em
  Mon.Not.Roy.Astron.Soc.} {\bf 318} (2000) 203,
  [\href{http://xxx.lanl.gov/abs/astro-ph/0001493}{{\tt astro-ph/0001493}}].

\bibitem{Peacock:2000qk}
J.~Peacock and R.~Smith, {\it {Halo occupation numbers and galaxy bias}},  {\em
  Mon.Not.Roy.Astron.Soc.} {\bf 318} (2000) 1144,
  [\href{http://xxx.lanl.gov/abs/astro-ph/0005010}{{\tt astro-ph/0005010}}].

\bibitem{Ma:2000yt}
C.-P. Ma and J.~N. Fry, {\it {What does it take to stabilize gravitational
  clustering?}},  \href{http://xxx.lanl.gov/abs/astro-ph/0005233}{{\tt
  astro-ph/0005233}}.

\bibitem{Berlind:2002rn}
A.~A. Berlind, D.~H. Weinberg, A.~J. Benson, C.~M. Baugh, S.~Cole, {\em
  et.~al.}, {\it {The Halo occupation distribution and the physics of galaxy
  formation}},  {\em Astrophys.J.} {\bf 593} (2003) 1--25,
  [\href{http://xxx.lanl.gov/abs/astro-ph/0212357}{{\tt astro-ph/0212357}}].

\bibitem{Cooray:2002dia}
A.~Cooray and R.~K. Sheth, {\it {Halo models of large scale structure}},  {\em
  Phys.Rept.} {\bf 372} (2002) 1--129,
  [\href{http://xxx.lanl.gov/abs/astro-ph/0206508}{{\tt astro-ph/0206508}}].

\bibitem{Seljak:2002zr}
U.~Seljak, {\it {Constraints on galaxy halo profiles from galaxy-galaxy lensing
  and tully-fisher/fundamental plane relations}},  {\em
  Mon.Not.Roy.Astron.Soc.} {\bf 334} (2002) 797,
  [\href{http://xxx.lanl.gov/abs/astro-ph/0201450}{{\tt astro-ph/0201450}}].

\bibitem{Sheth:1999mn}
R.~K. Sheth and G.~Tormen, {\it {Large scale bias and the peak background
  split}},  {\em Mon.Not.Roy.Astron.Soc.} {\bf 308} (1999) 119,
  [\href{http://xxx.lanl.gov/abs/astro-ph/9901122}{{\tt astro-ph/9901122}}].

\bibitem{Okumura:2012xh}
T.~Okumura, U.~Seljak, and V.~Desjacques, {\it {Distribution function approach
  to redshift space distortions, Part III: halos and galaxies}},
  \href{http://xxx.lanl.gov/abs/1206.4070}{{\tt arXiv:1206.4070}}.

\end{thebibliography}\endgroup

\end{document}